\documentclass[aps,superscriptaddress,twocolumn,showpacs,preprintnumbers,amsmath,amssymb,showkeys,textcomp]{revtex4}
\usepackage[cp1251]{inputenc}
\usepackage{amssymb,amsmath,wasysym}
\usepackage[mathscr]{eucal}
\usepackage{graphicx}
\usepackage{longtable}
\usepackage[dvips]{hyperref} 
\usepackage[cp1251]{inputenc}
\usepackage[english]{babel}
\usepackage{amssymb,amsmath}
\usepackage{amstext} 
\usepackage[dvips]{hyperref}
\usepackage[mathscr]{eucal}
\usepackage{longtable}
\usepackage{graphicx}
\begin{document}
\title{Quantum  Field Effects in Stationary Electron Spin Resonance Spectroscopy}
\author{Dmitri Yerchuck (a), Vyacheslav Stelmakh (b), Yauhen Yerchak (b),  Alla Dovlatova   (c)\\
\textit{(a) - Heat-Mass Transfer Institute of National Academy of Sciences of RB, Brovka Str.15, Minsk, 220072,  dpy@tut.by \\(b) - Belarusian State University, Nezavisimosti Ave., 4, Minsk, 220030, RB \\(c) M.V.Lomonosov Moscow State University, Moscow, 119899}}
\date{\today}
\begin{abstract}   It is proved experimentally on the example of electron spin resonance (ESR) studies of anthracites of medium-scale metamorphism, that by  strong electron-photon and electron-phonon interactions the  formation of the coherent system of the resonance phonons, that is, the transformation of a noncoherent microwave field into a coherent hypersound field takes place. 

The acoustic quantum  Rabi oscillations  were  observed for the first time in  ESR-spectroscopy.  Its Rabi  frequency value on the first  damping stage was found to be equal 920.6 kHz, being to be independent on the microwave power level in the range 20 - 6 dB [0 dB corresponds to 100  mW]. By the  subsequent increase of  the microwave power the stepwise transition to the new quantum phenomenon takes place,  just to the  phenomenon of nonlinear quantum  Rabi oscillations, characterised by splitting of the oscillation  group of lines  into two subgroups with doubling of the  total  lines' number and approximately the twofold decrease of linewidth of an  individual oscillation line, became to be equal the only  to $0.004 \pm 0.001$ G, that is, they are the most narrow lines, being registered by  the stationary ESR spectroscopy at all.  

  Along with the absorption process of EM-field energy the  emission process was observed in the standard registration regime with autofrequency control (AFC), at that the  absorption process was observed by the sweep of a static magnetic field from  the lower to  the higher values ($\frac{dH_0}{dt} > 0$) and the emission process was observed by $\frac{dH_0}{dt} < 0$. It was found, that  the emission process is the realization of  the acoustic spin resonance, the source of acoustic wave power in which is the  system of resonance phonons, accumulated in the samples by the registration  with AFC. 

It has been found, that the lifetime of coherent state of a collective subsystem of resonance phonons in anthracites  is very long and even by room temperature it is evaluated by the value exceeding 4.6 minutes.

The  model of new kinds of instantons was proposed. They are considered to be similar in the mathematical structure to Su-Schrieffer-Heeger solitons with "propagation" direction along time $t$-axis instead of space $z$-axis. The instantons of the first kind absorb microwave photons and relax mainly by  means of the generation of coherent hypersound phonons. The given prevailing coherent relaxation process determines rather narrow ESR-linewidth (equal to 0.18 G) in structurally disordered anthracite samples  to be the direct consequence of its coherence.   The instantons of the second kind absorb hypersound phonons and relax mainly by the generation of the coherent microwave radiation.

The proof that the superconductivity state in the anthracite samples studied is  produced  at the  room temperature in ESR conditions in the accordance with the theory of the quantised acoustic field,  has  experimentally been  obtained.

\end{abstract}
\maketitle 
\section{Introduction and Background}

The studies of optical properties of a number of condensed matter systems have shown, that there are phenomena, which can be explained the only within the frames of the concept of a quantized electromagnetic field (EM-field) interacting with given systems, which also have to be described quantum-mechanically. The brief review of the given phenomena along with recovering of the photon status to be genuine elementary particle and the proposal on the revival of  photons by absorption processes has been represented in \cite{PSPA}.

 Further,
the QED model for a multichain coupled
 qubit system proposed in \cite{Dovlatova_Yearchuck} has predicted that by strong electron-photon interaction the quantum nature of  EM-field can
 become apparent in any stationary optical experiment. In particular, a new quantum optics phenomenon — Rabi waves and Rabi
corpuscular-wave packet formation — which was theoretically predicted  in \cite{Slepyan_Yerchak} can give an essential contribution in the stationary spectral distribution of Raman scattering (RS) intensity,
 spectral distributions of infrared (IR), visible, or ultraviolet
 absorption, reflection, and transmission intensities. The given prediction was confirmed in \cite{Yerchuck_D_Dovlatova_A}, where the
additional lines
 corresponding to the Fourier transform of the revival
 part of  the time dependence of integral inversion, the detection
 of which in stationary RS measurements is determined by the
 formation and propagation of quantum Rabi corpuscular-wave packets, being to be
 a consequence of the quantum nature of the EM-field, have been registered and identified.

We have, however, to remark, that  the semiclassical theory can explain many  phenomena, observed in modern optics and in modern radiospectroscopy too. Moreover, in a number of the cases a quantum field description and a semiclassical description give the same results. It refers, for instance, to the work of \cite{Gromov} on so-called CW ESR (continuous wave electron spin resonance) spectroscopy, where the authors showed theoretically and experimentally
that sidebands in CW EPR spectra, which emerge by using of a radiofrequency radiation field by the detection of ESR signals, are
multiple photon transitions. 
The authors describe the process of a multiple photon transition
semiclassically in a toggling frame and by
using the function space of a Hamiltonian including the
quantized microwave and radiofrequency radiation fields.  It has been shown, that  the fully quantized description
 is equivalent to the semiclassical description with the Hamiltonian
in the k-th toggling frame, in which microwave and radiofrequency radiation fields are considered classically.

The known phenomena of the  display of the quantum nature of an EM-field, which cannot be described  semiclassically [spontaneous emission, Lamb
shift, Casimir effect,  entaglement and sqeezing of the states, laser emission linewidth broadening, complete statistics of laser photons \cite{Scully}, the emergence of longlived coherent states of absorbing centers \cite{Yerchuck_D_Dovlatova_A}, accompanying by the appearance of the
additional lines in stationary measurements] are referred  to an optical range of light frequencies. At the same time, in a microwave range of the  frequencies the phenomena with the  display of the quantum nature of an EM-field (which cannot be described  semiclassically) by the EM-field interaction with a matter were not established (to the best of our knowledge). It is displaying even in the terminology, for instance, in the term CW [continuous wave] ESR  spectroscopy. 

We will show  in the given work, that the effects of the EM-field quantization can be substantial in a microwave range of the  frequences too, indicating also, that the commonly accepted term CW ESR is not very well turned.

To understand, when the effects of the quantization are essential, let us consider some details of the quantization procedure, proposed in \cite{Dovlatova_Yerchuck} on the example, for instance, the time ($t$) local quantization.

It was supposed, that  EM-field is inside of a volume rectangular cavity, which is made up of perfectly electrically conducting walls. It was supposed also, that the field is linearly polarized and without restriction of a commonness it was choosed the one of two possible polarizations of an  EM-field electrical component $\vec{E}(\vec{r},t)$ along $x$-direction. Then the vector-function $E_x(z,t) \vec{e}_x$  was represented in the well known form of Fourier sine series
\begin{equation}
\label{eq1ab}
\vec{E}^{}(\vec{r},t) = 
E_x(z,t) \vec{e}_x = \left[\sum_{\alpha=1}^{\infty}A^{E}_{\alpha}q_{\alpha}(t)\sin(k_{\alpha}z)\right]\vec{e}_1 ,
\end{equation}
where $q_{\alpha}(t)$ is the amplitude of $\alpha$-th normal mode of the cavity, $\alpha \in N$, $k_{\alpha} = \alpha\pi/L$, $A^{E}_{\alpha}=\sqrt{2 \omega_{\alpha}^2m_{\alpha}/V\epsilon_0}$, $\omega_{\alpha} = \alpha\pi c/L$, $L$ is a cavity length along z-axis, $V$ is a cavity volume, $\epsilon_0 $ is the permittivity of the vacuum, $m_{\alpha}$ is parameter, which is introduced to obtain the analogy with mechanical harmonic oscillator, $\vec{e}_1$ is  the unit vector along x-coordinate direction.  Particular sine case of Fourier  series is the consequence of boundary conditions 
\begin{equation}
\label{eq1abc}
[\vec{n} \times\vec{E}]|_S = 0, (\vec{n} \vec{H})|_S = 0,
\end{equation}
which are held true for a perfect cavity considered. Here $\vec{n}$ is the normal to the surface $S$ of the cavity. 
There were obtained for the operator quantised field functions of  the electrical  and magnetic components $\hat{\vec{E}}^{}(\vec{r},t)$ and $\hat{\vec{H}}^{}(\vec{r},t)$ of  EM-field the following expressions
 
\begin{equation}
\label{eq34ab}
\begin{split}
&\hat{\vec{E}}^{}(\vec{r},t) = \\
&\{\sum_{\alpha=1}^{\infty} \sqrt{\frac{\hbar \omega_{\alpha}}{V\epsilon_0}} \left[\hat{a}{}^{+}_{\alpha}(t) + \hat{a}{}_{\alpha}(t)\right] \sin(k_{\alpha} z)\} \vec{e}_1,
\end{split}
\end{equation}
\begin{equation}
\label{eq35ab}
\begin{split}
&\hat{\vec{H}}^{}(\vec{r},t) = \\
&\{\sum_{\alpha=1}^{\infty} \sqrt{\frac{\hbar \omega_{\alpha}}{V\mu_0}} (-i) \left[\hat{a}{}^{+}_{\alpha}(t) - \hat{a}{}_{\alpha}(t)\right] \cos(k_{\alpha} z)\} \vec{e}_2,
\end{split}
\end{equation}
where $\hat{a}{}^{+}_{\alpha}(t)$, $\hat{a}{}_{\alpha}(t)$ are operators of a photon creation and an annihilation respectively, $\mu_0$ is the magnetic permeability of the vacuum, $\vec{e}_2$ is unit vector along $y$-coordinate direction. 

It is seen, that the quantised operator field vector-functions $E_x(z,t) \vec{e}_x$ and $\hat{\vec{H}}^{}(\vec{r},t)$ are the functions of wave vector $\vec{k}$. At the same time $\vec{k}$-space is a discontinuite space, which is the result of a finiteness of a cavity volume and correspondingly the necessity of the expansion in Fourier series instead of  Fourier integral expansion is  determined by the given  discontinuity of $\vec{k}$-space.  

Let us consider the discontinuity  extent of $\vec{k}$-space. It can be obtained from the expression  $k_{\alpha, L} = \alpha\pi/L$, considered to be the function of $\alpha$, and $L$. It is the following 

\begin{equation}
\label{eq36ab}
\begin{split}
\delta k_{\alpha, L} = \frac{\alpha\pi}{L}\delta\alpha - \frac{\alpha\pi}{L^2}\delta L,
\end{split}
\end{equation}
where $\delta L$ is the small increase of the cavity length, the minimal change of the value  $\alpha$ is equal to 1.
Hence we see, that  the discontinuity of $\vec{k}$-space is increasing by  cavity length decreasing, that is, if $\delta L = -|\delta L|$. If the change of $L$ is not small, then the value of $k_{\alpha, L}$ in dependence on $L$ is 
\begin{equation}
\label{eq36abg}
\begin{split}
&k_{\alpha, L}  = \alpha\pi[\ln{L}\Delta\alpha + \frac{1}{L}] + C ,\\
&C = - \alpha\pi\ln{L_0}\Delta\alpha, 
\end{split}
\end{equation} 
where the change of the number $\alpha$, equaled to $\Delta\alpha$ by the discreteness change is taken into consideration, $L_0$ is the initial cavity length.
Since the dependencies of $\delta k_{\alpha, L}$ and $k_{\alpha, L}$  on $L$ are substantional, we can expect that the quantization of EM-field will be also substantional in the microwave range, if the cavity length for the same frequency can be strongly diminished. It is possible in the case, when an electron-photon interaction is rather strong. The formation of an additional  cavity by a sample itself in the  measuring cavity of an ESR spectrometer can be an experinental criterion for the notion of the strong electron-photon interaction in the radiospectroscopy (see for details further).

 We have undertaken looking-for of appropriate samples for the given aim among carbon systems. Two factors have determined the given choice - our experience of the ESR study of a number carbon and carbon related systems and the theoretical consideration of the structure of EM-field in \cite{DAA}. The fundamental result, obtained by Dirac, that  the dynamical system, which consists of the ensemble of identical bosons is equivalent to the dynamical system, which consists of the ensemble of oscillators, was used in \cite{DAA}, in order to show, that the presence of the scalar  charge function $\rho(\vec{r},t)$ to be the peer scalar (pseudoscalar) characteristic of an electromagnetic  field along with vector force characteristics $\vec{E}(\vec{r},t)$, $\vec{H}(\vec{r},t)$ [that was established in  \cite{Dovlatova_Yerchuck}] agrees well with the charge neutrality of photons. The simplest analogue in the mathematical description of EM-field in the physics of condensed matter is the chain of bosonic (spin $S = 1$) carbon atoms, being to be the frame in \textit{trans}-polyacetylene structure. It has been shown, that 
neutral photons are topological relativistic solitons with nonzero spin value, which is equal to $\frac{1}{2}$ instead of the prevalent viewpoint,
that the photons possess by spin $S = 1$.   
It was argued, by the way, that 
the representation of  photons to be the result of the spin-charge separation effect in the rest massless "boson-atomic" structure of EM-field  makes substantially more clear the origin of the corpuscular-wave dualism \cite{DAA}, which is naturally explained by the  existence of corpuscles - photons, propagating along EM-field rays [light rays in the visible range] and by the own structure of the propagation medium - the rest massless "boson-atomic" structure of rays, where the Bloch-like waves can be formed.
Especially significant is that, that the  all aspects of the model are experimentally  entirely confirmed, see for details \cite{ABE}.

Let us remark, that the  affinity of "boson-atomic" structure of EM-field and carbon structures, especially carbon chain based structures, allows to understand qualitatively the key role just of the carbon in the formation of organic compounds and biosystems. The given formation seems to be supporting by EM-field, since  the  affinity of "boson-atomic" structure of EM-field and carbon structures can lead to a very high effectiveness of an interaction of EM-field with matter in the case of a carbon. In fact, the well known principle, which is popular in scientific media of chemists - similar substance is dissolved in similar substance - is realized.
 
Our looking-for  of appropriate systems with the strong electron-photon interaction in the radiospectroscopy range, taking into account the aforeindicated affinity of the structure of EM-field and carbon structures  was successful. The appropriate carbon system has been found, it was the anthracite. The study of the physical properties of anthracites represents also an
independent interest, since they are investigated very weak. 

It was found, that the samples in the form of parallelepiped with the length of the side along EM-field propagation direction, equaled to 0.3 cm and more, produce an additional cavity. Since the discreteness step of  $\vec{k}$-space was increased according to (\ref{eq36ab}) [the length of the cavity is 4.6 cm] in  15.3 times by the sample  length of the side along EM-field propagation direction, equaled to 0.3 cm [the evaluation has been done for the same value of $\alpha$], it has to be expected the display of quantum nature of a microwave field. It is really so, see the next Sections.

   Let us also remark, that the affinity of the structure of EM-field and a matter, having both the corpuscular structures, gives also  the physical substantiation of Slater principle.
Given general principle was proposed by Slater  already in 1924. It is – "Any atom
may in fact be supposed to communicate with other atoms all the
time it is in stationary state, by means of virtual field of radiation,
originating from oscillators having the frequencies of possible
quantum transitions. . ." \cite{Slater}. It means, that  interatomic bonds are supported by appropriate EM-field quanta and the presence of the given support is determining the main physicochemical properties of condensed matter substances. The application of Slater principle was  used by the analytical solution of the task of the interaction of a quantized EM-field with an axially symmetric quasi-1D multichain qubit system in \cite{Dovlatova_Yearchuck}, \cite{Yerchuck_D_Dovlatova_A} and in implicit form in \cite{SSH}, \cite{SSH_PRB}, \cite{Heeger_1988},  (see comment in \cite{Yerchuck_D_Dovlatova_A}).
 
The choise of an anthracite to be the subject for our ESR studies was also connected with a runaway concernment of QED and quantum field  theory (QFT) at all
 for  practical applications. They, in fact, become to be working instruments,  for instance, in an industrial spectroscopy control, in the
connection with  quickly developing new quantum physics branches like to cavity
quantum electrodynamics \cite{Berman}, in elaborations of quantum logic systems, including quantum computing \cite{Nielsen} and in many branches of nanotechnology.
 
It is significant for practical applications to have the systems, which allow to obtain quantum
bits  with long coherence times. 
The  results, presented by Loss and DiVincenzo in \cite{DiVincenzo} are the theoretial basic results, giving the algorithm to encode quantum
information using the spin states of semiconductor quantum dots.
Then were  understood that encoding of quantum
information using the spin states gives
 the key criteria \cite{DiVincenzoD} for quantum computation. In other words, they really can be natural
building blocks for quantum computation. It was shown, that for instance, it can lead to the
realization of 2-qubit operations, \cite{Petta}, \cite{Nowack}, \cite{Shulman}. However, the semiconductor
quantum dot qubits possess  by the limited lifetime and associated fidelity
of the quantum state. So, the coherence time,  determined by the time of the spin dephasing $ T_{2*}$ = 37 ns \cite{Koppens}, were improved to  $ T_{2*}$ = 94 ns (\cite{BluhmH}) using nuclear
spin bath control for quantum dot spin qubits
in GaAs/AlGaAs. A longer  $ T_{2*}$ = 360 ns has been achieved using
Si/SiGe quantum dots \cite{Maune}.

It is known, that 
 spin-based solid state quantum bits in the corrresponding single
crystals  are characterized by  more long coherence times in comparison with guantum dots,
while also offering the promise of scalability, and they are suggested to be natural
building blocks for quantum computation. Phosphorus donor nuclei in silicon have been known
since the 1950s to have some of the best spin coherence
properties in solids. The spin coherence time $T_2$ measured by   Hahn spin echo method for
donor electron spins in bulk Si:P has been reported to be equal
$\sim 60 ms$ \cite{Tyryshkin}. The given spin coherence time was till recently  the longest coherence
time measured in electron spin qubits in bulk mateials, and it greatly
exceeds the value reported, for instance, in GaAs quantum dots, which, being to be measured also by   Hahn spin echo method, is found to be equal to $ 30 \mu$s \cite{Koppens}. 

We wish to draw attention of the readershipt, that the coherence time is dependent on the pulse sequences technique developed for a bulk
magnetic resonance with the aim to measure the coherence times. It  can  be specified a $ T_{2}$  according to the
applied pulse sequence \cite{Dzurak}. Using a Hahn echo sequence the coherence
time of GaAs-based qubits has been extended to $ T^H_{2}$  = 440 ns
\cite{Koppens}, with  $ T^H_{2}$ = 30 $\mu$s achieved via pulse optimization \cite{Bluhm}, while
the use of a Carr-Purcell-Meiboom-Gill (CPMG) pulse sequence
has enabled $ T^{CPMG}_{2}$ = 200 $\mu$s \cite{Bluhm}. The physical nature of the given effect can be understandable, if to take into account, that the relaxation decoherence time of individual paramagnetic center (PC) is dependent on its coupling with the surrounging medium, and the various pulse sequences act upon given coupling in different ways.

By realizing of a quantum
dot qubits in isotopically enriched silicon $^{28}Si$, the dephasing
effect of the nuclear spin bath, presenting in the previous studies, was removed,
and it has been  shown \cite{Dzurak} that all  the  coherence times, earlier obtained, can be improved
by orders of magnitude. These long coherence times, in particular
the dephasing time $ T_{2*}$, lead to low control error rates and the
high fidelities that will be required for large-scale, fault-tolerant
quantum computing.

There is now much interest in fabricating of
spin-based devices in diamond, with potential applications in practically the same industry branches (quantum communication, quantum computation, and additionally in magnetometry).
Nitrogen-vacancy (NV) centers  appear promising  solid-state spin
qubits since they combine optical initialization and readout capabilities owing to also long electron spin coherence life times. They are more short, $\sim 1 ms$, in comparison with those ones in bulk Si:P, but they are realized 
 at room temperature \cite{Gaebel}, and there is an
ability to control coupling to individual nuclear spins.
However, it is necessary or at least advantageous to
couple NV centers to optical structures like to waveguides and microresonators, to enable
communication between distant qubits or to allow efficient extraction of emitted photons. Therefore, a reliable
method is needed to create NV centers with good spectral
properties in close proximity $(\apprle 100 nm)$ to a diamond
surface. In addition, the charge state of the NV centers
must be controlled.  The given tasks are the only under study at present.

We can indicate that exciting progress towards spin-based quantum computing has recently been 
made with qubits realized using nitrogen-vacancy centres in diamond and 
phosphorus atoms in silicon by another way. For example, long coherence times were made 
possible by the presence of spin-free isotopes of silicon   and carbon.
For  silicon, enrichment of the spin-zero $^{28}Si$ 
isotope drastically reduces spin-bath decoherence \cite{Witzel}. Experiments on bulk spin 
ensembles in $^{28}Si$ crystals have indeed demonstrated extraordinary coherence 
times \cite{Steger}, \cite{TyryshkinA}, \cite{Saeedi}. The similar results were obtained for carbon \cite{Bar-Gill}.
 However, 
despite promising single-atom nanotechnologies, there remained  till recently substantial 
challenges in coupling of qubits and addressing them individually.
Conversely, 
lithographically produced quantum dots have an exchange coupling that can be 
precisely engineered, but strong coupling to noise has severely limited their 
dephasing times and control fidelities. 

Two research teams working in the same laboratories have found distinct 
solutions to a critical challenge that has held back the realization of super 
powerful quantum computers. The teams created two types of quantum bits, or "qubits" 
-- the building blocks for quantum computers -- that each process quantum data 
with an accuracy above 99 percents \cite{Dzurak}, \cite{Morello}. 

The best aspects of 
both spin qubit schemes above described were combined in \cite{Dzurak}. The authors formed  gate-addressable quantum dot qubits in 
isotopically engineered silicon with a control fidelity of 99.6 percents, obtained via 
Clifford-based randomized benchmarking and consistent with that required for 
fault-tolerant quantum computing.  The qubit  dephasing time $ T_{2*}$  = 120 ${\mu}s$ 
and coherence time $T_2$ = 28 ms, both 
 larger than in other 
types of semiconductor qubits.
In \cite{Dzurak} was demonstrated a qubit that can be addressed and
tuned via a simple gate voltage.  The highly tunable quantum dot presented
in \cite{Dzurak} allows  to vary the internal electric field  resulting in a Stark shift that can tune the electron
spin resonance (ESR) frequency by  more than 8 MHz. Also, the long $ T_{2*}$ available
in isotopically enriched silicon results in a narrow ESR linewidth the only
of 2.4 kHz according to conclusons of authors. We have to remark that the  linewidth value was obtained indirectly, by means of Fourier transform of the first damping stage of quantum oscillations observed and it does not correspond to the  linewidth, if it would be registered by the stationarry ESR, see for details \cite{PSPA}.  According to \cite{Dzurak}, the qubit operation frequency [the electron g*-factor]   can be tuned 
by more than 3000 times. These results,
together with the inherent scalability of gated quantum dot qubits,
open the possibility for large-scale and gate-voltage-addressable
qubit systems that are compatible with existing 
manufacturing technologies. 

We wish  to give the short comments, concerning  ESR line parameters  along withthose ones represented in \cite{PSPA} also here. The relation  of the transient ESR free induction decay   envelope to the electron spin resonance line shape was obtained mathematically through a  Fourier transform. It seems to be correct in the case when the decay function f(t) satisfy to the conditions of the applicability of the  Fourier transform. Let us remember that they are the following: the decay function f(t) has to be summed up, the spectral function $f(\omega$) has to be bounded, uniformly continuous function, and $f(\omega) \rightarrow 0$ if $|\omega| \rightarrow \infty$. It is clear, that the Fourier analysis in  \cite{Dzurak} and in \cite{Morello} is correct the only in the case of classical Rabi oscillations and it becomes to be incorrect when Rabi oscilations have quantum character, that really can be concluded from the results presented in  \cite{Dzurak} and in \cite{Morello}.

 It is clear, that the Fourier analysis  is correct  in the case of usual spectroscopic transitions, realised by functions, being to be continuous functions on space coordinates and time,  the decay time in which is determined by ESR free induction decay being to be the result of the spin-spin and spin-lattice relaxation mechanisms, leading to the  transition of the absorbing system from an excited into ground state.  However, in the case of spectroscopic transitions under affect of the quantised EM-field  the decay process is discrete, that is, it is realised by quantum Rabi oscillations  and its connection with the shape of the ESR absorption lines becomes nontrivial, owing to its discreteness both in the time and in the space \cite{Slepyan_Yerchak}. Qualitative aspect of the description  of the line shape can be obtained by space averaging of propagating quantum Rabi oscillations. Then, we can restrict themselves to the consideration of the effect expected within the frames of the Jaynes-Cummings model (JCM)
\cite{Jaynes_Cummings}. It has been done in \cite{PSPA} and it has to be taken into account by the analysis of the results, presented in \cite{Dzurak} and in \cite{Morello}. We wish to accentuate that quantum character of Rabi oscillations
of coherent systems leads to the super-Lorentzian shape of the absoption lines, in distinction from the treatment in  \cite{Dzurak} and in \cite{Morello}. It was established theoretically in \cite{PSPA} and it has the experimental confirmation. In particular,
the  super-Lorentzian shape has been observed in natural diamond samples of type
Ia and IIa implanted with high energy ions of copper (63 MeV), neon (26.7 MeV), and nickel (335 MeV)  along the $\langle{111}\rangle$,  $\langle{100}\rangle$ and $\langle{110}\rangle$ crystallographic directions, \cite{Erchak}, \cite{Ertchak}, \cite{Ertchak_Stelmakh}.

When concern the practical applications, the team led by Dzurak \cite{Dzurak} has discovered a way to create an "artificial atom" 
qubit with a device remarkably similar to the silicon transistors used in 
consumer electronics - the metal-oxide-semiconductor field-effect transistors that is like to those, which are used in  laptops and phones. At the same time
MOSFETs constitute today's computer processors.

Meanwhile, Morello's team has been pushing the "natural" phosphorus atom qubit 
to the extremes of performance. The phosphorus atom 
contains in fact two qubits: the electron qubit and the nucleus qubit. With the nucleus qubit in 
particular, it  was achieved accuracy close to 99.99 percents. That means only one error 
for every 10000 quantum operations.

The given experiments are among the first in solid-state, and the first-ever in 
silicon, to fulfill the requirement of method effectiveness, which is only guaranteed if the errors occur less than 1 percent of the time. 

The high-accuracy operations for both natural and artificial atom qubits are  
achieved by placing each inside a thin layer of specially purified silicon, 
containing only the silicon-28 isotope. This isotope is perfectly non-magnetic 
and, unlike those in naturally occurring silicon, does not disturb the quantum 
bit.

Morello's research team also established a world-record "coherence time" for a 
single quantum bit held in solid state \cite{Morello}. Since the coherence time is a measure of how long quantum information
 can  be preserved  before it's lost, it means, the longer 
the coherence time, the easier it becomes to perform long sequences of 
operations, and therefore more complex calculations. 

The team was able to store quantum information in a phosphorus nucleus for more 
than 30 seconds \cite{Morello}. 
Let us represent some details of the given achievement. 

The spin of an electron or a nucleus in a semiconductor naturally implements 
the unit of quantum information — the qubit. In addition, because semiconductors 
are currently used in the electronics industry, developing qubits in 
semiconductors is a promising route to realize scalable quantum 
information devices. The solid-state environment, however, may provide 
deleterious interactions between the qubit and the nuclear spins of surrounding 
atoms, or charge and spin fluctuations arising from defects in oxides and 
interfaces. It was indicated above that for materials like to silicon or carbon enrichment of the spin-zero 
isotope drastically reduces spin-bath decoherence \cite{Witzel} 
 \cite{Steger}, \cite{TyryshkinA}, \cite{Saeedi}, \cite{Bar-Gill}. However, it remained unclear whether these would persist at the 
single-spin level, in gated nanostructures near amorphous interfaces. Just, in  the work \cite{Morello} is
presented the coherent operation of individual $^{31}P$ electron and nuclear spin 
qubits in a top-gated nanostructure, fabricated on an isotopically engineered  $^{28}Si$ 
substrate. The $^{31}P$ nuclear spin sets the new benchmark coherence time any single qubit in the solid 
state. We accentuate once again, that it was more than 30 s and  it was established by using
 Carr–Purcell–Meiboom–Gill (CPMG) impulse sequence in transient ESR experiments and reaches more than 99.99 percents  control fidelity above indicated. The electron spin CPMG coherence 
time exceeds 0.5 s, and detailed noise spectroscopy \cite{Alvarez} indicates that — contrary to 
widespread belief — it is not limited by the proximity to an interface \cite{Morello}. Instead,  according to the opinion of the authors of \cite{Morello}
decoherence is probably dominated by thermal and magnetic noises, being external to the 
device, and is thus amenable to further improvement. 
The world-record narrow ESR-lines, corresponding to coherent state of phosphorus donor PC with the linewidth value equaled to the only 1.8 kHz were obtained in \cite{Morello}. However,  the line contour with Gaussian shape and with the linewidth value aforesaid
 was obtained by Fourier transform of the free induction decay  envelope on finite time segment. Our analysis of the data presented in \cite{Morello} shows that they have observed quantised transient process and the result of the Fourier transform performed in accordance with the aforego;g remark does not correspond to the genuine line contour, which  would be obtained by usual stationary registration method \cite{PSPA}.

At the same time, all the  measurements of both groups were performed in a dilution refrigerator with a base very low
temperature of $T \approx$ 50 mK and even more low, that reduces strongly their significance and value for the practical usage.

It was predicted in the work \cite{QFTDST}, that there is the quite other way to obtain long-lived coherent states with the similar field of practical applications. The prediction is based on the joint quantum electrodynamics  and quantum deformation field (phonon field) theory, that is, in fact on quantum field theory of spectroscopic transitions, which represent itself the development of the quantum  theory, since it is beyond  the scope of probabilistic Born treatise of the quantum mechanics \cite{Born_M}. The only the given version of the quantum mechanics is used  at present, sometimes beyond the limits of  its applicability, see for instance examples, commented in \cite{PQMD}.

 The work \cite{QFTDST} represents the development of the 
known models, which capture the
salient features of the relevant physics in  the given  field. They are
the Jaynes-Cummings model (JCM) \cite{Jaynes_Cummings} for an one qubit
case and its generalization for multiqubit systems by
Tavis and Cummings \cite{Tavis}. Tavis-Cummings model was generalized in \cite{Slepyan_Yerchak} by taking into account  1D-coupling between qubits.   The QED-model for  the one chain coupled qubit system was generalized 
  for  the quasi-one-dimensional axially symmetric  multichain coupled qubit system  \cite{Dovlatova_Yearchuck}. 
  It is substantial, that in the model, proposed in \cite{Dovlatova_Yearchuck} the interaction of  the quantized EM-field with the multichain qubit system  is considered by taking into account both  intrachain and interchain qubit coupling without any restriction on their values.  

The work \cite{QFTDST}  in the aspect of the equations for the dynamics of spectroscopic transitions represents the most direct development of the works \cite{Yearchuck_Doklady} and \cite{Yearchuck_Yerchak_Dovlatova}. 
 It follows from theoretical results in \cite{Slepyan_Yerchak}, \cite{Dovlatova_Yearchuck} and from their experimental confirmation in 
 \cite{Yerchuck_D_Dovlatova_A}, that by the strong interaction of EM-field with matter the correct description of spectroscopic transitions, including the stationary spectroscopy, is achieved the only within the frames of  the full quantum consideration. It  concerns both optical and radio spectroscopies, that means, that  the full quantum consideration  has to be also undertaken by
 electron spin resonance  studies in the case of  the strong interaction of EM-field with spin systems. The results, obtained in \cite{QFTDST} show, that   the analogous conclusion can be
drawn for the case of the strong interaction of a phonon field with a spin system or an electron dipole system. In particular, it was theoretically shown, that  the relaxation of paramagnetic (or optical) centers in the case of the strong spin-phonon (electron-phonon) interaction can be described correctly the only within the framework of  the quantum field theory.

The aim of the work presented is to study experimentally quantum field effects in  the radiospectroscopy range, in particular, to confirm the prediction in \cite{QFTDST} of  the new quantum physics phenomenon - the formation of longlived coherent state of resonance phonons leading  to the appearance of quantum acoustic Rabi oscillations  and to propose the area of its  practical applications. 

\section{Experimental Technique}

ESR studies in anthracite samples of the  medium-scale metamorphism [in the own anthracite metamorphism scale, but not in the coal metamorphism scale, in which all anthracites are referred to the coals of high-scale metamorphism]  have been carried out. Let us remark, that the anthracite seems to be the individual carbon (on $\approx$ 99 percents) allotropic form, the structure of which is the only under study at present. [The suggestion on  the structure of anthracites, being to be the separate carbon allotropic form, follows from our experience of long-term studies of various carbon materials, and it is based, in particular, on their ESR characteristics, which  cannot be attributed to any known carbon allotropic form]. 

The ESR measurements were carried out on monolithic samples, pricked from the region of anthracite blocks with well-defined facetting in parallel with natural faces. Let us remark, that the angle between natural faces, which have appeared after a cleavage was usually about 15, 30, 60, 120 degrees. In some cases it was about 90 degrees. The samples with the most flat faces and with the angle in 90 degrees between them were selected. Additionally, filing with diamond broach file was used to obtain the samples in the form of parallelepipeds or cylinders.
 
ESR spectra were registered on X-band ESR-spectrometers   "RadioPAN SE/X-2544" both on the  standard spectrometer and the modified spectrometer, in  wich  a homodyne microwave channel was installed, allowing the phase-sensitive detection of the microwave signal. ESR spectra were registered at room temperature by using of $TE_{102}$ mode rectangular cavity.  The ruby standard sample was permanently placed in the cavity on its sidewall to realize the control procedure, proposed in \cite{Erchak} and consisting in the following. One of the  lines of ESR absorption by  $Cr^{3+}$ point paramagnetic centers (PC) in ruby  was used  for the correct relative intensity measurements of ESR absorption, for the calibration  of the relative amplitude value of magnetic component of the microwave field and for precise phase tuning of  a modulation field. The correct relative intensity measurements became to be  possible owing to unsaturating behavior of ESR absorption in ruby in the range of  the microwave power applied, which was  $\approx  100$ mW and $\approx 50$ mW in  modified and unmodified  spectrometers respectively in the absence of an attenuation. Unsaturable character of the absorption in a ruby standard was confirmed by means of the measurements of the absorption intensities in two identical ruby samples in dependence on the microwave power level. The first sample was standard sample,  permanently placed in the cavity, the second sample was placed in the cavity away from the loop of magnetic component of microwave field  so, that the intensity of the resonance line  was about 0.1 
of the intensity of corresponding line of the first sample. Both the samples were registered  simultaneously, however their absorption lines  were not overlapped owing to slightly different sample orientations. The  foregoing intensity ratio, equaled to 0.1,  was precisely preserved for all microwave power values in the range used. It follows hence, that really ruby samples are good standard samples in ESR spectroscopy studies. 

 The  100 kHz high frequency modulation of static magnetic field $H_0$ was used. The static magnetic field $H_0$ was sweeped in two directions from lower value to higher  value ($\frac{dH_0}{dt} > 0$) and from higher value to lower value ($\frac{dH_0}{dt} < 0$). Two regime of registration were used - with the automatic microwave frequency control (AF-mode) and without any automatic microwave frequency control, that is  by the constant microwave frequency (CF-mode), which was achieved by turning off of the  automatic frequency control (AFC) block. The frequency modulation of the microwave power with the value in 130 kHz was used in AFC unit. Its amplitude was less in all the range of microwave power used in approximately 3 times in modified "RadioPAN SE/X-2544" spectrometer in comparison with the standard "RadioPAN SE/X-2544" spectrometer.  Precise measurements of the static magnetic field value were realized by means of NMR (nuclear magnetic resonance) sensor, located in the interpole space, at that the projection of the center of the sensor-head  on the center of the cavity plane parallel to poles' surfaces was coinciding with the center of the sample location. The accuracy of the measurements  was determined by short-time stability of the  static magnetic field, which was rather high (better than $10^{-6}$). The change of the sensor-head position within the range of the most long side of the samples (not exceeding 7 mm) do not change NMR sensor readings indicating on the high homogeneity of the static field, allowing the measurements of g-values with an accuracy indicated in the paper (see further).

 Precise measurements of the  microwave frequency were realized by means of the digital microwave wavemeter, which was installed additionally to the standard "RadioPAN SE/X-2544" equipment. 

All the spectra were registered on the recorders representing themselves the standard blocks of the spectrometers above indicated without any computer digital conversion. 

\section{Quantum Field Description of Spectroscopic Transitions by Strong Electron-Photon and Electron-Phonon Coupling}

For the convenience of the wide readership, we present very briefly the results of the theory of spectroscopic transitions being to be applicable  without any restriction on the values of  electron-photon and electron-phonon coupling. In fact, the only the expression for the  Hamiltonian used and the equations themselves for the dynamics of spectroscopic transitions which are applicable also by strong electron-photon and electron-phonon coupling will be represented.
In the work \cite{Yearchuck_Yerchak_Dovlatova} the system of  difference-differencial equations for dynamics of spectroscopic transitions for both radio- and optical spectroscopy for the model, representing itself the 1D-chain of N two-level  equivalent elements coupled by exchange interaction (or its optical analogue for the optical transitions) between themselves and interacting with quantized EM-field and   quantized phonon field  has  been  derived. Naturally, the equations are true for any 3D system [that is, both for crystalline or non-crysralline 3D systems] of paramagnetic centers  or optical centers  by the absence of the exchange interaction. In the given case, the model presented  differs from Tavis-Cummings model \cite{Tavis} by the inclusion into consideration of the quantized phonon system, describing the relaxation processes from the quantum field theory position.  Seven equations for seven operator variables, describing a joint system \{field + matter\} were represented in the matrix form by three matrix equations. At the same time, the   Hamiltonian, given in \cite{Yearchuck_Yerchak_Dovlatova} describes in fact the only part of interaction with the phonon field, which corresponds  to $z$-component of the vector of the state. It is correct for the cases of relatively weak or moderate electron-phonon interactions.  

  However, it was shown in \cite{QFTDST}, that by strong electron-photon coupling and strong electron-phonon coupling a quite other picture of quantum relaxation processes becomes to be possible. Really, in \cite{QFTDST} is argued the following. The definition of the 
 wave function  of the chain system, interacting
with quantized EM-field  and with quantized lattice vibration field, to be the
vector of the state in Hilbert space over quaternion ring,  that is, the
quaternion function of the quaternion
argument, allowed to easily prove the Lorentz invariance of the quantum field operator equations for spectroscopic transitions, derived in \cite{Yearchuck_Yerchak_Dovlatova} and the possibility of the transfer to observables. In fact, in the work \cite{QFTDST}, the main role of  the spin vector  for the quantum state description, established in \cite{Yearchuck_Yerchak_Dovlatova} was taken into account. Since  the spin vector is the vector of the state [in Hilbert space over quaternion ring with the accuracy to a normalization factor] of a 1D quantum system, interacting with the quantized electromagnetic field \cite{Yearchuck_Yerchak_Dovlatova}, all the components of the vector of the state, that is, in  the equivalent description, the all components of  the spin vector, being to be peer components, have to be taken into consideration. It was done in  \cite{QFTDST}.

 Therefore, the following  Hamiltonian was obtained in a natural way
 \begin{equation}
\label{eq14}
\mathcal{\hat H} = \mathcal{\hat H}^C + \mathcal{\hat H}^F + \mathcal{\hat H}^{C F} + \mathcal{\hat H}^{Ph} + \mathcal{\hat H}^{CPh} ,
\end{equation} 
where 
${\mathcal{\hat H}^C}$ is  the chain Hamiltonian by the absence of the interaction with EM-field, ${\mathcal{\hat H}^F}$ and $ \mathcal{\hat H}^{Ph}$  are the photon and phonon field Hamiltonians correspondingly, ${\mathcal{\hat H}^{C F}}$ and  $\mathcal{\hat H}^{CPh}$ are, accordingly,  the Hamiltonians, describing the interaction between the quantized EM-field and an electronic subsystem of an atomic chain and  between the quantized phonon field and an electronic subsystem of an atomic chain.
Then the equations of the motion for spectroscopic transition operators $\hat {\vec {\sigma }}_l$,  for the quantized 
 EM-field operators $\hat{a}_{\vec k}$, $\hat{a}_{\vec k}^{ +}$ and for the phonon field operators $\hat{b}_{\vec q}$, $\hat{b}_{\vec q}^{ +}$  are the following.

\begin{equation}
\label{eq26}
\begin{split}
\raisetag{40pt}\frac{\partial}{\partial t} \left[\begin{array}{*{20}c}
{\hat\sigma^-_l}  \\
 \\
{\hat\sigma^+_l}  \\
\\
{\hat\sigma^z_l} 
\end{array} 
\right] = 2 \left\| g \right\| \left[\begin{array}{*{20}c}
{\hat F^-_l}  \\
 \\
{\hat F^+_l}  \\
\\
{\hat F^z_l} 
\end{array} 
\right] + ||\hat{R}^{(\lambda^z)}_{\vec{q}l}|| + ||\hat{R}^{(\lambda^\pm)}_{\vec{q}l}|| 
\end{split}
\end{equation}

\begin{equation}
\label{eq2}
\begin{split}
\raisetag{40pt}
&\frac{\partial}{\partial t} 
\left[\begin{array}{*{20}c}
 {\hat{a}_{\vec k^{}}} \\
 \\
 {\hat{a}_{\vec k^{}}^+} \\
\end{array} 
\right] = -i \omega_{\vec k^{}} ||\sigma_P^z|| \left[\begin{array}{*{20}c}
 {\hat{a}_{\vec k^{}}} \\
 \\ 
 {\hat{a}_{\vec k^{}}^+} \\
\end{array} 
\right] \\
\\
& + \frac{i}{\hbar}
\left[\begin{array}{*{20}c}
{-\sum\limits_{l = 1}^N (\hat\sigma_l^{+} + \hat\sigma_l^{-}) v_{l \vec k}^*} \\
\\
{\sum\limits_{l = 1}^N (\hat\sigma_l^{+} + \hat\sigma_l^{-}) v_{l \vec k}} \\
\end{array} \right],
\end{split}
\end{equation}

\begin{equation}
\label{eq30a}
\begin{split}
\raisetag{40pt}
&\frac{\partial}{\partial t} 
\left[
\begin{array}{*{20}c}
 {\hat{b}_{\vec k^{}}} \\
 \\
 {\hat{b}_{\vec q^{}}^+} \\
\end{array} 
\right] = -i \omega_{\vec q^{}} ||\sigma_P^z|| \left[\begin{array}{*{20}c}
 {\hat{b}_{\vec q^{}}} \\
 \\ 
 {\hat{b}_{\vec q^{}}^+} \\
\end{array} 
\right] 
 + \\
&\frac{i}{\hbar}
\left[
\begin{array}{*{20}c}
{-\sum\limits_{l = 1}^N \{\lambda^z_{\vec q l} \hat\sigma_l^{z} + \lambda^{\pm}_{\vec q l} (\hat\sigma_l^{+} + \hat\sigma_l^{-})\}} \\
\\
{\sum\limits_{l = 1}^N \{\lambda^z_{\vec q l} \hat\sigma_l^{z} + \lambda^{\pm}_{\vec q l} (\hat\sigma_l^{+} + \hat\sigma_l^{-})\}} \\
\end{array} \right],
\end{split}
\end{equation}
  where matrix $||\hat{R}^{(\lambda^z)}_{\vec{q}l}||$ is
\begin{equation}
\label{eq27}
\begin{split}
\raisetag{40pt}
||\hat{R}^{(\lambda^z)}_{\vec{q}l}|| = 
\frac{1}{i\hbar} \left[\begin{array}{*{20}c}{ 2 \hat{B}^{(\lambda^z)}_{\vec{q}l} \hat\sigma^-_l}  \\
 \\
{ -2 \hat{B}^{(\lambda^z)}_{\vec{q}l} \hat\sigma^+_l}  \\
\\
{0} \end{array} 
\right] 
\end{split}
\end{equation}
with  $\hat{B}^{(\lambda^z)}_{\vec{q}l}$, which is given by
\begin{equation}
\label{eq28}
\hat{B}^{(\lambda^z)}_{\vec{q}l} = \sum\limits_{\vec{q}}[(\lambda^z_{\vec{q}l})^* \hat{b}^{+}_{\vec{q}} + \lambda^z_{\vec{q}l} \hat{b}_{\vec{q}}].\end{equation} 
Matrix $||\hat{R}^{(\lambda^\pm)}_{\vec{q}l}||$ is
\begin{equation}
\label{eq29l}
\begin{split}
\raisetag{40pt}
||\hat{R}^{(\lambda^z)}_{\vec{q}l}|| = 
\frac{1}{i\hbar} \left[\begin{array}{*{20}c}{ -\hat{B}^{(\lambda^\pm)}_{\vec{q}l} \hat\sigma^z_l}  \\
 \\
{  \hat{B}^{(\lambda^\pm)}_{\vec{q}l} \hat\sigma^z_l}  \\
\\
{\hat{B}^{(\lambda^\pm)}_{\vec{q}l} (\hat\sigma^+_l - \hat\sigma^-_l)} \end{array} 
\right],  
\end{split}
\end{equation}
where $\hat{B}^{(\lambda^\pm)}_{\vec{q}l}$ is
\begin{equation}
\label{eq30j}
\hat{B}^{(\lambda^\pm)}_{\vec{q}l} = \sum\limits_{\vec{q}}[(\lambda^\pm_{\vec{q}l})^* \hat{b}^{+}_{\vec{q}} + \lambda^\pm_{\vec{q}l} \hat{b}_{\vec{q}}].\end{equation}

The parameters  $\lambda^z_{\vec q}$ and $\lambda^\pm_{\vec q}$ are the electron-phonon coupling constants, which characterise respectively the interaction of an electron subsystem of jth chain unit, corresponding to  $z$- component of its  vector of the state (or $S^z_j$) and the interaction of an electron subsystem of jth chain unit, corresponding to  $\pm$- components of its  vector of the state (or  $S^+_j$- and $S^-_j$ components of the spin of jth chain unit). It seems to be understandable, that they can be different in the general case. Moreover, in order to take into account the interaction with both equilibrium and nonequilibrium phonons  the electron-phonon coupling constants have to be complex numbers.

The operator $\hat{\vec{\sigma}}_l$ 
\begin{equation}
\label{eq4}
\begin{split}
\raisetag{40pt}
\left[\begin{array}{*{20}c}
{\hat\sigma^-_l}  \\
 \\
{\hat\sigma^+_l}  \\
\\
{\hat\sigma^z_l} 
\end{array} 
\right] = \hat{\vec{\sigma}}_l =  \hat\sigma^-_l  \vec e_ +  +  \hat\sigma^+_l \vec e_ - +  \hat\sigma^z_l\vec e_z
\end{split}
\end{equation} is the vector-operator of spectroscopic transitions for $l$th chain unit, $l = \overline{2,N-1}$ \cite{Yearchuck_Yerchak_Dovlatova}, which was represented in the matrix form.
Its components, that is,  the operators 
\begin{equation}
\label{eq6a}
{\hat\sigma_v}^{jm} \equiv {\left|j_v \right\rangle} {\left\langle m_v \right|} 
\end{equation} are set up in correspondence to the states ${\left|j_v \right\rangle}$, ${\left\langle m_v \right|}$, where $v = \overline{1,N}$, 
$j = \alpha, \beta$, $m = \alpha, \beta $. For instance, the relationships for commutation rules are
\begin{equation}
\label{eq9a}
[\hat {\sigma}_v^{lm}, \hat {\sigma}_v^{pq}] = \hat {\sigma }_v^{lq} \delta_{mp} - \hat {\sigma }_v^{pm}\delta_{ql}. 
\end{equation} 
Further
\begin{equation}
\label{eq5}
\begin{split}
\raisetag{40pt}
\left[\begin{array}{*{20}c}
{\hat F^-_l}  \\
 \\
{\hat F^+_l}  \\
\\
{\hat F^z_l} 
\end{array} 
\right] = \hat {\vec F} =  \left[ {\hat {\vec {\sigma}}_l \otimes \hat {\vec {\mathcal{G}}}_{l - 1,l + 1}} \right],
\end{split}
\end{equation}
where the vector operators $\hat {\vec {\mathcal{G}}}_{l - 1,l + 1}$,  $l = \overline{2,N-1}$, are given by the expressions
\begin{equation}
\label{eq6}
\hat {\vec {\mathcal{G}}}_{l - 1,l + 1} = \hat {\mathcal{G}}_{l - 1,l + 1}^-  \vec e_ +  + \hat {\mathcal{G}}_{l - 1,l + 1}^ +  \vec e_ - + \hat {\mathcal{G}}_{l - 1,l + 1}^z \vec e_z,
\end{equation}
in which 
\begin{subequations}
\label{eq7}
\begin{gather}
\hat {\mathcal{G}}_{l - 1,l + 1}^-  = -\frac{1 }{\hbar} \sum\limits_{\vec k}\hat{f}_{l \vec k} - \frac{J }{\hbar }(\hat\sigma _{l + 1}^- + \hat\sigma _{l - 1}^-) , \\
\hat {\mathcal{G}}_{l - 1,l + 1}^+ = -\frac{1}{\hbar} \sum\limits_{\vec k}\hat{f}_{l \vec k} - \frac{J}{\hbar }(\hat\sigma _{l + 1}^+ + \hat\sigma _{l - 1}^ + ), \\
\hat {\mathcal{G}}_{l - 1,l + 1}^z = - \omega_{l} - \frac{J}{\hbar }(\hat\sigma _{l + 1}^z + \hat\sigma _{l - 1}^z ).
\end{gather}
\end{subequations}
Here the operator $\hat{f}_{l \vec k}$ is
 \begin{equation}
\label{eq8}
\hat{f}_{l \vec k} = v_{l \vec k} \hat{a}_{\vec k} + \hat{a}_{\vec k}^{+} {v^*}_{l \vec k}.
\end{equation}
In the relations (\ref{eq7}) $J$ is the exchange interaction constant in the case of magnetic resonance transitions or its optical
analogue in the case of optical transitions, the function $v_{l \vec k}$ in (\ref{eq8}) is
\begin{equation}
\label{eq9}
 v_{l \vec k} = - \frac{1}{\hbar} p_l^{jm} (\vec e_{\vec k} \cdot \vec e_{\vec P_{l}}) \mathfrak{E}_{\vec k} e^{ - i \omega_{\vec k}t + i \vec k \vec r},
\end{equation}
where $p_l^{jm}$ is the matrix element of the operator of magnetic (electric) dipole moment $\vec P_{l}$ of $\textit{l-th}$ chain unit between the states $\left| {j_{l}} \right\rangle$ and 
$\left| m_{l} \right\rangle$ with $j \in \{\alpha, \beta\}$,  $m \in \{\alpha, \beta\}$, $j \neq m$, $\vec e_{\vec k}$ is the unit polarization vector, $\vec e_{\vec P_{l}}$ is the unit vector along $\vec P_{l}$-direction, $\mathfrak{E}_{\vec k}$ is the quantity, which has the dimension of a magnetic (electric) field strength, $\vec k$ is the quantized EM-field wave vector, the components of which get a discrete set of values, $\omega_{\vec k}$ is the frequency, corresponding to ${\vec k}$th mode of EM-field,  $\hat{a}^+_{\vec k}$ and $\hat{a}_{\vec k}$ are the EM-field  creation and  annihilation operators correspondingly. In the suggestion, that the contribution of a spontaneous emission is relatively small, $p_{l}^{jm} = p_{l}^{mj} \equiv p_{l} $, where $j \in \{\alpha, \beta\}$,  $m \in \{\alpha, \beta\}$, $j \neq m$, 
 $\hat{b}_{\vec q}^+$, $\hat{b}_{\vec q}$ are the creation and
annihilation operators of the phonon with impulse ${\vec q}$ and with
energy $\hbar \omega_{\vec q} $ correspondingly. In equations (\ref{eq2}) and (\ref{eq30a})
$\|\sigma_P^z\| $ is Pauli $z$-matrix,  $\left\| g \right\|$ in the equation (\ref{eq26}) is the diagonal matrix,
numerical values of its elements are dependent on the basis choice.

The right hand side expression  in (\ref{eq5}) is  the vector product of quantities, being to be vector operators.   It can be 
calculated in accordance with the expression
\begin{equation}
\label{eq12}
\left[ {\hat {\vec {\sigma}} _l \otimes \hat {\vec {\mathcal{G}}}_{l - 1,l + 1} } \right] = \frac{1}{2} \left| {\begin{array}{*{20}c}
 {\vec e_- \times \vec e_z} & {\hat{\sigma}_l^-} & {\hat {\mathcal{G}}_{\,\,l - 1,l + 1}^-} \\
 {\vec e_z \times \vec e_+} & {\hat{\sigma}_l^+} & {\hat {\mathcal{G}}_{\,\,l - 1,l + 1}^+} \\
 {\vec e_+ \times \vec e_-} & {\hat{\sigma}_l^z} & {\hat {\mathcal{G}}_{\,\,l - 1,l + 1}^z} \\
\end{array}} \right|',
\end{equation}
 that is, by using of known expression for an  usual vector product with the additional coefficient ${\frac{1}{2}}$ the only, which is appeared, since
the products of two components of two vector operators are replaced by anticommutators of corresponding 
components. The given detail is mapped by the symbol $\otimes$ in (\ref{eq5}) and by the symbol $'$ in determinant (\ref{eq12}).

Thus, the QFT model for the dynamics of spectroscopic transitions in a 1D multiqubit exchange coupled system was generalized by taking into account, that the  spin vector is proportional to the quaternion vector of the state of any relatiistic quantum system [the joint matter + EM-field system
has to be referred to relativistic systems] in Hilbert space
defined over quaternion ring. It leads to the requirement, that all the spin
components has to be taken into account.

The new quantum
phenomenon was predicted in \cite{QFTDST}. The prediction results
from the structure of the equations derived and it consists
in the following. The coherent system of the resonance
phonons, that is, the phonons with the energy, equaled
to the resonance photon energy can be formed by a resonance, that can lead to an appearance along with   Rabi oscillations determined by spin (electron)-photon coupling with the frequency $\Omega^{RF}$ of Rabi oscillations determined by spin (electron)-phonon coupling with the frequency $\Omega^{RPh}$. In other words, the QFT model prediction of the quantum oscillation character of the relaxation gives the new insight on the nature of relaxation processes in condensed matter. It has qualitatively different character in comparison with the classical exponential relaxation, described by phenomenological and semiclassical Bloch models. Moreover, if $\mid\lambda^{\pm}_{\vec q l}\mid < g_{\vec q l}$, the second Rabi oscillation process will be observed by the stationary state of two subsystems \{EM-field + magnetic (electric) dipoles\}, that is, it will be registered in quadrature with the first Rabi oscillation process.  It can be experimentally detected even by traditional stationary spectroscopy methods.

Let us remark, that the theory above represented is applicable for the systems, in which  the atomic vibrations can be described  within the frames of a phonon model. It seems to be essential for the work presented, that the notion of the phonon, introduced in the physics by Tamm I.E for the description of lattice vibrations in crystalline solids, can be applicable also for amorphous solids including glass structures for the acoustic vibrations, that is, just for the vibrations which are essential in ESR spectroscopy, see, for instance \cite{Ph.Enc.}. Moreover, the notion of the phonon is applicable for the description of vibrations in quantum liquids \cite{Ph.Enc.}. Therefore, the given theory can also be  applicable for any disordered optical or magnetic resonance systems,  electronic subsystems of which can be described within the frames of quantum liquids.

\section{Results and Discussion}

\subsection{Classification of Anthracites and Some ESR-Characteristics of  Small Volume Anthracite Samples}

It has been found, that physical and  physicochemical properties of anthracites are dependent on their metamorphism degree. Since  the classification of anthracites was absent, the task has been appeared to make up for a given deficiency. It seemed to be quite appropriate to use for the solution of the given task an ESR-method, being to be the most powerful method for the study of a structure of substances on an atomic and molecular level. Really, the ESR studies performed have shown, that there is the correlation of a metamorphism extent of anthracites with an extent of a display of Dyson (or Dyson-like) effect \cite{Dyson} and with its peculiarities.  It is known, that Dyson  effect can be characterized by the asymmetry parameter of resonance line $A/B$, where A and B are the values of amplitudes of low field and high field wings respectively of the derivative of a resonance line.  Dyson  effect is expressed relative;y weakly in  anthracites of the low  metamorphism degree, $A/B$ is of order of 2 or less, however not less than 1. The most strongly Dyson  effect is expressed formally in the usual form in anthracites of the medium-scale  metamorphism degree, $A/B$ is of order of 20 or even  more. In the samples of the high metamorphism degree,$A/B$ value  becomes to be less than 1. It means that asymmetry of the ESR lines, observed in the samples of the high metamorphism degree cannot be described within the frames of Dyson theory. In the given case we have the only Dyson-like effect, the detailed study of which is the subject of the subsequent studies.

Therefore, Dyson  effect and Dyson-like effect give the clear express method for the determination of the  metamorphism degree of anthracites and given effects can serve for their classification. 

Let us remark that above considered classification was fulfilled on the samples of the relatively small volume (less than 30 $mm^3$). It ensures  a smallness of the changes of Q-factor and the resonance frequency of a measuring cavity by the sweep of a direct current (DC) field through the resonance. Registered spectra were identical in the given case by the registration in both the AF- and CF-modes.
Nevertheless, even in the given case the anisotropy of the spectral characteristics relatively  the direction of magnetic component $H_1$ of microwave field has been established for the samples representing themselves the plates in their geometry.
So, for the square plate with the thickness 0.5 mm and with the other two side length in 5 mm the asymmetry extent $A/B$ = 2.4 $\pm 0.02$, if $H_1$ is transversely to plate surface. At the same time, if $H_1$ was parallel to plate surface the asymmetry extent $A/B$ was near 1. It was equal to  $1.06 \pm 0.02$ and $1.04 \pm 0.02$ for the plate surface position transversely and in parallel to the DC magnetic field $H_0$
correspondingly. Further, it has been observed an antisymbasis in the correlation 
of angular dependencies of an  asymmetry extent $A/B$ and a linewidth $\Delta H_{pp}$ (peak-peak value in the signal deriative). For instance, for above indicated sample  if $H_1$ was parallel to plate surface the linewidth $\Delta H_{pp}$ was equal to $0.43  \pm 0.01$  if $H_1$ was transversely to plate surface $\Delta H_{pp}$ was reduced in $1.7\pm 0.05$ times. All the measurements have been done in the given sample by the same microwave power value, equaled to 50 mW.

It is interesting that the similar anisotropy of ESR characteristics relatively $H_1$-direction was observed in carbynoid films \cite{Ertchak_J_Physics_Condensed_Matter} and in diamond single crystals modified by high energy ion implantation \cite{Ertchak}, \cite{Ertchak_Stelmakh}. It corresponds to the  appearance of ordering in electronic subsystems of   carbynoid films and in modified regions located near surface  in
diamond single crystals. It means that anthracite samples studied are characterised by some ordering in the electronic subsystem too. It can be suggested, that they represent themselves spin glass.

\subsection{Experimental Observation of  Quantum Rabi Oscillations in Stationary ESR Studies}

 The most interesting, that the ESR-spectra can be registered by using of 100 kHz high frequency (HF) modulation canal at the  modulation amplitude ($H_m$) with the value $H_m = 0$, Figure 1. The other registration conditions are the following: scan range was equal to 2 G, scan time was 8 minutes, time constant by the registration on the recorder was 0.3 s, microwave power level was 6 dB, the power value at 0 dB corresponds to 100 mW. Any  digital treatment or accumulation of the spectral data  were not used.  It is seen from Figure 1, that the spectrum has the  very unusual form for the stationary ESR-spectroscopy. It is similar to some extent to the spectra, registered by nonstationary classical transient ESR-spectroscopy. At the same time, the functional dependence is another. The spectral distribution of nutation signals in classical transient ESR-spectroscopy is described usually by continuous Bessel functions. The spectrum, presented in  Figure 1, has  the time-discrete character and  consist of two oscillation groups of very narrow resonance lines in the range [3322.49 - 3322.87] G and [3322.88 - 3323.5] G, at that the second less intensive group consist of two subgroups in the ranges [3322.88 - 3323.24] G and [3323.24 - 3323.5] G correspondingly. The number of the lines in the ranges [3322.49 - 3322.87] G and [3322.49 - 3322.87] G  is the same. Hence, it seems to be evident, that  the subgroup [3323.24 - 3323.5] G is registered the only partly. The linewidth ${\Delta}H$ of the near central line with the maximal amplitude in the left group [we designate it by term  main group (MG)] is equal to $0.01 (\pm 0.002)$ G. The linewidth of the other lines is  the same within limits of the accuracy above indicated, that is, it is equal to $0.01 (\pm 0.002)$ G. The lines, presented in  Figure 1, seem to be the most narrow among all the lines, registered in the stationary ESR-spectroscopy at all. It allows to determine $g$-value (that is, g-factor in the Zeeman term of the spin-Hamiltonian for absorbing centers) very precisely. It is equal to 2.002780  $(\pm 0.000005)$. It is the maximal precision of the g-value determination in all the world practice of the stationary ESR-measurements. The accuracy of the measurements  achieved  was determined by the high short-time stability and the high homogeneity of the  static magnetic field, indicated above in the second Section and also by the high frequency stability at a fixed static magnetic field position, which was controlled by the precise digital microwave wavemeter.  

We have to remark that the g-value obtained from the position in the middle  of spacing between the 11th and 12th lines, that in the center of gravity of the main oscillation group relatively the number of lines [it is equal to 23] is assumed to be coinciding with the position of the maximal absorption amplitude in the line of usual ESR-spectra. The determination of g-value at DC magneic field, corresponding to the center of gravity of the main oscillation group [in the position at the middle  of spacing between the 11th and 12th lines] althogh the intensity of the 11 and 12 lines is not maximal in the spectrun, represented in Figure 1 (or in equivalent Figure 3), seems to be confirmed by the comparison with the spectra registered at lower microwave power, Figure 4 to Figure 7. It is  seen the clear tendency to increasing of the symmetry extent with  the decrease of the microwave power level. At the attenuation, equaled to 20 dB, the MG-spectrum, observed in the scan field range [322.68 - 3322.90] G becomes to be practically symmetric relatively the center of gravity and the 11th and 12th lines have the maximal amplitude. 

 The corresponding positions between the 11th and 12th lines  in the right and left subgroups  of the second oscillation group are given by effective
$g$-values 2.00253 $(\pm 0.00002)$ and 2.00232 $(\pm 0.00002)$. 

The structure of the spectrum observed is very similar to the picture, representing itself the theoretical time evolution for an one qubit system  in James-Cummings model  \cite{Jaynes_Cummings}, Figure 2, indicating on  the quantum Rabi oscillation origin of the given spectrum. Really, the spectra in both the groups in Figure 1 can be described by almost sinusoidal  dependence  with the envelopes, the shape of which is close to the theoretical shape of oscillation groups in JCM, Figure 2, however it is asymmetric to some extent in distinction from JCM. The qualitative difference between the time evolution, given by JCM and the spectral distribution, represented in  Figure 1, along with that, that the amplitude distribution is asymmetric to some extent, consists  also in that, that the left group of lines in JCM, that is the main group  in our designation, has the cosinusoidal dependence (the evolution
is starting from  the maximum), Figure 2, at the same time, the main group of lines in  Figure 1 or in  Figure 3, in wich the same oscillation process is represented in dependence on the time for  the comparison with Figure 2, has sinusoidal character. Let us remark, that the spectral distribution,  presented in  Figure 1, is also the time evolution process, here the time is expressed in magnetic field units owing to the dependence of the static magnetic field (DC field) on the time by its sweep. Moreover,  there exists the bijective mapping by both the representations.  However, the given dependence is slightly different in comparison with the  dependence, which could be obtained by the registration with the standard transient magnetic resonance spectroscopy. The difference seems to be connected with  some increase   
of the distance between the energy levels in qubits by the sweep   of the static magnetic field, which is determined by the increase of Zeeman splitting of energy levels with the increase of the strengh of the static magnetic field (see futher for the details of the substantiation of the possibility of the registration of the given transient process on a stationary ESR spectrometer and how to take into consideration the nonconstant qubit levels' splitting). Let us remark, that the  possibility to
explain the spectrum observed by any technical inaccuracy of
spectrometer functioning, for instance, by means of an usual
amplitude modulation of possible technical noises (determined, for instance, by functioning  of the static magnetic field stabilization unit or the  microwave frequency stabilization unit and so on)  is excluded, since the envelope of the signals by an amplitude modulation is described by the sinusoidal dependence, we have the nonsinusoidal envelope, which is similar to the form of the theoretical envelope in oscillations' packets in Figure 2, described by Gauss function. 

\begin{figure}
\includegraphics[width=0.5\textwidth]{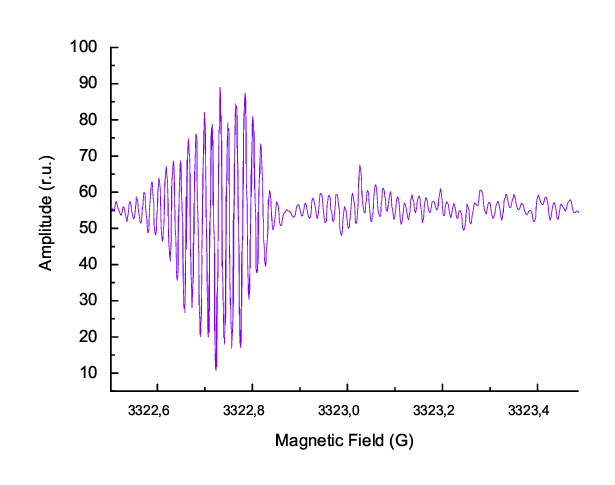}
\caption[Quantum Rabi oscillation process indicating on the formation of the coherent system of the resonance phonons]
{\label{Figure1} Quantum Rabi oscillation process, P = 6 dB}
\end{figure}

\begin{figure}
\includegraphics[width=0.47\textwidth]{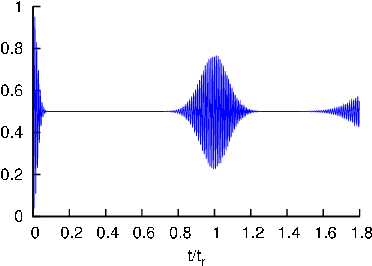}
\caption[Theoretical evolution for  JCM-system, initial state $|\psi\rangle \otimes |\alpha\rangle$ of which is direct product of atomic ground state and coherent field state with $\overline{n} = 50$]
{\label{Figure2} Theoretical evolution for  JCM-system, initial state $|\psi\rangle \otimes |\alpha\rangle$ of which is direct product of atomic ground state and coherent field state with $\overline{n} = 50$}
\end{figure}

\begin{figure}
\includegraphics[width=0.5\textwidth]{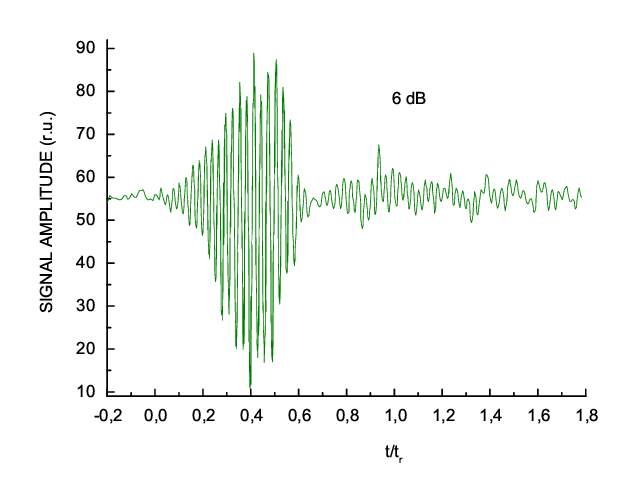}
\caption[Quantum Rabi oscillation process]
{\label{Figure1a} Quantum Rabi oscillation process in dependence on the time, P = 6 dB}
\end{figure}

Moreover, the value of $g$-factor of the center of the first group agrees well with g-factor of ESR line of the anthracite sample with relatively small volume, registered by usual conditions with the nonzero amplitude of 100 kHz high frequency modulation [however, it was determined now substantionally more precisely]. The clearly pronounced dependence of the amplitude and the shape of the first (left)-group-lines and the dependence of the amplitude, the shape and the position of the second-group-lines  on the microwave power level, Figure 4 to  Figure 7,  is the additional argument for the reality of the detection of the spin system response at $H_m = 0$. It was established by the detailed   comparison of Figure 1 [or Figure 3] with Figures 4 to 7, that along with the asymmetry extent in the distribution of the amplitude in the main oscillation group, which is increasing with the increase of the magnetic
component of microwave field, there are the dependencies of g-value, of the characteristic amplitude, of the mean value of splitting between the lines in the  oscillation groups  and of the mean-square deviation of
splitting between the lines  on the magnetic
component of microwave field, Figures 8 to 11, in which the aforesaid dependencies are represented for MGs.

 The analysis of the shape of the oscillation envelope is represented for the main group of lines in the quantum Rabi
oscillation process at P = 10 dB on the  Figure 12. It is seen from the  given Figure, that the envelope shape can be approximated by the function, which is near to the Gauss distribution function, naturally, it cannot be approximated by the sine or cosine functions, which can be expected for the signal attributed to possible technical inaccuracies.
 
The character of the amplitude distribution in the second group of lines is also changed, see Figure 6, where the spectrum of oscillation process, observed at the microwave power level in 10 dB is represented. It  is seen from Figure 6 that the amplitude distribution cannot be described by Gauss function. There is appeared trend to the uniform distribution, however not for  the positions of individual lines, but the the positions of subgroups consisting of four lines, see for details further.

 The distribution of the amplitude in the main oscillation group  at 20 dB attenuation, Figure 13, the DC field range [3322.68 - 3322.90] G, is, how it was above indicated, practically symmetric. However, what is especially interesting, at the given power level is very clear registered the group of lines, which is foregoing to the main oscillation group. It is partly represented in the given Figure, the DC field range is [3322.56 - 3322.68] G. The group of lines, being to be  foregoing to the main oscillation group, which was observed at  20 dB attenuation, is represented fully on the Figure 14. The full spectrum at  20 dB attenuation is represented on the Figure 15. It is seen from Figure 15, that the intensity in the group of lines, being to be  foregoing to the main oscillation group is substantially exceeding the intensity in the main and  revival groups of lines at the microwave power level indicated. Consequently, it is reazonable to suggest, that  the excitation, corresponding to the group of lines, being to be  foregoing to the main oscillation group was existing before scan of DC field in the straight direction at 20 dB attenuation, that is, it can be determined by  the previous DC field scan, just by  return scan of DC field at 15 dB attenuation, for which the signal is naturally more strong. The given
suggestion is confirmed  by the comparison of the characteristics  of the  oscillation groups' spectrum, which are registered at 15 dB attenuation with those ones for  the excitation, corresponding to the group of lines, being to be  foregoing to the main oscillation group by the spectrum registration at 20 dB attenuation. Really, mean splitting value in the group before main group at 20 dB   is 0,01092 G, compare with mean splitting value 0,01062 G for revival group RG I at 20 dB and with mean value of splitting between the lines in the main group at 15 dB, which is 0,01098 G ] and with mean value of splitting between the lines in the RG I + RG II group at 15 dB, which is equal to 0,01097 G. At the same time the mean value of splitting between the lines in the main group at 20 dB [lines 14-36] is 0,00852 G. Therefore, the groups before main group at 20 dB are RG groups, which are really determined by previous DC field scan from more high to more lower position [which was however not registered]. The distance between RG groups at 20 dB, localised correspongingly on the  right and  on the left from MG at 20 dB determines the hysteresis value relatively DC field scan direction, which  is equal approximately 1.123 G. 

In fact, the  appearance  of the excitation, corresponding to the group of lines, being to be  foregoing to the main oscillation group  is the display of memory effects, which are observed in carbon materials for the first time. The given conclusion is confirmed by that, that by the first record, which has been done at 6 dB,  the excitation corresponding to the group of lines, being to be  foregoing to the main oscillation group,  is absent. We have given  the description  of the excitation corresponding to the group of lines, being to be  foregoing to the main oscillation group at 20 dB attenuation, since in the given case the memory excitation is the most pronounced, however it is also presenting by other power levels, excluding only the first record at 6 dB aforesaid. The observation of the memory effect seems to be one the msin results obtained by the given study. It is in fact the direct proof, that absorbing centers are excifed into coherence state with vwery long coherence times. Low boundary of the  coherence time value can be evaluated from the value of the total scan time, which is equal how it was above indicated 8 minutes.  It is not shorter than part of scan time required for the registration of the group of lines, being to be  foregoing to the main oscillation group, yjat is not shorter than 4.6 minutes. We see that it exceeds substantially the world record coherence time in 30 seconds, which was found in \cite{Morello}.

 The frequency of the oscillations in the first group was the same in the power range from 6 to 20 dB in the zeroth appoximation. More precise measurements have shown, that there is some dependence of the oscillation frequency  on the magnetic component of microwave field.  However, instead of strong linear dependence indicating on the growth  of the oscillation frequency with increasing of the magnetic component of microwave field, predicted by both classical model and JCM, the only weak dependence was observed, Figure 10. At that, it is nonlinear and, moreover, some decrease [instead of linear increase] was reliably registered in the power range from 6 to 20 dB. It seems to be rather interesting result. At the same time at 5 dB, Figure 16, and by higher microwave power level the picture was quite another. It was observed  splitting of the main group  lines into two groups with the same oscillation frequencies, but  being now different from  the initial frequency, the value of splitting is $\approx$ 0.1 G. The envelopes of the given groups differ markedly from the envelopes, which have the oscillations, represented in  Figure 1, Figure 3, Figure 4 to Figure 7. 

We accentuate once again, that the character of the amplitude distribution in the second group of lines is strongly differs from the character of the amplitude distribution in the main group of lines, see Figure 6, where the spectrum of oscillation process, observed at the microwave power level in 10 dB is represented. It  is seen from Figure 6, that the amplitude distribution cannot be described by Gauss function. There is appeared  the trend to the uniform distribution, however for the subgroups consisting of four lines [the  distribution of individual lines in the subgroups remains nonuniform].

Given arguments and the comparison with Figure 2 allow to conclude, that really the spectrum in Figure 1 represents itself the quantum Rabi oscillation picture, however the deviations from the simple JCM are evident.
\begin{figure}
\includegraphics[width=0.5\textwidth]{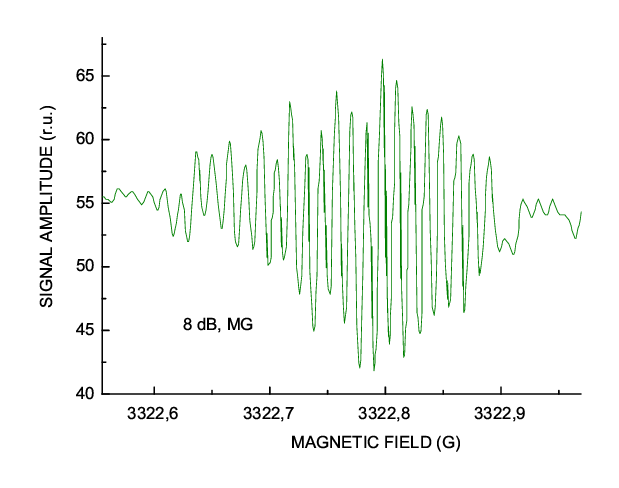}
\caption[Quantum Rabi oscillation process, the main group of lines, P = 8 dB]
{\label{Figure3a} Quantum Rabi oscillation process, the main group of lines, P = 8 dB}
\end{figure}
\begin{figure}
\includegraphics[width=0.5\textwidth]{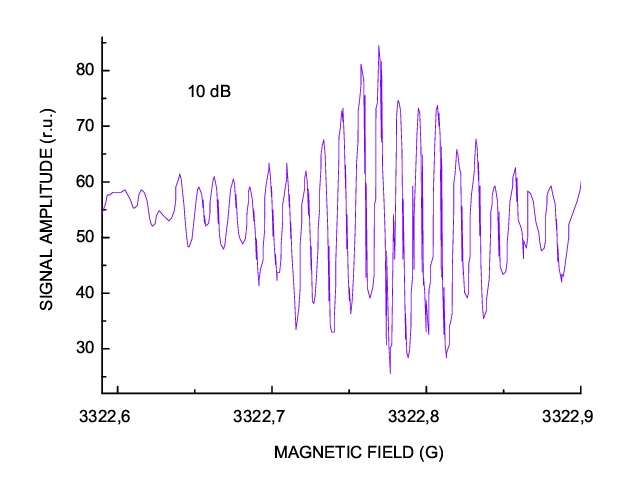}
\caption[Quantum Rabi oscillation process, the main group of lines, P = 10 dB]
{\label{Figure3b} Quantum Rabi oscillation process, the main group of lines, P = 10 dB}
\end{figure}
\begin{figure}
\includegraphics[width=0.5\textwidth]{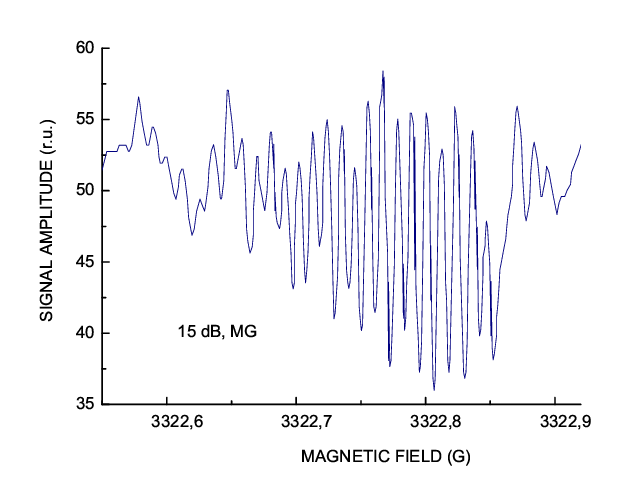}
\caption[Quantum Rabi oscillation process, main group of lines,  P = 15 dB]
{\label{Figure4a} Quantum Rabi oscillation process, the main group of lines,  P = 15 dB}
\end{figure}

\begin{figure}
\includegraphics[width=0.5\textwidth]{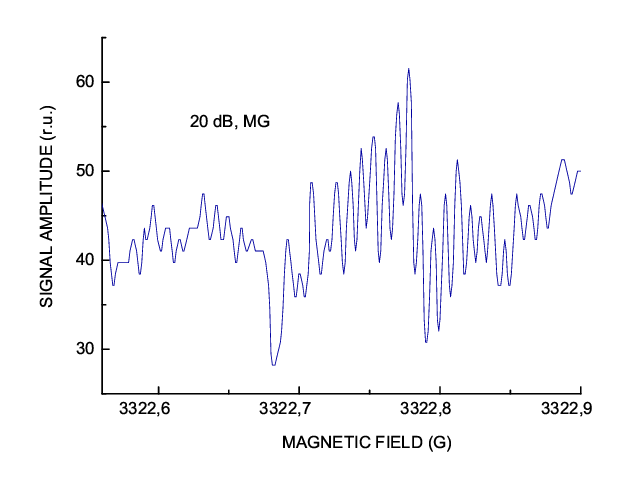}
\caption[Quantum Rabi oscillation process, main group of lines,  P = 20 dB]
{\label{Figure5a} Quantum Rabi oscillation process, the main group of lines,  P = 20 dB}
\end{figure}
\begin{figure}
\includegraphics[width=0.5\textwidth]{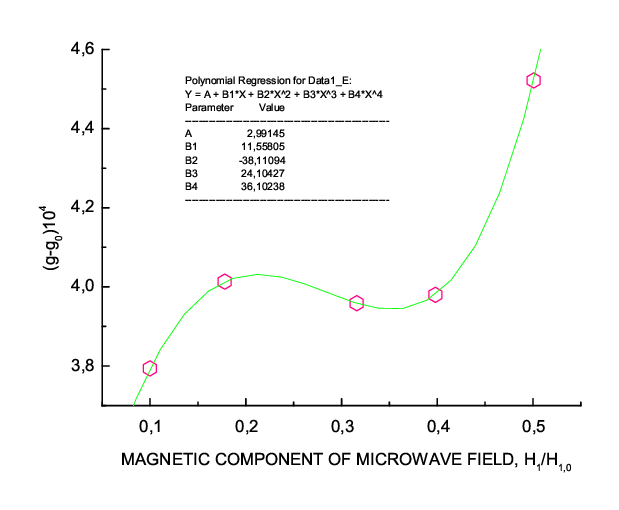}
\caption[The dependence of g-value of the main  oscillation group on the magnetic component of microwave field]
{\label{Figure4cm} The dependence of g-value of the main  oscillation group on the magnetic component of microwave field}
\end{figure}
\begin{figure}
\includegraphics[width=0.5\textwidth]{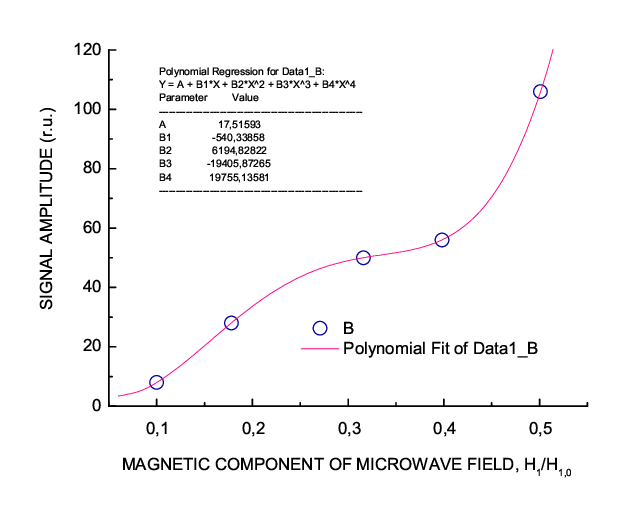}
\caption[The dependence of the characteristic amplitude of the main  oscillation group on the magnetic component of microwave field]
{\label{Figure4cl} The dependence of the characteristic amplitude of the main  oscillation group on the magnetic component of microwave field}
\end{figure}

\begin{figure}
\includegraphics[width=0.5\textwidth]{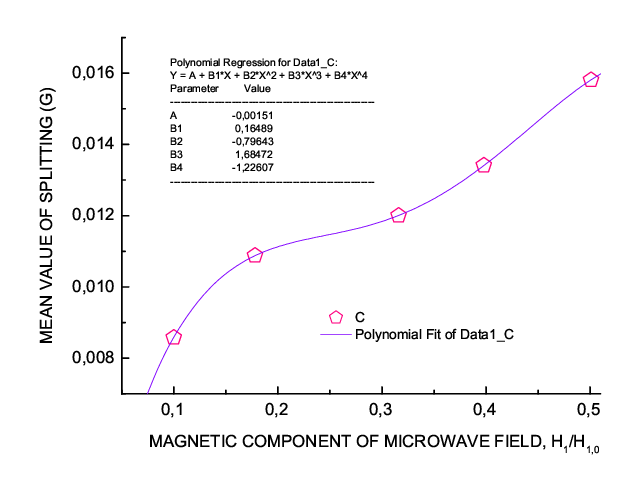}
\caption[The dependence of the  mean value of splitting between   the lines in the main  oscillation group on the magnetic component of microwave field]
{\label{Figure4cw} The dependence of the  mean value of splitting between   the lines in the main  oscillation group on the magnetic component of microwave field}
\end{figure}
\begin{figure}
\includegraphics[width=0.5\textwidth]{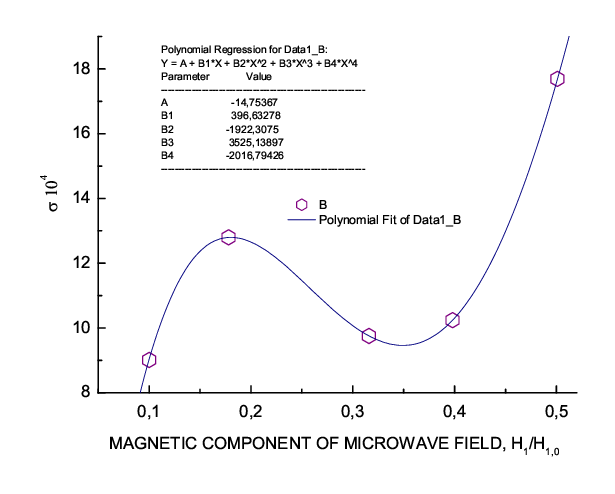}
\caption[The dependence of the  mean-square deviation of splitting between   the lines in the main  oscillation group on the magnetic component of microwave field]
{\label{Figure4cz} The dependence of the  mean-square deviation of splitting between   the lines in the main  oscillation group on the magnetic component of microwave field}
\end{figure}

\begin{figure}
\includegraphics[width=0.5\textwidth]{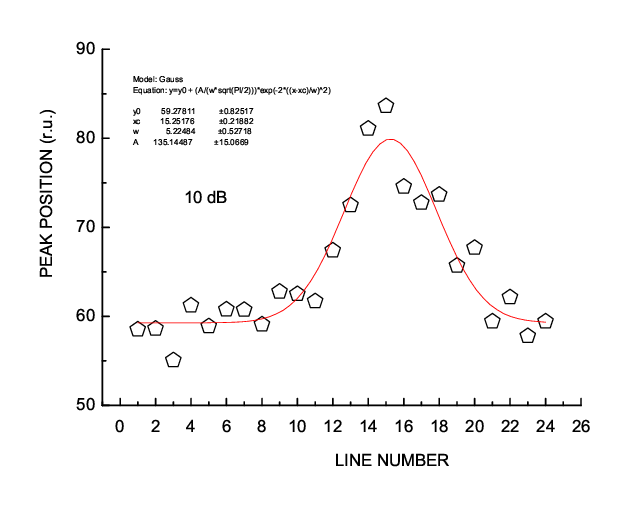}
\caption[The shape of the envelope function in the amplitude distribution  for   the main group of lines in the quantum Rabi oscillation process at P = 10 dB]
{\label{Figure3abp} The shape of the envelope function in the amplitude distribution  for   the main group of lines in the quantum Rabi oscillation process at P = 10 dB}
\end{figure}
\begin{figure}
\includegraphics[width=0.5\textwidth]{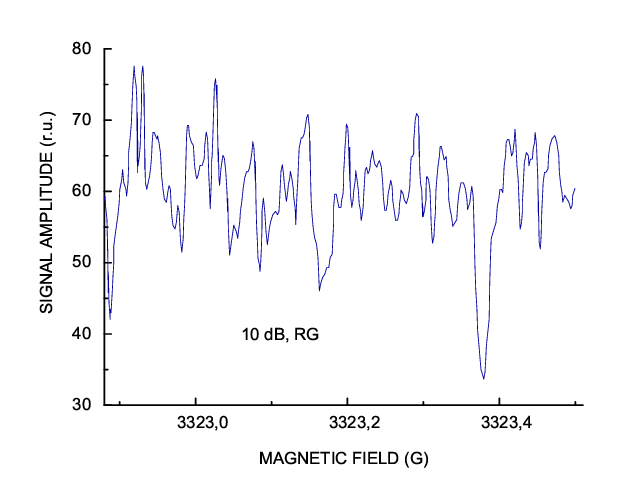}
\caption[Quantum Rabi oscillation process, the main group of lines, P = 10 dB]
{\label{Figure3ab} Quantum Rabi oscillation process, the revival group of lines, P = 10 dB}
\end{figure}
\begin{figure}
\includegraphics[width=0.5\textwidth]{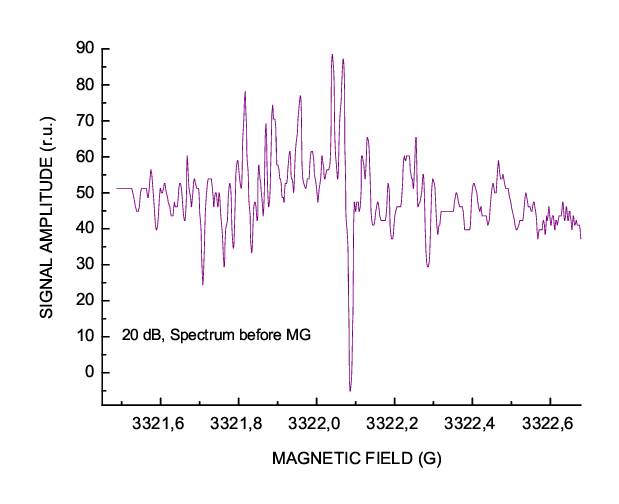}
\caption[Quantum Rabi oscillation process, the group of lines before the main group,  P = 20 dB]
{\label{Figure5aw} Quantum Rabi oscillation process,  the group of lines before the main group, P = 20 dB}
\end{figure}
\begin{figure}
\includegraphics[width=0.5\textwidth]{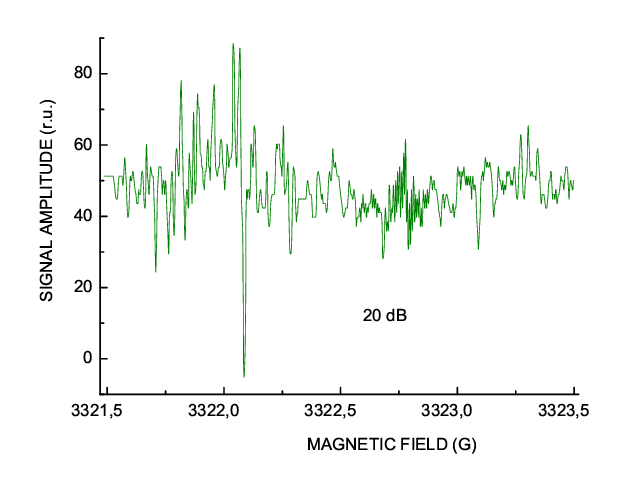}
\caption[Quantum Rabi oscillation process, P = 20 dB]
{\label{Figure5awq} Quantum Rabi oscillation process,   P = 20 dB}
\end{figure}

\begin{figure}
\includegraphics[width=0.5\textwidth]{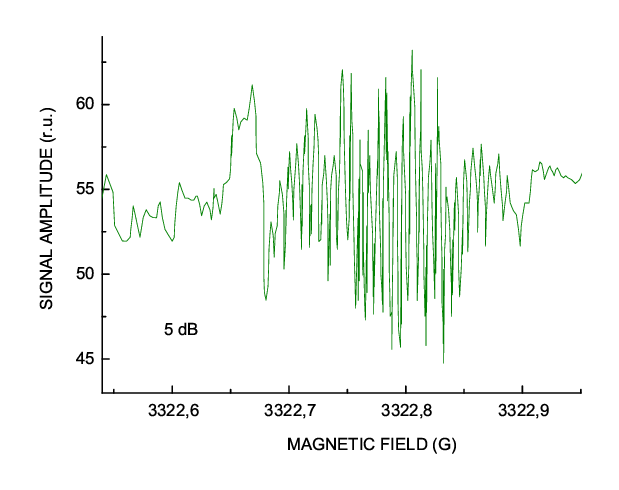}
\caption[Quantum Rabi oscillation process, main group of lines,  P = 5 dB]
{\label{Figure6a} Quantum Rabi oscillation process, the main group of lines,  P = 5 dB}
\end{figure}
 Let us touch on JCM in some details in order  to understand the physics of the differences observed. Central role in JCM plays the sum with infinite summation limit, which represents on a scale [-1,1] the degree of excitation of two-level system resulting of interaction between the atom dipole or spin and single mode quantized EM-field. It is
\begin{equation}
\label{eq31m}
\begin{split}
\langle\hat{\sigma}^{z}(t)\rangle = -\exp[-|\alpha|^2]\sum_{n=1}^{\infty}\frac{|\alpha|^{2n}}{n!} \cos{2g\sqrt{\overline{n}}t},
\end{split}
\end{equation} 
where $|\alpha\rangle$ is fully coherent state of the field, taking place at $t = 0$, at that $|\alpha|^2$ = $\overline{n}$, that is, it is  the average number of photons
in the field, $g$ is coupling constant between field  and atom (spin). The sum in  (\ref{eq31m}) cannot be expressed  exactly in analytical form. For very short times and very large $\overline{n}$ behavior of $\langle\hat{\sigma}^{z}(t)\rangle$ is determined by $\cos{2g\sqrt{\overline{n}}t}$.
Cummings \cite{Cummings} has shown, that by resonance and for intermediate time $t$ values the cosine Rabi oscillations damp quickly (so called collapse takes place). Given damping can be described by Gaussian envelope
\begin{equation}
\label{eq32q}
\exp[-\frac{1}{2}(gt)^2].
\end{equation}
 It is substantial, that it  not depends on field intensity unlike to semiclassical Rabi oscillation damping process and it is determined entirely  the only by coupling constant $g$. We have really observed  a rather good agreement with JCM in the description of the first quantum damping process (that is, collapse process), corresponding to the main group of lines, envelope of which seems to be very near to Gaussian [more strictly, the theory gives the product of Gaussian and linear dependence by modulation method of the signal detection, see futher]. However the good agreement with JCM aforesaid takes plsce the only at low microwave power level equaled to  20 dB, Figure 7. With  an increase of the microwave power level the asymmetry in amplitude distribution of individual oscillations relatively center of gravity is appeared, that is, along abscissa axis. The asymmetry in amplitude distribution of individual oscillations along ordinate axis is observed in all power range used. It is seen,  that asymmetry extent depends nonlinearly on microwave power level. The  envelope of the main group of lines at 10 dB is more symmetric than  the envelope of the main group of lines at 15 dB, compare Figures 5 and 6. Moreover,  the amplitude distribution of the main group of lines at 10 dB  can be described by the theoretical dependence, Figure 12, while the deviation from the theoretical dependence at 15 dB is substantial.

The authors of the work \cite{Eberly} have found, that JCM contains so-called revival process with revival time $t_r$, given by the expression
\begin{equation}
\label{eq33}
t_r = \frac{\pi}{g^2} \sqrt{\Delta^2 + 4 g^2 \overline{n}}, 
\end{equation}
where $\Delta$ is a deviation of a  field mode frequency from resonance value. Revival process takes place at all time values, satisfying the relation $t$ = $k t_r$, $k \in N $. It is seen from  (\ref{eq33}), that revival time depends on $\overline{n}$ and it is proportional to oscillating field amplitude at $\Delta = 0$ like to the dependence of Rabi frequency on the magnetic component of microwave field in transient ESR spectroscopy by semiclassical consideration. 

Let us illustrate mathematically the dynamics of spectroscopic transitions in JCM. The initial state for single qubit system can be defined  by direct product of the two level  matter (atomic, spin and so on) subsystem state $|\psi\rangle$ and coherent EM-field state $|\alpha\rangle$. It is the following
\begin{equation}
\label{eq34}
|\Psi(0)\rangle = |\psi\rangle \otimes |\alpha\rangle,
\end{equation}
where $|\psi\rangle$ is 
\begin{equation}
\label{eq35}
 |\psi\rangle = c_1 |\psi_1\rangle + c_2 |\psi_2\rangle \end{equation}
and $|\alpha\rangle$ is 
\begin{equation}
\label{eq36}
|\alpha\rangle =
\exp[\frac{-|\alpha|^2}{2}]\sum_{n=1}^{\infty}\frac{|\alpha|^{n}}{n!}|n\rangle, \alpha = \sqrt{\overline{n}} e^{i\phi}.
\end{equation}
The scalar field functions $|\psi_1\rangle$ and $|\psi_2\rangle$ in (\ref{eq35}) describe correspongingly the initial ground state and the excited state in the two level matter system,  $c_1 $ and  $ c_2$ are constant coefficients, characterising the contribution of the the initial ground state and the excited state in the current state $|\psi\rangle$ of the  two level matter system. 
 The Rabi oscillations of the probability, that the qubit is in the initial state in the first damping stage (collapse), on
a time scale of $t_c \simeq \frac{\sqrt{2}}{g}$ and then revive at $t_r \simeq  \frac{2\pi\sqrt{\overline{n}}}{g}$ are  shown in Figure 2 for $c_1 = 1$, $ c_2 = 0$ by plotting 
\begin{equation}
\label{eq37}
\sum_{n=1}^{\infty}|\langle\psi_1, n|\Psi(t)\rangle |^2,
\end{equation}
where $\langle\psi_1, n| = \langle\psi_1|\otimes \langle n|$  is the ground  state for the
qubit  with $n$ photons in the cavity. Our measurements of the dependence of the ESR spectral characteristics on the microwave power (photon number) is also showed, that, in the correspondence with JCM, the positions of the lines of left side, that is, main group are not dependent on photon number in  the zeroth approximation, while the positions of the lines of right hand side group are  dependent on photon number in the same approximation. At the same time, there is some dependence of the positions of the lines of the  main group in the first  approximation, Figure 8. It is reliably detected and it is see, that it is strongly nonlinear.
  
The aforesaid difference between our results and JCM (the evolution
is starting from minimum instead of maximum) can be easily explained, if we present the expermental proof of  the Rabi oscillations'   origin. The value of the oscillations' frequency $\Omega_{Ph}$ can be evaluated in the following way. Taking into account the quantum character of the oscillations observed we can use the expression of JCM (\ref{eq32}) for the first stage damping process with an accuracy to factor, containing the linear dependence on the time and which is appeared by the application of the modulation method of the signal detection [see for details further]. Then we can use the fit of the envelope of  the first stage damping process, represented in Figure 12. Using statistics mean value for the splitting between the lines in MG group at 10 dB equaled to 0,012 G, we have $\Delta H =  5.225 \times 0,012$ = 0.063 G. From the value of  $\Delta H$ of the  envelope of the first damping stage expressed in frequency units by means of well known relation $\Delta\nu$ = 1.39961 g $\Delta H$ we obtain the evaluation of $\Omega_{Ph}$, if to multiply the value obtained on the number  $n$ of oscillations in the range of $\Delta H$. So, we have 
     $\Omega_{Ph}$ = $1.39961  \times 0.06285  \times 2.00278 n $ = $0.17619 n$ = $ 0.17619 \times 5.225$ MHz = 0.9206 MHz. 
It is  remarkable, that the value of  $\Omega_{Ph}$  equaled to 920.6 kHz is remaining  approximately the same in the range of microwave power level 20 - 6 dB [some deviation can be determined from the data, presented in Figure 10]. The given value is comparable  in the same microwave power range with the Rabi frequency values in analogous systems, excited by microwave power. So, for instance, for Si-P3 centers in Si the Rabi frequency changes from $\approx 1 \times 10^6 rad s^{-1}$ to $\approx 4.5 \times 10^6 rad s^{-1}$  \cite{Yerchak_Stelmakh} in dependence of the level of microwave power used. We see, that the range of Rabi frequencies in standard transient ESR-spectroscopy is approximately the same. It means, that we really have observed the quantum Rabi oscillations, which at the same time are not connected with  a  microwave absorption process immediately. It becomes to be evident, that the Rabi oscillations observed have to be identified with the quantum relaxation oscillations, predicted theoretically in \cite{QFTDST}, that is the quantum Rabi oscillations, appearing in the result of the interaction of resonance phonons with the spin subsystem of absorbing centers. The prediction of sinusoidal character (instead of cosinusoidal character in JCM) in the first time-ranged packet of oscillations, that is, on the first damping stage, is also confirmed (see for details further). Then it becomes to be understandable the absence of the pronounced linear dependence of the oscillation frequency on the magnetic component of microwave field [in the zeroth approximation] in the range of microwave power level 20 - 6 dB. The Rabi oscillation frequency is determined by the number of phonons.  In its turn, the number of phonons by linear generation process, which seems to be realised in the range of microwave power level 20 - 6 dB is determined the only by the number of absorbing centers, which is constant and independent on the microwave power level, that is, independent on  the  photon number [in zeroth approximation]. The photon number is very large, substantionally exceeding the absorbing centers' number and its increase does not lead to phonon number increase. So, we see, that there is no contradictions with JCM in the qiven aspect. Moreover, it can be considered being the direct proof of the phonon origin of Rabi oscillations observed.

The  appearance of the two Rabi oscillations' subsystems at microwave power level equaled to 5 dB, Figure 16,  and higher  is in fact the direct indication, that resonance phonon generation process becomes  to be nonlinear, since doubling is typical characteristics of nonlinear physics. The mean value of linewidth  of oscillation lines was   decreased stepwise
 approximately in 2 times and  spacing between the lines was  decreased stepwise
in comparison  with spacing at  6 dB to the value being very near to  the value 
at 20 dB. Naturally, the nonlinear Rabi oscillations' process  is beyond the scope of JCM. It is the subject for the subsequent studies.

From the value of  $\Delta\nu$ can be obtained the relaxation time in the system \{spins of absorbing centers + phonons\}, characterising  the rate of the first damping stage process.  It is determined by the relation 
\begin{equation}
\label{eq37a}
 T^{Ph}_2 = \frac{ln2}{\pi\Delta\nu}.
\end{equation}
Hence $T^{Ph}_2$ = $1.25\times 10^{-6}$ s.

Let us remark that  quantum Rabi oscillations in the system \{spins of absorbing centers + phonons\} has been observed for the first time. It is new phenomenon in quantum physics, indicating that quantum relaxation processes can have the  oscillation character.
The bifurcation of the quantum Rabi oscillations is also new phenomenon in quantum physics, more strictly, in  nonlinear quantum physics, which has been also observed for the first time.

The comparison with the model for dynamic of spectroscopic transitions, given in  \cite{QFTDST} is correct. At the same time, it is required to understand, how the resonance  quantum relaxation process,  above considered, can be registered at $H_m = 0$.

 Let us give some details, explaining the  possibility of the registration of the resonance signal by quantum relaxation process. Quantum Rabi relaxation oscillations, that is phonon-assisted (or in equivalent definition acoustic) quantum Rabi oscillations lead to the  oscillation of the magnetization of the sample magnetic subsystem. The very strong  in their amplitudes ESR-signals, registered in usual registration regime, that is with nonzeroth HF-modulation amplitude, can indicate, that some type of magnetic ordering takes place in the samples studied. Taking into account the disordered character of carbon structure in anthracite form, it is reasonable to suggest, that spin glass magnetic ordering takes place. Then, alternating magnetic flux, created by the sample itself in result of  an acoustic quantum Rabi oscillation appearance, leads to the emergence of an electromotive force on modulation  spools and to a modulation current. It is proportional to 100 kHz Fourier component of alternating (with $\Omega_{Ph}$  frequency) magnetic flux in its Fourier series on time segment, corresponding mainly to the evolution time for the left [MG] group of lines in Figures 1, 3 to 7. Therefore, the sample itself creates nonzeroth modulation field with small  amplitude (it is evident, that it is not exceeding  the linewidth value in $\simeq 0.005$ G, which was registered st 5 dB).  The registration is possible on a stationary ESR spectrometer, since, by the detection using  100 kHz modulation, the time dependent acoustic quantum Rabi oscillations will have  100 kHz Fourier component, and they will be registered like to usual ESR signals. The registered acoustic Rabi oscillation signals represent, therefore, the derivative of the product of the Gaussian envelope function and harmonic function, filling up 
the time-discrete oscillation packets, which in accordance with \cite{QFTDST} has to be sinusoida, 
\begin{equation}
\label{eq32a}
\exp[-\frac{1}{2}(x)^2]\sin{2\sqrt{\overline{n}}x},
\end{equation}

 where $x = gt$, and they will be consisting of two parts, the relative contribution of which will be determined  by
the ratio of value of $g$ and $2\sqrt{\overline{n}}$. Let us remark, that the
coupling constant $g$ is in fact $\mid\lambda^{\pm}_{\vec q l}\mid$ in (\ref{eq30}), (\ref{eq30a}), that is
$g$ $\equiv$  $\mid\lambda^{\pm}\mid$ (the independence of coupling constant on
$\vec q$ and on a site number $l$ seems to be reasonable approximation). The value of $\mid\lambda^{\pm}\mid$ was evaluated from the relation $t_c \simeq \frac{\sqrt{2}}{g}$. It is equal to $\mid\lambda^{\pm}\mid$ = $4\times{10^3 s^{-1}}.$ The
value of phonons in coherent state $\overline{n}$ is evaluated to be
equal 10 (per one absorbing center). Taking into account, that $\mid\lambda^{\pm}\mid$ $\gg$ $2\sqrt{\overline{n}}$, the
signal component with harmonic dependence, coinciding
with harmonic dependence in initial expression, given by (\ref{eq32a}), will be strongly dominating. The observed sinusoidal character of filling of the first wave packet  instead of
cosinusoidal character confirms, therefore, the theoretically predicted $\pi/2$ difference in  registration phase of the phonon-assisted oscillation process
relatively the photon-assisted
oscillation process. It is the consequence
of  the known retardation effect of an any phonon subsystem.

Let us remark, that the notion of the phonon, introduced in the physics by Tamm I.E for the description of lattice vibrations in crystalline solids, remains  also correct for the description of atomic  vibrations  in amorphous solids including glass structures for the acoustic  branch of vibrations, that is, just for the vibrations, which are essential in ESR spectroscopy, see, for instance \cite{Ph.Enc.}. Moreover, the notion of the phonon is applicable for the description of vibrations in quantum liquids, \cite{Ph.Enc.}.

Now we will show that the model, which is suitable for the theoretical description of acoustic quantum Rabi oscillations on the initial damping stage at low microwave power levels is mathematically equivalent to JCM model.  Really, the
qualitative comparison with known theory is achieved, if to
omit the part from Hamiltonian (\ref{eq14}), corresponding to EM-
field and $z$-relaxation term and to set N, that is, qubit number \cite{QFTDST}, equaled to 1.
Then, we will have in rotating wave approximation (with the accuracy to $\pi/2$ phase change) the Hamiltonian for the  phonon subsystem, which is mathematically equivalent to JCM
Hamiltonian, the difference consists in that, that by the  replacement of  the photon subsystem into 
phonon subsystem $\pi/2$ phase change takes place. It  explains the resemblance of the experimental dependence for  main group, presented in Figure 7, and theoretical dependence in Figure 2. The definite resemblance for  main groups of oscillation signale is retained with the increase of the microawe power level, Figures 6, 5, 4, 3 or 1. However it is seen the increase of asymmetry in amplitude distribution 
\begin{figure}
\includegraphics[width=0.5\textwidth]{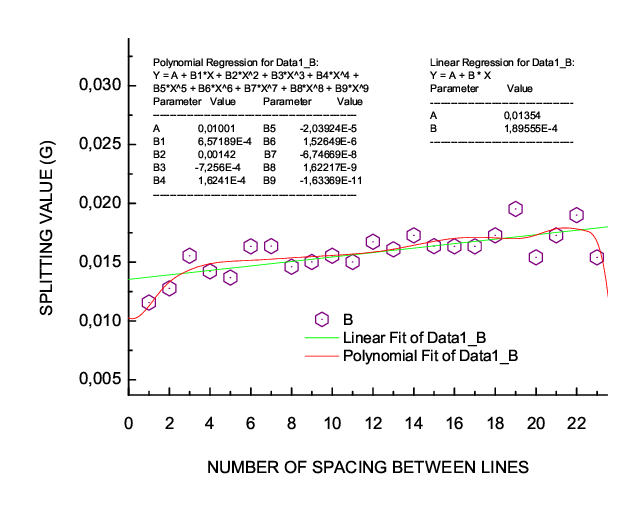}
\caption[The fit of the splitting between the lines in the main oscillation group in the spectrum, presented in Figure 1]
{\label{Figure4n} The fit of the splitting between the lines in the main oscillation group in the spectrum, presented in Figure 1}
\end{figure}

\begin{figure}
\includegraphics[width=0.5\textwidth]{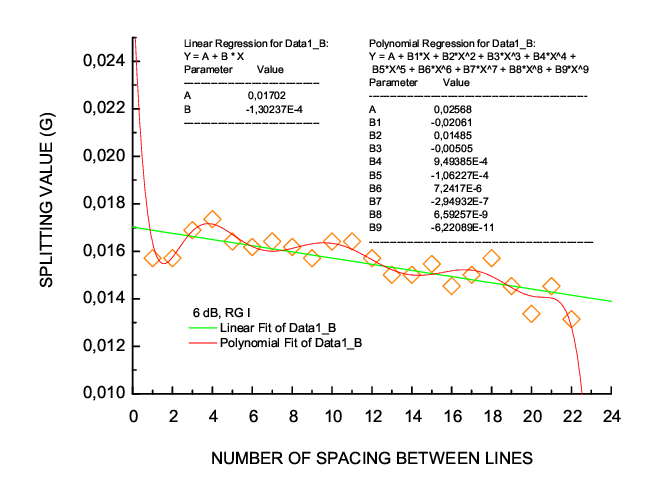}
\caption[The fit of the splitting between the lines in the first revival oscillation group in the spectrum, presented in Figure 1]
{\label{Figure4} The fit of the splitting between the lines in the first revival oscillation group in the spectrum, presented in Figure 1}
\end{figure}
\begin{figure}
\includegraphics[width=0.5\textwidth]{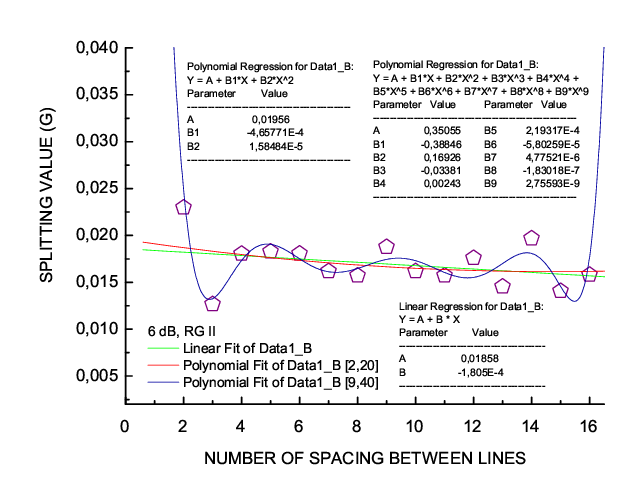}
\caption[The fit of the splitting between the lines in the second revival oscillation group in the spectrum, presented in Figure 1]
{\label{Figure4c} The fit of the splitting between the lines in the second revival oscillation group in the spectrum, presented in Figure 1}
\end{figure}

\begin{figure}
\includegraphics[width=0.5\textwidth]{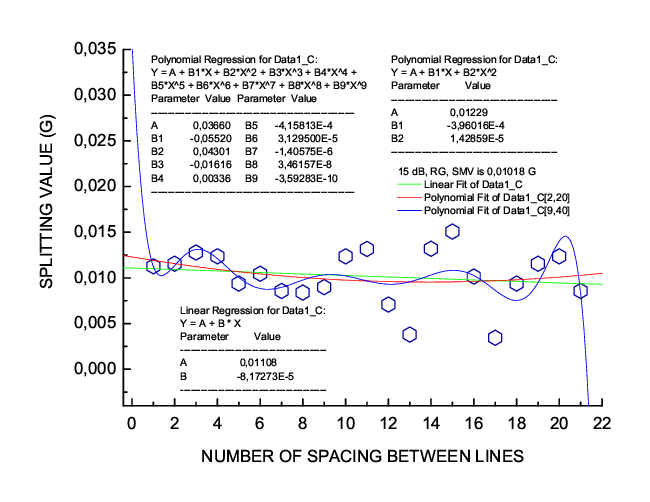}
\caption[The fit of the splitting between the lines in the first revival oscillation group in the oscillation spectrum, registered at 15 dB]
{\label{Figure4cg} The fit of the splitting between the lines in the first revival oscillation group in the oscillation spectrum, registered at 15 dB}
\end{figure}
\begin{figure}
\includegraphics[width=0.5\textwidth]{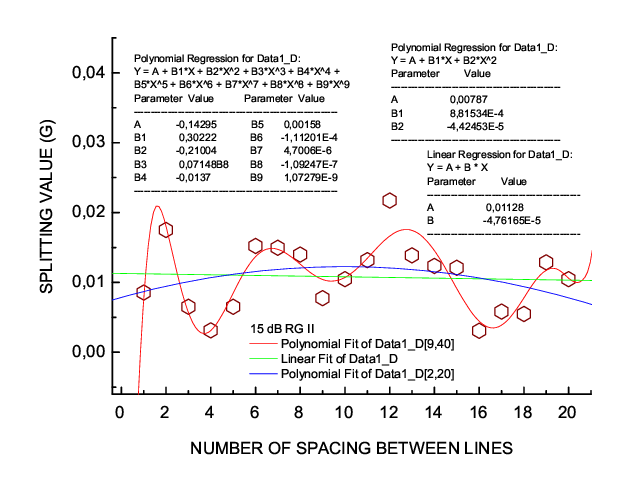}
\caption[The fit of the splitting between the lines in the second revival oscillation group in the oscillation spectrum, registered at 15 dB]
{\label{Figure4cj} The fit of the splitting between the lines in the second revival oscillation group in the oscillation spectrum, registered at 15 dB}
\end{figure}
\begin{figure}
\includegraphics[width=0.5\textwidth]{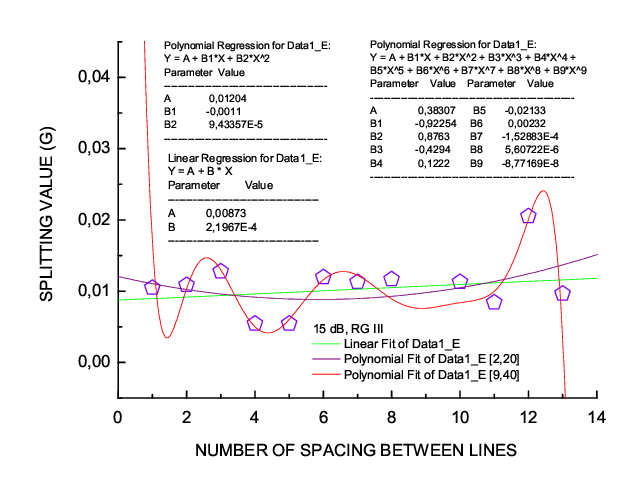}
\caption[The fit of the splitting between the lines in the third revival oscillation group in the oscillation spectrum, registered at 15 dB]
{\label{Figure4ck} The fit of the splitting between the lines in the third revival oscillation group in the oscillation spectrum, registered at 15 dB}
\end{figure}

 The distribution of the spectral positions of individual lines in phonon-assisted Rabi oscillation process was analysed, using linear and polynomial fits of experimental data. The results for fit of the splitting between the lines in the
 oscillation groups in the spectrum, registered at 6 dB  attenuation of the microwave power art presented in Figures 17 to 19. It is seen  from Figure 17, that both linear and polynomial fits for MG lines' position gives almost the same result, if to  exclude from consideration the initial and end data points, it is linear in the given range.  at the same time, the positions of the individual lines in revival groups, Figures 18, 19, cannot be describd by linear dependence. The polynomial fit shows, that the formation of subgroups, consisting of four lines with different separation and, consequently, with different frequencies takes place. The given process becomes to be more pronounced with the decrease of the microwave power level. It is seen from   Figures 20-22, where  the fit of the splitting between the lines in the
 revival oscillation groups in the oscillation spectrum,
registered at 15 dB, is presented. The spectrum was registered in more wide range in comparison with the spectrum at 6 dB and three revival groups have been observed [ the third - partly].

 The linear dependence of splitting, presented in Figure 17, can be explained by the  dependence of the frequency $\Omega_{Ph}$  of phonon-assisted Rabi oscillations on the static $H$-field value, since by the change of static $H$-field the distance between the energy levels in qubits is also changes. Expanding  the  dependence $\Omega_{Ph}(H)$ in Taylor series and restricting by linear term (it is correct, since $H$-change is small), we obtain the agreement with experiment. At the  same time  the aforegoing spectral  properties of revival groups cabboy be explainedin analogous way. Moreover they indicatr on the inapplicability of both one qubit JCM and  Tavis- Cumming model for n, $n \in N$, independent qubits, describing the interaction of quantised EM-field with $n$ noninteracting between themselves absorbing centers. It is the strong indication that  the interaction of quantised EM-field and correspondingly quantised acoustic field [that is, phonon field] with $n$ interacting between themselves absorbing centers takes place.

It seems to be reasonable to compare the results obtained with the model proposed by 
Slepyan, Yerchak, Hofmann, Bass [SYHB-model] \cite{Slepyan_Yerchak}, in which the interacyion between absorbing centers is taking into account.
In the work \cite{Slepyan_Yerchak} was 
developed a theory of Rabi oscillations
in a periodical 1D chain of two-level quantum dots (QD) with tunneling
coupling, exposed to quantum light. The role of
interdot coupling and Rabi oscillations on each other was
considered in details. The following conclusions have been done
from the studies:

1.The interdot tunneling in the QD chain
exposed to quantum light leads to the appearance of spatial
modulation of Rabi oscillations and in the appearance of the phenomenon of Rabi waves propagation.
It is shown, that Rabi waves can propagate if the
light mode wave vector has nonzero component along the
chain axis. Characteristics of the Rabi waves depend strongly
on relations between parameter of electron-photon coupling,
frequency deviation and transparency factors of potential
barriers for both of levels of individual QDs.

2.Traveling Rabi wave represent the quantum state of
QD chain dressed by radiation, that is, joined states of electron-hole (e-h)
pair and photons. The qualitative distinction of these states
from the similar states of single dressed atom is the space-time
modulation of dressing parameter according to the traveling
wave law. The propagation of traveling Rabi wave
looks like supported by periodically inhomogeneous nonreciprocal
effective media, whose refractive index is determined by electric field distribution.

 For the  quantum description of the given Rabi oscillation propagation process the authors introduce the  quasiparticles of a new type - rabitons, which can be considered being to be the 
generalization of Hopfield polaritons for the case of indirect
quantum transitions.

3.Two traveling Rabi modes with different frequencies
of Rabi oscillations corresponds to the given value of wave number.
The range of Rabi oscillation frequencies is limited by the
critical value, which is different for both of the modes. The QD chain
is opaque in the regime of Rabi oscillation frequencies below
the critical value. At the same time it is shown, that the critical frequencies and dispersion
characteristics of Rabi modes depend on number of photons.

4.The formation of different types of Rabi wave packets was considered. It has been found, that they represent themselves
 arbitrary superpositions of four partial subpackets with
different amplitudes, frequency shifts, and velocities of a motion.
Two of subpackets correspond to the contribution of
excited initial state and two others caused by the ground
initial state contribution. It was established that Rabi wave packets transfer energy,
inversion, quasimomentum, electron-electron, and electron-photon
quantum correlations along the chain. The number of
subpackets can be diminished in specific circumstances.

5.For the case of QD chain considered in the work cited, it is found, that Rabi oscillations qualitatively change the electron tunneling
picture in the given chain. In contrast to the case of the absence of
electron-photon coupling, the movement of initially ground state
subpacket is governed by tunneling transparency of excited
energy level and vice versa. Thus, Rabi oscillations can
stimulate  tunneling through low-energy level and suppress
it through high-energy one.

6.It has been established, that Rabi wave packet movement along the QD chain alters
the light statistics. Particularly, it was predicted for the QD chain, exposed
to coherent light the drastic modification
of the standard collapse-revival phenomenon: collapses and
revivals appear in different areas
of the chain space. 

It seems especially significant the conclusion of the authors of \cite{Slepyan_Yerchak}, that the phenomenon of
Rabi waves' formation can take place in a number of other distributed
systems strongly coupled with electromagnetic field.
For the example they indicate on the possibility of
Rabi waves' formation in  superconducting circuits based on Josephson
junctions, which are currently the most experimentally advanced
solid-state qubits. According to the opinion of the authors of \cite{Slepyan_Yerchak} the
qubit-qubit capacitance coupling in the chain of qubits
placed inside a high-Q transmission-line resonator will be
responsible for the Rabi waves propagation similar to described in the paper above cited. Let us also remark, that the theory  elaborated in \cite{Slepyan_Yerchak} can be considered to be the theory of quantum space radiative transport of a radiation field, in particular, a light field, by its interaction with matter.

It was drawn attention in  \cite{PSPA}, that the analysis of the results obtained in the work \cite{Slepyan_Yerchak} allows to predict the additional new quantum phenomenon -  the two stage process of a photon absorption, which can have rather long times, even more long than the relaxation times of the excited atomic states. So, in the first stage the rabiton formation takes place. The given process can be rather fast. The second stage is determined by a lifetime of rabitons, in which photons, being to be not absorbed coexist with absorbing matter excitons. Really, the state vector of the
"QD-chain+light" system, that is, the state vector of the
rabiton was represented in \cite{Slepyan_Yerchak} by direct product of the
eigenstates of isolated QDs and photon number states in the following form
\begin{equation}
\label{state_vector}
| {\Psi (t)}\rangle =\sum\limits_n \sum\limits_p (A_{p,n}(t) |a_p,n \rangle +B_{p,n}(t) | b_p,n \rangle ) ,
\end{equation}
where $| b_p,n \rangle = | b_p \rangle\otimes|n \rangle$, $| a_p,n \rangle = | a_p \rangle\otimes|n \rangle$,  in which  $|n \rangle$ is the light Fock state with $n$  photons, $|a_p \rangle$, $| b_p \rangle$ are
one-electron orbital wave-functions on the $p$-th QD in
the excited and ground states respectively, $B_{p,n}$, $A_{p,n}$ are the  probability amplitudes, $p \in N$.

 The probability
amplitudes are determined from the following equations

\begin{align}
\label{system_diskr_a}
 &\frac{\partial A_{p,n}}{\partial t}=-\frac{i\omega_0}{2}A_{p,n}+i\xi_1(A_{p-1,n}+A_{p+1,n})\\
\nonumber&-i g\sqrt{n+1}B_{p,n+1} e^{i(kpa-\omega t)} -i\Delta\omega B_{p,n}\sum\limits_m A_{p,m}B_{p,m}^*,   \\ \rule{0in}{4ex}
\label{system_diskr_b}
\nonumber&\frac{\partial B_{p,n+1}}{\partial t}=\frac{i\omega_0}{2}B_{p,n+1}+i\xi_2(B_{p-1,n+1}+B_{p+1,n+1})\\
&-i g\sqrt{n+1}A_{p,n} e^{-i(kpa-\omega t)} -i\Delta\omega A_{p,n+1}\sum\limits_m A_{p,m}^*B_{p,m},
\raisetag{35pt}
\end{align}
where
\begin{equation}
\label{dep_shift}
\Delta\omega=\frac{4\pi}{\hbar{}V}{ \mu}(\widetilde{{
\underline N}}{ \mu})
\end{equation}
 is the local-field induced depolarization shift, $\underline N$ in which
is the depolarization tensor, $ k =
(\omega/c) \cos \alpha$ is the axial wave number, $\alpha$ denotes the angle
between the light propagation direction and QD-chain,
$\omega$ is angular frequency, $\xi_{1;2}$ are the electron tunneling
frequencies for the excited ($\xi_{1}$) and ground ($\xi_{2}$) states of
the QDs, $g = (\vec\mu \vec{\mathfrak E})/\hbar$ is the interaction constant, $\vec\mu$ is
the QD dipole moment, $\vec{\mathfrak{E}} = \sqrt{2\pi\hbar\omega/V_0} \vec{e}$, $V_0$ is the
normalizing volume, $\vec{e}$ is the unit polarization vector. Obtaining the given equations it was taken into account that the interaction  can cause the transitions between the states $| a_p,n \rangle$, $| b_p,n+1 \rangle$  only. It is seen from Eqs.\eqref{system_diskr_a}--\eqref{system_diskr_b}, that two competitive mechanisms
manifest themselves  additionally to the ordinary Jaynes-Cummings dynamics: the local-field induced nonlinearity and quantum diffusion being to be the consequence of the interdot tunneling.

It is concluded  in \cite{PSPA}, that it seems to be  substantial for the practical applications, that the 
rabiton  state is quantum coherent state. It also significant, that photons in the given bound state are moving with rather small velocities in comparison with $c$, that is, with the light velocity in vacuum.  The extrapolation of the rabiton velocity to zero means that photon can exist in the state with the zeroth energy and zeroth impulse. In fact, pinning of photons by absorbing centers is taking place. It is substantial  to clarify, which characteristics remain for the rest zero mass photons in the given pinned state. It is spin, which in correspondence with results of \cite{Dovlatova_Yerchuck} is the most fundamental characteristics of quantum states. It is reasonable  to suggeste that the characteristic for topological solitons of SSH-family shape is also retained. It however remains to be unclear, can the photon  in pinned state obtain  a mass or not.

 It has been considered in \cite{Slepyan_Yerchak} the 
  analytical solution of the equations {\eqref{system_diskr_a}--\eqref{system_diskr_b} making use the dispersionless approximation. It was approximated the  initial spatial distributions  by the Gaussian beams
\begin{equation}
\label{d_sh}
C_{A,B}\exp[-(x-d_{A,B})^2/2\sigma_{A,B}^2,
\end{equation}
 where $C_{A,B}$,  $d_{A,B}, \sigma_{A,B}$ are normalization constants, positions of the beams and their widths, respectively (the indexes $A, B$ refer to excited and ground state of the electron in the QD-chain, respectively). 

 It has been found 
\begin{widetext}
\begin{align}
\label{appr_sol_a}
\nonumber A_n(x,t)&=e^{i(kx-\omega t)/2}e^{-\lambda t} \left\{\left[A_{n}(x+v_2^+t,0)\zeta_2^-e^{-i\nu_2^+t}+A_{n}(x+v_2^-t,0)\zeta_2^+e^{-i\nu_2^-t}\right]e^{-ikx/2}\right.\\
 &+\left.\left[B_{n+1}(x+v_1^-t,0)\eta_1e^{-i\nu_1^-t}-B_{n+1}(x+v_1^+t,0)\eta_1e^{-i\nu_1^+t}\right]e^{ikx/2}\right\},\\
 \label{appr_sol_b}
\nonumber B_{n+1}(x,t)&=e^{-i(kx-\omega t)/2}e^{-\lambda t} \left\{\left[A_{n}(x+v_2^-t,0)\eta_2e^{-i\nu_2^-t}-A_{n}(x+v_2^+t,0)\eta_2e^{-i\nu_2^+t}\right]e^{-ikx/2}\right.\\
&+\left.\left[B_{n+1}(x+v_1^+t,0)\zeta_1^+e^{-i\nu_1^+t}+B_{n+1}(x+v_1^-t,0)\zeta_1^-e^{-i\nu_1^-t}\right]e^{ikx/2}\right\},
\end{align}
\end{widetext}
where the velocities $v^{\pm}_{1,2}$ are 
\begin{align}
\label{group_v_a}
v^{\pm}_1&=v^{\pm}(h_1^{(0)})=-2\xi_1a^2k\zeta_1^{\mp},\\
 \label{group_v_b}
 v^{\pm}_2&=v^{\pm}(h_2^{(0)})=2\xi_2a^2k\zeta_2^{\pm},
\end{align}
the frequencies $\nu_{1,2}^{\pm} = \nu_{1,2}(n,\pm k/2)$ can be found from 
\begin{equation}
\label{disp_law}
\nu_{_{1,2}}(n,h) = -\frac{1}{2}\left[\vartheta_1(h)+\vartheta_2(h)\mp \Omega_n(h)\right]
\end{equation}
 after implementing the replacement $h=\pm k/2$ therein. 

Here,

 \begin{align}
\label{theta}
\vartheta_{1,2}(h)=\xi_{1,2}[2-a^2(h\pm k/2)^2], 
\end{align}

\begin{align}
\label{omega}
\Omega_n(h)=\sqrt{\Delta_{eff}^2+4g^2(n+1)},\\
\label{delta_eff}
\Delta_{eff}(h) = \Delta -\vartheta_1(h)+\vartheta_2(h).
\end{align}

  $\Delta=\omega_0-\omega$.

The values  $\zeta_{1,2}^{\pm}$, $\eta_{1,2}$ were introduced to denote the amplitude factors, respectively:
\begin{align}
\label{zeta}
\zeta_{1,2}^{\pm}&=\frac{\Omega_n(h_{1,2}^{(0)}) \pm \Delta_{eff}(h_{1,2}^{(0)})}{2\Omega_n(h_{1,2}^{(0)})}, \
\\
\label{eta}
\eta_{1,2}&= \frac{g\sqrt{n+1}}{\Omega_n(h_{1,2}^{(0)})}. \ \
\end{align}
Generally,  from \eqref{appr_sol_a}, \eqref{appr_sol_b} follows, that any probability amplitude in the Rabi-wave packet \textit{is made of  four components}. Each of them represents the separate subpacket, which is characterized by the own velocity of movement $v^{\pm}_{1,2}$, partial amplitude factors  $\zeta_{1,2}^{\pm}$, $\eta_{1,2}$, and frequency shifts $\nu_{1,2}^{\pm}$. Two of them (first and second terms in expressions \eqref{appr_sol_a},\eqref{appr_sol_b}) correspond to the excited initial state and two another (third and fourth terms in \eqref{appr_sol_a},\eqref{appr_sol_b}) correspond to the ground initial state.
 
The authors  of \cite{Slepyan_Yerchak} argue, that it is essential, that the velocities  $v^{\pm}_{1,2}$ and the frequency shifts $\nu_{1,2}^{\pm}$ depend  on the photon number $n$. It means, that the spatial propagation of wavepacket  is accompanied by the change of quantum light statistics (for example, by the assumption, that the light is initially in the coherent state, Poisson photon distribution transforms with propagation to the sub-Poisson or super-Poisson statistics). The authors indicate also, that the region of asymptotic solution \eqref{appr_sol_a}, \eqref{appr_sol_b} is limited by the appearance of diffraction spreading.
 
 It was remarked in \cite{Slepyan_Yerchak}, that under the some specific conditions the number of subpackets could decrease. Two mechanisms of decreasing are possible:  the first mechanisms is the tending to zero the subpacket amplitude and the second mechanisms is the confluence of the subpackets because of velocities' synchronism. 
 
It is seen from \eqref{group_v_a}, \eqref{group_v_b}, if 
\begin{equation}
\label{synchr_cond}
\Delta_{eff}(h_{1,2}^{(0)})=0,
\end{equation}
for one pair of the subpackets the velocity synchronism condition  $v^{+}_{1,2}=v^{-}_{1,2}$  is fulfilled.  Equation\eqref{synchr_cond} is similar to the ordinary synchronism condition  $\Delta=0$ in the single two-level system.  The velocity synchronism condition for given QD-chain parameters can be fulfilled by fitting the value of the detuning  $\Delta$. In the  given case three subpackets exist instead of four ones. 

If the system is initially prepared in the stationary state, and the synchronism condition is fulfilled in the corresponding point (for example  $B_{n+1}(x,0)=0$ for any $n$ and $\Delta_{eff}(-k/2)=0$), only one subpacket is preserved. Then the expression for the inversion density becomes is: 
\begin{equation}
\label{inv_density_2}
w(x,t)=a\sum \limits_n A_n^2(x+\xi_2a^2kt,0)[1-2\sin^2(g\sqrt{n+1}t)].
\end{equation}
The relaxation factor $\lambda$ for simplicity  was supposed to be zero. Expression \eqref{inv_density_2} sums up contributions of different photonic states (terms with different $n$). Each contribution is the product of two multipliers describing different optical processes. The second multiplier represents the temporal law of Rabi oscillations with the frequencies $\nu_n = 2g\sqrt{n+1}$, occurring in the isolated QD, while the first one indicates the movement of the region of Rabi oscillations along the QD-chain with velocity  $v=\xi_2a^2k$. Remarkably, that in the case of the exact synchronism $v$ does not depend on $n$, whereby in this partial case the quantum statistics of light is not distorted with propagation of the wavepacket. In particular, the initial coherence of quantum light persists in time.  It is immediately follows from the equation
\begin{equation}
\label{int_inversion}
\widetilde{w}(t)=\frac{1}{a}\int \limits_{-\infty}^{\infty}w(x,t)dx.
\end{equation} 
and normalization condition that the integral inversion $\widetilde{w}(t)$ associated to the inversion density \eqref{inv_density_2} oscillates in time identically to the Rabi-oscillations in the single two-level system  (see Eq.(6.2.21) in \cite{Scully} at $\Delta = 0$).

Thr given result allows to understand, why   Rabi oscillations  in the system with interaction between absorbing centers studied could be analysed on the first damping stage within the frames of JCM. Really, on the given stage any desynchronising effect is absent.
yearchuck@gmail.com
The movement of the electron along the QD-chain in the absence of electron-photon coupling is caused entirely by interdot tunneling through the corresponding energy level.  It is shown also in \cite{Slepyan_Yerchak}, that Rabi oscillations lead to a qualitatively new effects in the tunneling. According to the equations \eqref{appr_sol_a}, \eqref{appr_sol_b}, the movement of the initially ground-state subpacket is governed by the tunneling transparency of the excited energy level and vice versa. It implies that the tunnel transition of the Rabi subpacket occurs only through the opposite energy level after leaving the initial level bu means of  the Rabi-jump.  If one of the barriers becomes absolutely opaque, the corresponding pair of subpackets does not move: $v_{1,2}^{\pm}\rightarrow0$ at $\xi_{1,2}\rightarrow0$. If inequality $\xi_1>>\xi_2$ is satisfied, the next surprising mechanism takes place: Rabi oscillations induce  an abnormally high effective tunneling transparency for the initially ground-state subpacket and suppress it for initially   excited-state subpacket. It takes place even for $n=0$ ($B_n(x,0)=0$ for all $n$), whereby it was concluded that the tunneling may be suppressed by the photon vacuum.  
It  was considered the spatial propagation of vacuum Rabi oscillations. It is well known that such type of Rabi oscillations exist in initially excited two-level system strongly coupled with zero-photon light mode, as a result of spontaneous emission \cite{Scully}. The similar effect takes place in the QD-chain, but in contrast to uncoupled two-level systems, temporal oscillations accompanied by its spatial movement. The spontaneous emission support couples single mode zero-photon state with one-photon state only, thus, this excitation can be imagined as a wave beam characterized by the monochromatic Rabi-frequency spectrum and continuous spatial spectrum at the same time. 
 The existence of the momentum exchange between the photon and the e-h pair  is another necessary condition of Rabi subpackets motion:  $v_{1,2}^\pm   \to 0$ at $k \to 0$. 

From comparison of experimental results above presented with SYHB-theory follows the reasonable suggestion, that the character of amplitude and position distribution of oscillation lines in revival groups is determined by responce on the first  Gaussian beam which is formed, being to be result of the first stage damping process. At  the same time we have in fact  an independent task, since the  initial spatial distribution was approximated   by the Gaussian beam in  \cite{Slepyan_Yerchak}, while the first  Gaussian beam which is formed, being to be result of the first stage damping process, describes  the  initial temporal  distribution. To solve the given task wecan take into account the symmetry of Minkowski space, that is, its homogeneity relatively all the coordinates, being to be equal in rights.  homogeneity.

So using the symmetry of Minkowski space for the task above formulated we can consider the chain along t-coordinate.

The vector of the state of the joint system \{resonance  phonons + spin subsystem of absorbing centers\} can be represented mathematically analogously [ however with quite different physical content] to the expression (\ref{state_vector}), that is

\begin{equation}
\label{st_vector}
| {\Psi (x)}\rangle =\sum\limits_n \sum\limits_p (\mathfrak A_{p,n}(x) |\mathfrak a_p,n \rangle +\mathfrak B_{p,n}(x) |\mathfrak b_p,n \rangle ) ,
\end{equation}
where $| \mathfrak b_p,n \rangle = | \mathfrak b_p \rangle\otimes|n \rangle$, $| \mathfrak a_p,n \rangle = |\mathfrak a_p \rangle\otimes|n \rangle$,  in which  $|n \rangle$ is the resonance phonon  Fock state with $n$  phonons, $|\mathfrak a_p \rangle$, $|\mathfrak b_p \rangle$ are
 scalar field functions of  the excited and ground states respectively of  the absorbing center, which is excited at  $p$-th position in discrete time lattice  with spacing $t_1$ = $t'_1 c$, corresponding to the time $t'_1$, which is necessary for the phonon generation by simultaneous absorbing center excitation, $\mathfrak B_{p,n}$, $\mathfrak A_{p,n}$ are the  probability amplitudes, $p \in N$.

 Then  we obtain the following system of differential equations
 \begin{align}
\label{system_dis_a}
 &\frac{\partial \mathfrak A_{p,n}}{\partial x}=-\frac{i\omega_0}{2}\mathfrak A_{p,n}+i \xi'_1(\mathfrak A_{p-1,n}+\mathfrak A_{p+1,n})\\
\nonumber&-i \mathfrak g\sqrt{n+1}\mathfrak B_{p,n+1} e^{i(kx-\omega\frac{pt_1}{c})} - i\Delta\omega \mathfrak B_{p,n}\sum\limits_m\mathfrak A_{p,m}\mathfrak B_{p,m}^*,   \\ \rule{0in}{4ex}
\label{system_dis_b}
\nonumber&\frac{\partial\mathfrak B_{p,n+1}}{\partial x}=\frac{i\omega_0}{2}\mathfrak B_{p,n+1}+i\xi'_2(\mathfrak B_{p-1,n+1}+B_{p+1,n+1})\\
&-i \mathfrak g\sqrt{n+1}\mathfrak A_{p,n} e^{-i(kx-\omega \frac{pt_1}{c})} -i\Delta\omega \mathfrak A_{p,n+1}\sum\limits_m \mathfrak A_{p,m}^*\mathfrak B_{p,m},
\raisetag{35pt}
\end{align}
where $\xi'_{1;2}$ are the electron hopping
frequencies for the excited ($\xi'_{1}$) and ground ($\xi'_{2}$) states of
the absorbing centers, $\mathfrak g $ is the electron-phonon coupling  constant.
The system above presented can be solved in the manner, described in \cite{Slepyan_Yerchak}. However, a priori, taking into account  the mathematically analogous structure of the differential equations, it follows
  from (\ref{system_dis_a}, \ref{system_dis_b}), that any probability amplitude in the Rabi-wave revival groups 
  \textit{is made of  four components}. Each of them represents the separate subpacket, which is characterized by the own velocity of movement $v^{\pm}_{1,2}$, partial amplitude factors  $\zeta_{1,2}^{\pm}$, $\eta_{1,2}$, and frequency shifts $\nu_{1,2}^{\pm}$. Therefore, the suggestion  on the key role of the first damping stage group of oscillation lines, determining the structure of subsequent revival groups, is mathematically proved.
On the other hand, it is also direct proof of the existence of the interaction between absorbing centers.

\subsection{Formation of Long-Lived Coherent States  and Coherent Emission Effect}

 To confirm the conclusion on  the coherent character of a relaxation (which is like [in the sence of its coherence] to a maser emission character) of the anthracite spin system, excited by the resonance  interaction  with a  microwave field [it is illustrated by the spectra, observed at $H_m = 0$, Figure 1 (or 3), Figures 4  to 7], and to  establish the structure of the centers, which are responsible for the  given phenomenon, we have undertaken the study of ESR-responce at various values of HF-modulation amplitude. The spectra, registered in AF-regime by 100 kHz HF-modulation of static magnetic field with amplitude $H_m = 0.01$ $G$ are presented in Figures 23 to 26.
\begin{figure}
\includegraphics[width=0.5\textwidth]{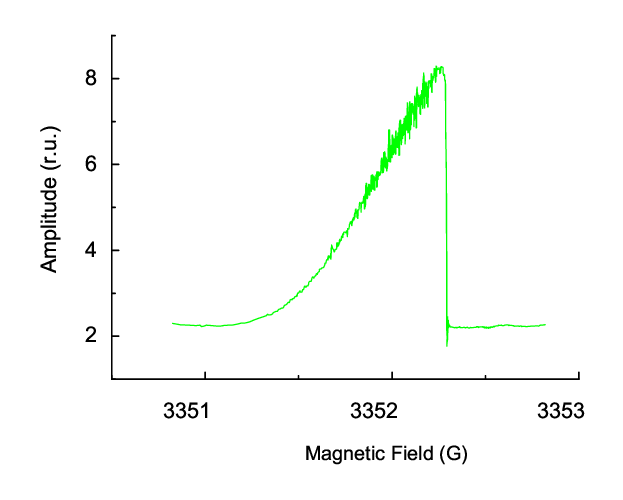}
\caption[ESR spectrum of anthracite sample by the registration with automatic microwave frequency control, $\frac {dH_0}{dt} > 0$]
{\label{Figure5} ESR spectrum of anthracite sample by the registration with automatic microwave frequency control, $\frac {dH_0}{dt} > 0$}
\end{figure}
\begin{figure}
\includegraphics[width=0.5\textwidth]{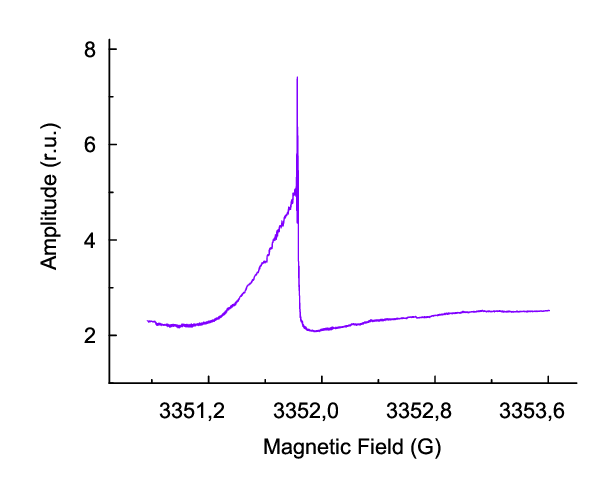}
\caption[ESR spectrum of anthracite sample by the registration with automatic  microwave frequency control, $\frac {dH_0}{dt} < 0$]
{\label{Figure6} ESR spectrum of anthracite sample by the registration with  automatic microwave frequency control, $\frac {dH_0}{dt} < 0$}
\end{figure}
\begin{figure}
\includegraphics[width=0.5\textwidth]{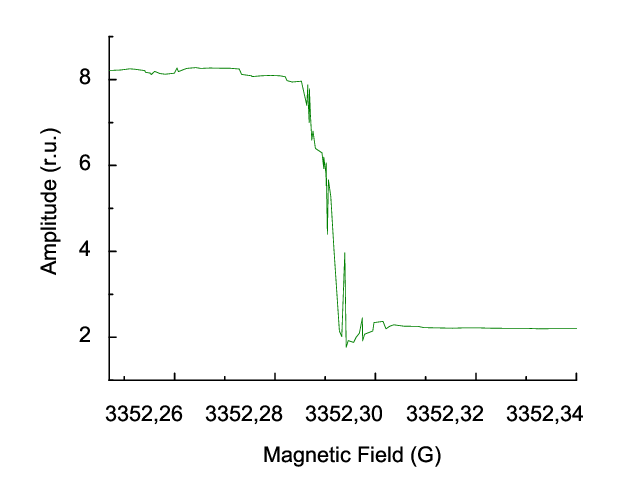}
\caption[Central part of ESR spectrum of anthracite sample by the registration with automatic microwave frequency control, $\frac {dH_0}{dt} > 0$]
{\label{Figure7} Central part of ESR spectrum of anthracite sample by the registration with automatic microwave frequency control, $\frac {dH_0}{dt} > 0$}
\end{figure}

\begin{figure}
\includegraphics[width=0.5\textwidth]{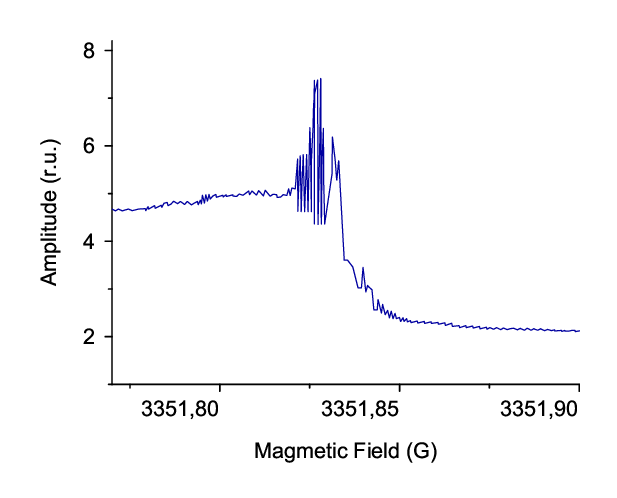}
\caption[Central part of ESR spectrum of anthracite sample by the registration with automatic microwave frequency control, $\frac {dH_0}{dt} < 0$]
{\label{Figure8} Central part of ESR spectrun of anthracite sample by the registration with automatic microwave frequency control, $\frac {dH_0}{dt} < 0$}\end{figure}

 The lines observed have, like to Figure 1 (or 3), Figures 4  to 7, also unusual shape  for  the absorption derivative. The spectra are also very different by the change of the sweep direction, compare Figures
23 and 25 with  Figures 24 and 26, which all were registered  with automatic microwave frequency control without change of the other registration conditions. It is seen from Figures 23, 24, that for both the static field sweep directions one of the spectrum wings has stepwise shape, which usually is characteristic by AFC  cycle skip. At the same time, cycle skip does not takes place, that is confirmed by more detailed representation of the central part of the spectra in Figures 23 and 24, see Figures  25 and 26 correspondingly, and by the simultaneous (parallel) registration of the dependence of the frequency of  the measuring cavity, that is, the  registration of the time dispersion, Figure 27.

\begin{figure}
\includegraphics[width=0.5\textwidth]{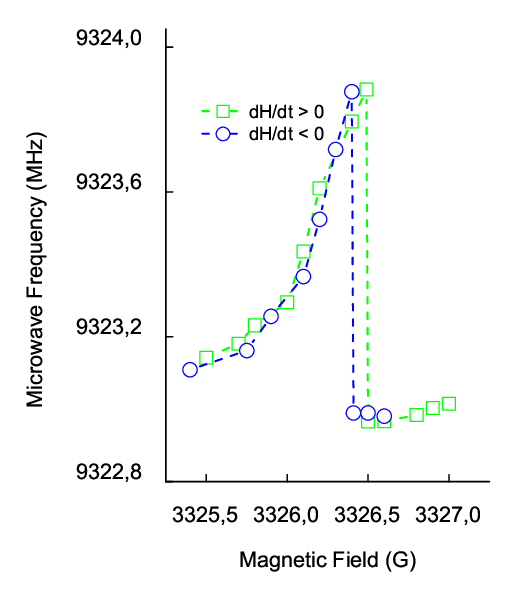}
\caption[Dependence of the frequency of measuring cavity by the ESR absorption ($\frac {dH_0}{dt} > 0$) and emission ($\frac {dH_0}{dt} < 0$) spectrum registration of anthracite sample   with automatic microwave frequency control]
{\label{Figure9}Dependence of the frequency of measuring cavity by the ESR absorption ($\frac {dH_0}{dt} > 0$) and the emission ($\frac {dH_0}{dt} < 0$) spectrum registration of anthracite sample   with automatic microwave frequency control}\end{figure}

 It is seen from Figure 27, that microwave frequency jumps are not exceeding 1 MHz, that is, they are on the order of value less, than the adjustment range of AFC  unit, which is $\simeq$ 10 MHz. It is seen also  from Figure 27, that the  time dispersion responce has practically the same amplitude for both  the sweep directions in distinction from the amplitude and intensity  characteristics of the  spectra, presented in Figures 23 and 24, where the amplitude and the intensity of resonance responces at $\frac {dH_0}{dt} > 0$ markendly exceed the amplitude and the intensity of the  resonance responces at $\frac {dH_0}{dt} < 0$. So, the  amplitude ratio is equal to  2, the intensity ratio is evaluated to be equal $\approx 4.6$. It is interesting, that the step position is different for $\frac {dH_0}{dt} > 0$ and $\frac {dH_0}{dt} < 0$, that is, the phenomenon of the hysteresis in ESR response is taking place. All the  differences for both  the sweep directions disappear by the registration  without automatic microwave frequency control. The spectrum, observed in the given case is the same within limits of the  experimental accuracy for both the sweep directions, it is  presented in Figure 28.

\begin{figure}
\includegraphics[width=0.5\textwidth]{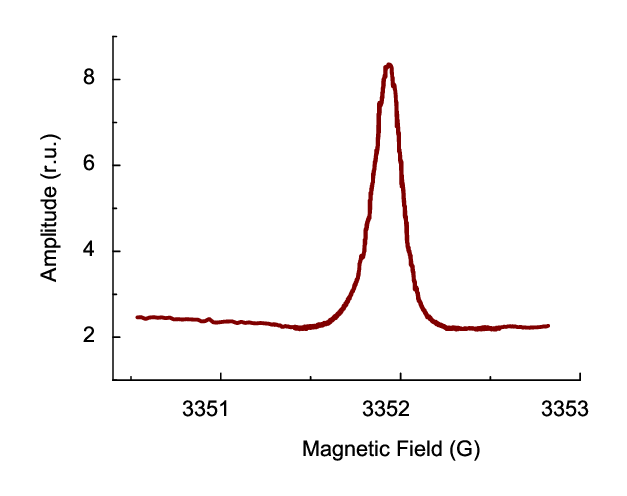}
\caption[ESR spectrun of anthracite sample without automatic microwave frequency control, scan range $\Delta H_0$ = 5 G, scan time $t_0$ = 4 minutes, time constant $\tau$ = 0.1 s, modulation amplitude $H_m$ = 0.01 G, modulation frequency $f_m$ = 100 kHz, $\frac {dH_0}{dt} > 0$]
{\label{Figure10} ESR spectrun of anthracite sample without automatic microwave frequency control, scan range $\Delta H_0$ = 5 G, scan time $t_0$ = 4 minutes, time constant $\tau$ = 0.1 s, modulation amplitude $H_m$ = 0.01 G, modulation frequency $f_m$ = 100 kHz, $\frac {dH_0}{dt} > 0$}\end{figure}

 We see, that by the registration in CF-regime, the
absorption line is registered with  the very pronounced Dyson
effect \cite{Dyson}. It has Dyson shape, which is characteristic
for ESR absorbtion in metals and it is determined by
the prevailing space dispersion contribution \cite{Erchak_Zaitsev_Stel'makh}, which
is appeared in conductive media. Dispersion-like shape
of an absorption ESR-response is also the indication, that
the ESR centers are or strongly mobile themselves or their spin  subsystem is characterised by the very strong spin energy transfer process [the given process seems to be physically different from so-called spin diffusion process]. Let us remark, that in the case of the static
PCs in a conductive media the space dispersion contribution
and the absorption contribution to the  resulting ESR responce in thick samples (in comparison with the  scin depth) is 1 to 1  \cite{Erchak_Zaitsev_Stel'makh} and the line observed has Dyson shape with asymmetry extent $\frac{A}{B}$ = $2.55$ (unlike to the asymmetry extent in the  samples, used in our work).
 So we see, that the model of immobile PCs without spin energy transfer process in their spin  subsystem (or without strong spin diffusion) is not the case. On Dyson effect with the strong mobility of  PC  or strong spin energy transfer process in their spin  subsystem (strong spin diffusion)  indicates  the evident deviation of line shape from Lorentz or Gauss usual dispersion line shapes.  Really, experimental value of  the asymmetry extent $\frac{A}{B}$ is equal to $23.4 \pm 0.5$. (Compare with $\frac{A}{B}$ = $8$ and $\frac{A}{B}$ = $3.5$ for usual Lorentz and Gauss dispersion line shapes correspondingly). The strong spin energy transfer process means in its turn, that in anthracite samples the electron-(spin)-photon interaction is really strong. Strong electron-(spin)-photon coupling at rather
large amount of PCs  represents the sufficient condition for the 
formation of the second cavity in measuring cavity (large amount of PCs is the necessary condition, which was formulated earlier, but it is insufficient). The
qualitative explanation of the given conclusion is the following. The microwave
power, that is microwave  field photons penetrating through the samplle surface in the
sample, remain [are captured and entirely absorbed] in
the sample like to the  microwave power in  a measuring cavity
with metal walls. Any transmission of a microwave power
through the sample does not take place. It is the necessary and sufficient condition for the formation of the
second cavity in a measuring cavity. In other words, the
sample itself becomes to be the second cavity, that clearly
was confirmed by the  direct observation of the additional narrow peak on an oscillogramme of the generation zone of the microwave generator  (the coresponding Figure  is not represented in the given work). By the application of the AF-mode registration the microwave generator frequency follows  the changes  in the frequency of the sample-cavity, determined by the time dispersion of the sample-cavity. The changes in the  frequency of the order of 1 MHz, Figure 27,  are relatively large and the resonance phonons with starting frequency remain in coherent state in the sample [they cannot interact  immediately with microwave photons, which have now the other energy value]. In fact, the  lifetime of a coherent state of resonance phonons is determined by their interaction with nonresonance phonons and it can be very long, if the given interaction is relatively weak. It takes place in the anthracite samples studied and it seems to be the consequence of their disordered structure. By the sweep back of the  static magnetic field $H_0$ ($\frac{dH_0}{dt} < 0$) the phenomenon of an  acoustic spin resonance (ASR) is realised, which is characterised by an own Rabi oscillation process, presented in Figures 1, 3 to 22. It is especially substantial, (let us accentuate once again) that an acoustic Rabi oscillation process has  the quantum nature with the discrete dependence on the time ( the time sweep is connected with 
the  static magnetic field sweep). 
 The phenomenon of  the acoustic spin resonance observed  differs from the  well known phenomenon of acoustic spin resonance, since it can be realised by strongly coherent resonance phonons, being to be the source of the acoustic excitation. Therefore, the spectra represented in Figures 24 and 26  are the demonstration of the  microwave emission.
 
The appearance of the steps in Figures 23 to 26 is explained in the following way. 
The spectral bandwidth of the microwave cavity in the experiments described is of the order of 1 MHz. In fact, the conditions of impulse-wise modulation of  the microwave power were realised by a rather large deviation of a resonance frequency, see Figure 27, in AF-mode registration. The given conclusion is confirmed by the observation of 100-kHz Fourier components of transient oscillation processes, Figures 24, 26.

The conclusion on the photon emission is confirmed by the polarity of the derivative of ASR responce. Really,
it is substantial, that the observed polarity of the derivative of signals observed by usual tuning on the absorption registration with autofrequency control  does not depend on the sweep direction of the  static magnetic field. At the same time, when ESR spectra by $\frac{dH_0}{dt} > 0$ and by $\frac{dH_0}{dt} < 0$ are tuned for the absorption registration, the polarities of qiven spectra have to be different \cite{Weger}. Consequently, one of the spectrum is an emission spectrum. It is 
evident, taking into consideration the energy conservation law, that the emission, corresponding to ASR phenomenon,  is observed by $\frac{dH_0}{dt} < 0$. It is understandable, that by the registration in CF-mode the process of an accumulation of resonance phonons, for which the EM-field is the main heat reservoir, will be not realised. It is the consequence of effective back process, since  resonance phonons have the same energy with resonance photons and the only ESR responce without ASR responce can be registered in full accordance with  the experiment, see Figure 28.   

Let us also accentuate, that, although the emission, Figures 24, 26, is taking place, it is not maser effect (how it was mentioned above), since the amplification is absent (maser effect seems to be in principle impossible for the two-level- system). Then the hysteresis effect (compare Figure 23 and 24), which is observed by the registration in AF-mode, can be explained in a natural way like to  Stokes shift in optics of luminescence spectra relatively the absorption spectra,
where the emission spectrum is shifted to more low frequencies in comparison with excitation spectrum being to be the
consequence of interaction with nonresonance phonons. Given conclusion is confirmed by the study of the dependence of the value of radiospectroscopic Stokes shift
analogue [which was identified in radiospectroscopy for
the first time] in ESR-ASR spectra of anthracite sample,
Figure 29, and by the study of the dependence of the
effective linewidth in ESR spectra for both AF and CF
registration modes, Figure 30, on the amplitude of 100
kHz high frequency modulation of static magnetic field. Really, passage velocity through the resonance is determined by product of $\omega_m H_m$, where $\omega_m$ is modulation frequency value.
It means, that by increase of passage velocity 
the direct interaction of spin-photon subsystem with non-
resonance phonons [ the given interaction seems to be playing the same
role, that the interaction with phonons by a Stokes shift
formation in luminescence studies] can be "turned off",
then emission and absorbtion ESR-spectra have to be registered without Stokes-analogue shift. Therefore, the dependence of the Stokes-analogue shift value on the amplitude of
100 kHz high frequency modulation of the  static magnetic
field has to have the steplike view (determined by tanh in its
approximation) and Stokes-like shift has to vanish at high
modulation amplitude side of the given dependence. It really
takes place, see Figure 29.

\begin{figure}
\includegraphics[width=0.5\textwidth]{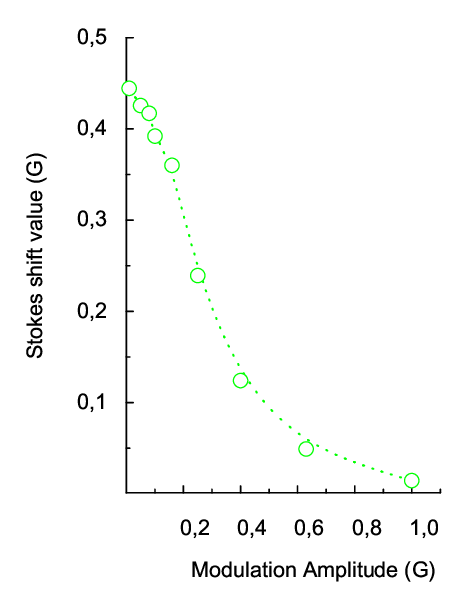}
\caption[Dependence of the Stokes-like shift value in ESR-ASR spectra of anthracite sample on the amplitude of 100 kHz high frequency modulation of the static magnetic field]
{\label{Figure11}Dependence of the Stokes-like shift value in ESR-ASR spectra of anthracite sample on the amplitude of 100 kHz high frequency modulation of the  static magnetic field}\end{figure}

 Moreover, Figure 29 allows to
estimate the lifetime of the  coherent state in the spin-photon subsystem from the $H_m$-value (in frequency units) in inflection of the given curve. It gives the value $T_s = 1.30\times 10^{-6}$ s,
which gets to the typical transverse $T_2$-value range for PCs
in many solid state systems. On the other hand, it seems to be the strong argument of the participation of phonon subsystem in spin-spin interaction processes, in particular, like to those ones in BCS-mechanism of the superconductivity.
 
The possibility to "turn
off" the direct interaction of spin-photon subsystem with
nonresonance phonons from the spin-spin interaction in the AF-mode registration regime  means in its turn, that the dependences of the effective
linewidth in ESR spectra by AF and CF registration modes  have to be qualitatively different.
It is actually so, see Figure 30.

\begin{figure}
\includegraphics[width=0.5\textwidth]{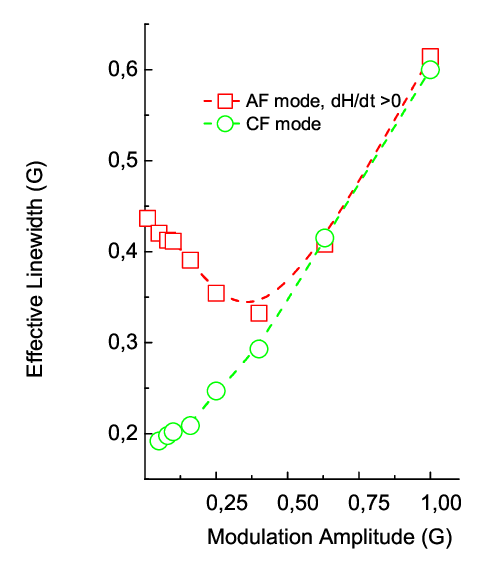}
\caption[Dependence of the effective linewidth in ESR spectra of anthracite sample on the amplitude of the  high frequency modulation of the static magnetic field for both AF and CF registration modes]
{\label{Figure12}Dependence of the effective linewidth in ESR spectra of anthracite sample on the amplitude of the  high frequency modulation of the static magnetic field for both AF and CF registration modes}
\end{figure}

 In AF-mode  at small modulation amplitudes the contribution of nonresonance phonons in spin-spin relaxation mechanisms can be essential, and it has to lead and it really
leads to broadening of the registered effective ESR line, at that, it
seems to be evident, that effective linewidth value has to increase
with  moving along abscissa axis to the left and to decrease (without
taking into account the modulation broadening) with 
moving to the right from the $H_m$-value, corresponding to the inflection value in Figure 29 (equaled to 0.274 G).
 The explanation proposed will
be true, if the genuine linewidth value is equal or exceeds the
vaue 0.274 G. It is evident from CF-mode registration,
that the genuine linewidth value is less than 0.274 G, it is
equal to 0.18 G. It means, that the  linewidth at inflection
(that is, at $H_m$ = 0.274 G) will be modulationary broadened. The competition of modulation broadening and narrowing being to be the consequence of decrease of the contribution  of nonresonance phonons into relaxation
   processes  by $H_m$ > 0.274 G shifts the
minimum position to the value of $H_m$ $\approx 0.4 G$, see Figure 30,
in AF-mode dependence of the effective linewidth. Taking into account the value of modulation broadening at
$H_m$ = $\approx 0.4 G$, obtained from CF-mode dependence of the
effective linewidth, we can also obtain the independent evaluation of the value of $T_s$. It is equal to $T_s = 1.36 \times 10^{-6}$ s. Some difference of the values of $T_s$, obtained from
experimental dependences, presented in Figures 29 and
30, is within the limits of the experimental inaccuracy, estimated to be equal to $10^{-7}$ s. The value of $T_s$ allows
to evaluate the range for the time of a spin polarization
transfer, corresponding to so-called spin diffusion time
$T_D$  in Dyson effect. In the case of normal skin effect the
value of $\frac{A}{B} = 23.4 \pm 0.5 $ means, that $\frac{T_D}{T_s} < 10^{-4}$ and, consequently, $T_D < 1.3 \times 10^{-10} s$. Like to conclusions from ESR studies in nanotubes (NTs) incorporated in diamond single
crystals \cite{Ertchak}, \cite{Ertchak_Stelmakh}, we can conclude, that the absorbing
centers in anthracite are characterized by two times of
spin-lattice relaxation, by $T_1$, which is comparable with the
relaxation time of usual point PC, for instance, in diamond single crystals ($10{-3} s - 10{-4}$ s) and by very short
time $\tau$ ($\tau$ < $10^{-10}$ s) of conversion of the energy of the
spin system into the mechanical energy. It can be considered
to be consequence of the fundamental in magnetic resonance
phenomenon of gyromagnetic coupling between magnetic
and mechanical moments. Therefore, $\tau$-process is a very
fast process in comparison with any process, realized by
means of the  pure magnetic interaction, the relaxation times
of which are not shorter, than $10{-10}$ s \cite{Abragam}. The possibility of the given conversion can be realized naturally, if the
energy transfer activation by (or from) absorbing PC is
very small. It is typical for mobile topological solitons
like them ones, identifed in \textit{trans}-polyacetylene (t-PA) \cite{Heeger_1988} 
 and in carbon NTs \cite{Ertchak}, \cite{Ertchak_Stelmakh}, incorporated in diamond single crystals. The main difference
in the properties of PC in t-PA and in carbon NTs on
the one hand and absorbing centers in anthracite samples on the
other hand consists in that, that in the given case the energy
transfer from immobile centers is realized by rather fast
formation of the  resonance phonon system. Let us remark, that the characteristic times for the given process above evaluated 
($\lesssim{10{-10}}$  s) are more long (from order upto
two orders of the values) in comparison with $\tau$-process times in
t-PA \cite{Weinberger}, \cite{Nechtschein} and in carbon NTs \cite{Ertchak}, \cite{Ertchak_Stelmakh} incorporated in diamond single crystals.

The key to establish the structure of ESR centers in anthracite can be found by
the analysis of the central part of ESR spectra by the registration with automatic microwave frequency control with nonzeroth modulation amplitude,
presented in Figures 25 and 26. It is seen from Figure 26,
that the acoustic Rabi oscillation process is also takes place,
although spectral resolution is not very good, since the
effective modulation amplitude is greater (up to 4 times), than by the registration of the spectrum, represented in Figures 1, 3 to 7, the scan
range was also 4 times more long by the same scan time
and the registration time constant. We see, Figure 25, also, that there is
mapping of the main, that is, photon-assisted transient ESR-oscillation process, determined by an interaction of the spin subsystem with microwave field photons.
However, the only Fourier low frequency component, equaled to 100
kHz, is registered, the shape of which can  be not
corresonding to the genuine shape of the oscillation process, obtained by a standard transient ESR-spectruscopy technique.
The most interesting is the shape of  the background line.
In both the spectra the shape of the  background line has the
step-like form and can be approximated by tanh-function. At the same time, background lines in Figures 25 and 26 being to be approximated by tanh-functions have different numerical parameters in the given approximation (it is clearly seen , that the background line, presented in Figure 25 is substantionally steeper). It allows to exclude the possible radiotechnical effect of the role of the  cavity bandwidth on the shape of the background lines observed. The shape of the background lines  seems to be determined by the time distribution of the dynamical spin polarization of absorbing centers.
It can mean, that the model of the ESR centers has to
be similar to Su-Schrieffer-Heeger (SSH) model of topological solitons in \textit{trans}-polyacetylene  \cite{Heeger_1988}, in which however the dimerization coordinate is the time $t$ instead of the space coordinate $z$. The $g$-factor
value, which is very near to the $g$-factor value of neutral solitons in t-PA testifies in favour of the given proposal. It seems
to be reasonable to suggest, that the structural element of anthracite samples includes the corresponding element of the carbon backbone of t-PA. At the same time, the  anthracite is the structurally
disordered material. It means, that the  topological solitons has to
be pinned and they have to give usual ESR spectra. To
explain qualitatively the appearance of step-like spectral functions in
AF registration regime [at that it is substantionally, that
they are characterized by different parameters for absorption and emission spectra] and the possibility to register acoustic
Rabi oscillations, we have to consider the fully quantized
joint \{EM-field + magnetic dipole + resonance phonon
field\} system. EM-field and the  field of resonance phonons produce the radiation communication in the direction of EM-field propagation and in the time. It means, that we have
1D chains along $t$-axis in Minkowski space in the given joint system, at that, the chains of
two kinds are produced in the time - with the time spacing $t_1$, characterising the process of the resonance phonons' generation in ESR condition by the direct  sweep  of the  static magnetic field, that is  by $\frac{dH_0}{dt} > 0$ and with the time spacing $t_2$, characterising the process of the resonance photons' generation in ASR condition by the reverse  sweep direction of the  static magnetic field, that is,  by $\frac{dH_0}{dt} < 0$. The formation of the 1D chains along $t$-axis in Minkowski space is confirmed experimentally, the proof of which is given  in the foregoing Subsection. Moreover, in the foregoing Subsection  is given the proof of the existence of the interaction between the absorbing centers. 1D chains along $t$-axis produce in its turn the ordered 1D chains in disordered sample along the direction of the EM-field propagation. At that, in the case of the uniform distribution  of the
absorbing centers in the sample volume the spacing parameter will be constant along  the chain position number. The formation of the 1D chains along $t$-axis in  EM-field propagation direction seems to be the result of the affinity of the structure of EM-field and carbon systems, described in \cite{DAA}. The full experimental confirmation of the QFT model of EM-field is given in \cite{ABE}. Therefore, in the presence of EM-field in structurally disordered  anthracite sample structure the ordered 1D chain structures are produced. The value of $g$-factor allows to suggest that initially they are metalline with half filled $c$-band. Then, in result of $t$-Peierls transition the $t$-dimerisation takes place with $t$-dimerisation  parameter $\delta t$ of  1D chain structures, which in the same time become to be twofold degenerated. Like to SSH- model the twofold degeneration  leads to the formation of $t$-analogue of SSH topological solitons - instantons. They are the centers in ESR conditions, which   absorb photons and generate the resonance phonons. In the case
of ASR the instantons of the other kind are produced. They are the centers, which   absorb phonons and generate the resonance photons.

 Therefore, we introduce into consideration the new quantum physics objects - topological instantons of two kinds.
 The instantons
of the first kind absorb the photons of a radiospectroscopy range 
and relax mainly by means of the generation of coherent
hypersound phonons. Given prevailing coherent relaxation process determines rather narrow ESR-linewidth
(equal to 0.18 G) in structurally disordered anthracite
samples to be direct consequence of the relaxation mechanism coherence. The instantons of the second kind absorb hypersound phonons and relax mainly by the generation
of coherent microwave photons.

 It is understandable,
that the instantons of both the kinds can free, that is
without additional activation energy, "propagate" along
$t$-axis like to SSH-solitons' moving along space $z$-axis. The very
short value of the spin-lattice $\tau$-relaxation, being  $\lesssim{10^{-10}}$  s, which is required for the  formation of resonance phonon system, above evaluated is the
strong experimental support in the favour of the model
proposed. Therefore, let us accentuate once again, that the
given  qualitative model allows to explain the narrowing of ESR lines
in structurally disordered materials like to the samples studied, at that, they are in more than 50 times narrower in
anthracites in comparison with ESR lines in a  very good ordered related systems - carbon nanotubes, produced by
high energy implantation of diamond single crystals \cite{Ertchak}, \cite{Ertchak_Stelmakh}. It seem to be the first experimental detection of
instantons (to our knowledge) at all.
The models of known instantons in the literature are considered
theoretically to be a special kind of vacuum oscillations
in gluon field theory and also in gravitation field theory \cite{Radjaraman}. They have quite another origin.

Therefore, we have observed the new kind of coherent
states. They are modes of the resonance phonon field. The
coherence time value at room temperature in the samples studied is   exceeding 4.6 minutes, that is, it is  the maximal known value at
present. The substantial advantage for possible applications is that, that the responce on the external
EM-field   scan is very strong. At the same time, the detection
of a single spin is  a rather technically difficult task.

It is well known, that many complex disordered systems in nature are existing in the state which does not equilibrate with their environment
below certain temperature. The most wide class of disordered systems existing in nonequilibrium state consists of various glasses, from window glass to spin
glass \cite{Struik}. The glassy metastable frustrated state gives rise to novel and unusual behavior
like to memory effects, ageing and nonlinear dynamics. 
\begin{figure}
\includegraphics[width=0.5\textwidth]{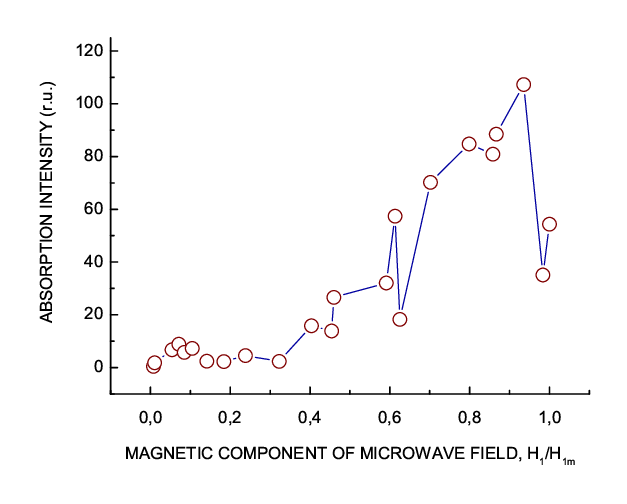}
\caption[Dependence of the absorption intensity in ESR spectra of anthracite sample on the amplitude of the   magnetic component of microwave field in CF registration mode]
{\label{Figure12ac} Dependence of the absorption intensity in ESR spectra of anthracite sample on the amplitude of the   magnetic component of microwave field in CF registration mode}
\end{figure}
\begin{figure}
\includegraphics[width=0.5\textwidth]{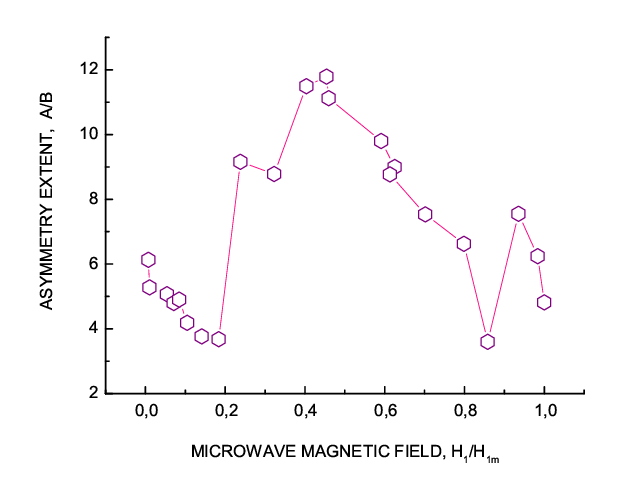}
\caption[Dependence of the asymmetry extent in ESR spectra of anthracite sample on the amplitude  of the   magnetic component of microwave field in CF registration mode]
{\label{Figure12a}Dependence of the asymmetry extent in ESR spectra of anthracite sample on the amplitude  of the   magnetic component of microwave field in CF registration mode}
\end{figure}
\begin{figure}
\includegraphics[width=0.5\textwidth]{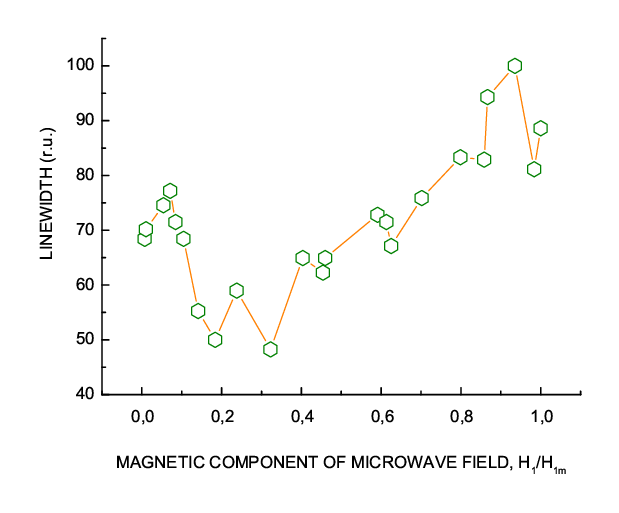}
\caption[Dependence of the asymmetry extent in ESR spectra of anthracite sample on the amplitude  of the   magnetic component of microwave field in CF registration mode]
{\label{Figure12ab}Dependence of the effective linewidth in ESR spectra of anthracite sample on the amplitude  of the   magnetic component of microwave field in CF registration mode}
\end{figure}

 The glass family
 includes additionally a number of specific system - domain walls \cite{Repain},
two dimensional electron gas \cite{Vaknin} and Abrikosov vortices in type II superconductors under
magnetic fields (see \cite{Du} and references therein). It means, that all type II superconductors can display memory effects.
The glassy vortex state is usually prepared by field-cooling (FC) the sample containing
quenched disorder across the superconducting transition temperature Tc. A memory
phenomenon in superconductors is appeared if  to apply  a perturbation
which suddenly disturbs the superconductor. After removing the perturbation, subsequent
response shows the memory effects \cite{Du}. Experiments on different conventional
superconductors \cite{Henderson} and in Y-Ba-Cu-O single crystals \cite{Valenzuela} demonstrated the presence of a
history effect similar to those ones in Heisenberg spin glasses.
Here we report  In strongly
anisotropic high Tc superconductors an entirely different manifestation of the memory effect takes place. The  magnetic field directed parallel to the layers after its penetration in the sample represents itself a system of Josephson vortices (JV), which are qualitatively
distinct from Abrikosov vortices (AV) appearing when the magnetic field is perpendicular
to the layers (or in less anisotropic materials like to  most conventional superconductors and
optimally doped Y-Ba-Cu-O). 

It is well known that for sufficiently strong disorder the AV
form vortex glass \cite{Blatter}. In JV system a randomly distributed set of defects of the layered
structure (microresistences \cite{Barone}), is analogous to the quenched disorder, which leads to pinning of  AV in the
conventional vortex glass. It is argued in in \cite{D_Shaltiel} that pinning of JV has to be less effective than
 pinning of the AV since the JV are coreless. However, accordig to results represented in  \cite{Barone} the JV system
can also be pinned.
 The formation of the glassy state of the JV  in 
$Bi_2Sr_2Ca_1Cu_2O_\delta$ (Bi2212) and $Bi_2Sr_2Ca_2Cu_3O_\delta$ (Bi2223) single crystals was demonstrated in \cite{D_Shaltiel}, \cite{Shaltiel}. The memory
properties were studied by measuring the microwave power absorption (MPA) of the sample
subjected to colinear DC and AC magnetic fields parallel to the layers. MPA method is
sensitive for the  registration of weakly pinned JV, while strongly pinned JV do not contribute to the
signal. In the experiments, described in \cite{D_Shaltiel}, \cite{Shaltiel} the JV glass was prepared by field cooling (FC) the
sample from above Tc to Tm = 4.2K subjected simultaneously to the DC magnetic field Hm parallel to the
a-b crystal plane. At Tm = 4.2K the external field was decreased to zero and after
applying a low frequency AC field it was subsequently increased again. No MPA signal was
detected until the external magnetic field reaches the initial field Hm. The peak of the MPA
signal was appearing just above Hm, demonstrating the new memory effect. The dependence of the
signal on deviation of magnetic field H - Hm was the same in the wide range of the cooling fields.

Therefore, the observation of magnetic  memory effects can be considered to be  the strong indication that the state of the system studied is superconducting state.
Concerning the anthracite samples studied it is also the strong argument in the favour of the suggestion  of  their referring to spin glass systems.
 
There have been studied the dependencies of parameters [intensity, asymmetry extent and linewidth] of the ESR signal, registered in CF-mode, on  the   magnetic component of the  microwave field. They are represented on Figures 31 to 33. They are very unusual, being to be strongly nonlinear and being to be characterised by the appearance of the stepwise regions of the parameters' change. 

The stepwise changes in the dependencies of ESR parameters on  the   magnetic component of the  microwave field were observed in radiospectroscopy for the first time. So, the initial part of the intensity dependence, the segment $[H_1^0, H_1^1 = 0.08 H_{1m}]$, where $H_1^0$ and $H_{1m}$ are, correspondingly, the  minimal and the maximal  values of the   magnetic component of the  microwave field used,  can be described by the linear dependence of the ESR signal. Then, the some decrease of the intensity takes place on the segment $[0.08 H_{1m}, 0.16 H_{1m}]$. On the segment $[ 0.16 H_{1m}], 0.40 H_{1m}]$ the intensity is weakly dependent on the microwave power, although weakly  pronounced peaks with minima at $0.18 H_{1m}$  and $0.32 H_{1m}$ can be identified. Then, on the segment  $[0.40 H_{1m}, H_{1m}]$, the practically
linear growth of the ESR signal intensity is  observed, being, however, overlapped
with sudden almost stepwise changes. Minima of sudden stepwise change of the intensity were observed at the values of the   magnetic component of the  microwave field, equaled to 
 $0.45 H_{1m}$, $0.62 H_{1m}$, $0.86 H_{1m}$,  $0.98 H_{1m}$, Figure 31. The dependence of  the linewidth on
the   magnetic component of the  microwave field has the same character, moreover, the positions of peak minima are strictly coinciding with those ones, observed on
the dependence of the intensity on
the   magnetic component of the  microwave field, Figure 32. The difference consists in that, 
that the peaks at $0.18 H_{1m}$  and $0.32 H_{1m}$ are essentially more pronounced.
The dependence of asymmetry extent on
the   magnetic component of the  microwave field has the similar character, Figure 33, however, the positions of peak maxima at $0.32 H_{1m},  0.45 H_{1m},  0.86 H_{1m}, 0.98 H_{1m}$  corresponds to the positions of  peak minima on the dependencies of  the ESR signal intensity and linewidth  on
the   magnetic component of the  microwave field. The only minimum at $0.62 H_{1m}$
 corresponds to the position of  peak minimum on the dependencies of  the ESR signal intensity and linewidth  on
the   magnetic component of the  microwave field.

The presence of sudden stepwise change  is characteristic for  the current-voltage
dependencies, observed in superconductors \cite{Ustinov}.  

Let us consider the given property in some details.
We are starting for the  convenience  of a  readership  from the definition of a fluxon and its physical characteristics.
A fluxon in a long Josephson junction carries a magnetic
flux equal to 1 flux quantum $\Phi_0$ = $h/2e$ = $2.07\times 10^{-15}$ Wb. The magnetic
flux is generated by a circulating supercurrent, often
called Josephson vortex, located between two superconducting
films separated by a few nanometer thin
layer of  an insulator. Mathematically,
the fluxon corresponds to a $2\pi$ kink of the
quantum phase difference $\varphi$ between the two superconducting
electrodes of the junction. The perturbed
sine-Gordon equation which describes the dynamics
of the system \cite{Parmentier}, \cite{McLaughlin}, in normalized form, is
\begin{equation}
\label{fluxon}
\varphi_{xx} - \varphi_{tt}  - \sin\varphi = \alpha \varphi_{t}  - \beta\varphi_{xxt} -\gamma,
\end{equation}
where the time $t$ is measured in units of $w^{-1}_0$,  $w_0$ is the
Josephson plasma frequency, the spatial coordinate $x$
is measured in units of the Josephson penetration depth
$\lambda_j$, $\alpha$ is a dissipative term [dissipation is realised by means of quasi-particle tunneling,
normally assumed ohmic], $\beta$ is a dissipative term, taking into account 
the surface resistance of the superconductors, and
$\gamma$ is a normalized bias current density. To account for
the behavior of a real system, the equation \ref{fluxon} must be solved
together with appropriate boundary conditions, which
depend on the junction geometry, and take into account
the magnetic field applied in the plane of the junction \cite{Parmentier}, \cite{McLaughlin}, \cite{Pedersen}, \cite{ParmentierR}.
A unique property of real Josephson transmission lines (JTLs) is that the parameters
$\alpha$, $\beta$, and $\gamma$ are rather small. The equation \ref{fluxon} with zero r.h.s. has the soliton solution  in the form
\begin{equation}
\label{fluxon_solution}
\varphi = 4 \arctan[\exp(\frac{x-ut}{\sqrt{1-u^2}})],
\end{equation}
 where $u$ is the soliton velocity. The given solution
  describes a $2\pi$-kink
moving with a velocity $u$. According to the perturbation
theory by McLaughlin and Scott \cite{McLaughlin}, the velocity $u$
is determined by a balance between the losses, governed
by $\alpha$ and $\beta$, and the energy input owing to the bias
$\gamma$. The velocity $u$ is measured in units of $\overline{c}$, where $\overline{c}$
is the velocity of light in the junction, also called the
Swihart velocity. The velocity $u$, according to the perturbational approach,
 can be  approximately given by the expression
\begin{equation}
\label{velocity}
u = \frac{1}{\sqrt{1+(\frac{4\alpha}{\pi\gamma})^2}}.
\end{equation}
It is seen, that the velocity 
$u$ may assume values
0 < $u$ < 1, at that  for low values of the bias current
$ u \sim \gamma$, while for large values of $\gamma/\alpha$  a saturation
occurs, in result of which the velocity approaches unity. In other words,
a fluxon behaves in the given case like a relativistic particle with respect
to the Swihart velocity. If $u \rightarrow 1$, the magnitude
of the local magnetic field $ \varphi_{x}$  in the center of the
fluxon increases and its width decreases according to
the Lorentz contraction.

A fluxon moving in a JTL of length $L$  generates the
DC voltage $V = \Phi_0 \overline{c} u/L$, \cite{Ustinov}. In experiments with long
Josephson junctions, the signatures of fluxon motion
are the so-called zero-filed steps (ZFSs) at voltages, \cite{Ustinov}
\begin{equation}
\label{ZFS}
V_n = n\frac{\Phi_0 \overline{c}}{L},
\end{equation}
 which appear in the current-voltage
characteristics ($I-V$ curve) of the junction. The zero-filed steps 
 were first observed by Fulton and Dynes in 1973
\cite{Fulton}. Fulton and Dynes suggested that the step index $n$  is equal to
the number of fluxons oscillating in the junction. The characteristic feature  of the current-voltage dependencies with the
ZFS is the equidistant positions of ZFSs.
 The
ZFS appeaance is the consequence of the dependence of the average fluxon
velocity $u \sim V$ on the driving force $\gamma \sim I$ \cite{Ustinov}.

Let us compare the positions of the steps on  the Figures 31 to 33 with the positions of the steps, determined by  the expression (\ref{ZFS}).  The distances between the individual steps are the following - $(0.14 \pm 0.01) H_{1m}$, $(0.13 \pm 0.01) H_{1m}$, $(0.17\pm 0.01) H_{1m}$, $(0.24\pm 0.01) H_{1m}$, $(0.12\pm 0.01) H_{1m}$. We see, that the first three steps are equidistant within  the accuracy of the measurements, some  deviation of the distance between the third and the fourth steps from the equidistancy, equaled to $0.03 H_{1m}$,  takes place. The only the deviation of the distance between the  fourth  and the fifth steps from the equidistancy, equaled to $0.10 H_{1m}$, seems to be  rather large. Then, the distance between  the fifth and the sixth steps is almost corresponding to the initial distance between the steps. It seems to be reasonable to suggest, that the observed  steps on  the Figures 31 to 33 are 
ESR analogues  of the ZFSs, observed in  the current-voltage
characteristics. They can be called ESR-ZFSs. The deviation of the distance between  the steps from the equidistancy, beginning  from the  fourth step, can be explained by the nonlinearity of the mechanism of the resonance phonon   generation, taking place in the samples studied at magnetic component value, equaled to $0.562 H_{1m}$ and greater. Just the resonance phonon   generation process is crucial for the transition of the sample into superconducting state in ESR conditions, see further for details.

There is, however,  the  other mechanism for the appearance of the steps on the current-voltage
characteristics  of the Josephson junctions in  the nonequidistant step range. It is the possibiity of Cherenkov radiation.
The idea of  Cherenkov radiation by a soliton
moving in a Josephson junction has been discussed
in several theoretical works \cite{Kivshar}, \cite{Mints}, \cite{Kurin}. The
first experimental evidences for Cherenkov radiation
have been obtained by Goldobin et al in \cite{Goldobin}. 
The mechanism of the Cherenkov radiation phenomenon is very general:
Cherenkov radiation can be generated, if the fluxon velocity
$v = u \overline{c}$ is equal to the phase velocity $\omega/k$ of
linear electromagnetic waves. The given condition can be
satisfied if the fluxon velocity $v$ exceeds the lowest
phase velocity of linear waves in the junction.  The given possibility seems  to be realised in the range [$0.62 H_{1m}$ - $H_{1m}$]. Really, in the favour of  the given idea  indicate the studies of the asymmetry extent of the resonance signals observed in CF registration mode on the HF-modulation amplitude, Figure 34.

\begin{figure}
\includegraphics[width=0.5\textwidth]{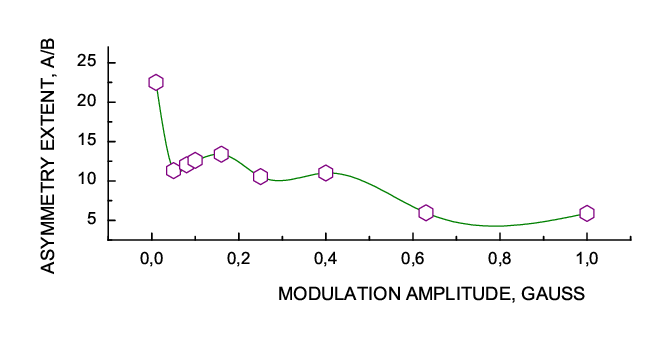}
\caption[Dependence of the asymmetry extent value in ESR spectra of anthracite sample registered in CF  mode on the amplitude of 100 kHz high frequency modulation of the static magnetic field]
{\label{Figure11Q}Dependence of the asymmetry extent value in ESR spectra of anthracite sample registered in CF  mode on the amplitude of 100 kHz high frequency modulation of the  static magnetic field}\end{figure}

It is seen from Figure 34, that the given dependence is very unusual, it has oscillation character. The positions of minima are the following: 0.053 G, 0.282 G, and 0.775 G. It is clear, that the oscillations are nonsinusoidal - the oscillations' period is increased with the modulation amplitude increase. It is the first difference between the dependencies, represented in Figure 34 and in Figure 3 in \cite{Goldobin}, where
the phase gradient profiles for various points of the current-voltage characteristic (IVC)
 are represented. If the soliton (fluxon) velocity $u$ exceeds the phase velocity of emitted linear electromagnetic out-of-phase
waves $c_-$ [both the solitary wave (fluxon) and the linear electromagnetic in-phase and out-of-phase waves are the solutions of the same sine-Gordon equation,\cite{Ustinov}], an
oscillating wake arises behind the moving soliton. It is clear seen from  Figure 3b. Just The given wake is interpreted being to be the display of Cherenkov radiation.
With increasing of the soliton velocity, the wavelength of the
radiation increases  and the amplitude
and length of the wake grow quickly, Figure 3c  in \cite{Goldobin}. 
 At some velocity $u$ the Cherenkov radiation wake extends
over all length $L$ of a long Josephson junction, so that soliton  "sees" its own radiation
wakes after turning around the junction.  The interaction of the soliton
with its Cherenkov radiation wake results in the appearance
of resonances on the IVC of the system at $u > c_-$.
 They were  called  in \cite{Goldobin}
 Cherenkov steps.
There is the second difference between Figure 34 and Figure 3 in \cite{Goldobin} - the direction of  the abscissa axis is opposite. 

Let us give the explanation of the resemblance of dependencies on the Figures above indicated and their differences. It is well known, that any change of the phase of  the magnetic component of microwave radiation by its propagation into the sample, that is, the presence of the space dispersion [which always takes place in conductive media], leads by its classical consideration to Dyson effect, that is, to the deviation of an  $A/B$ ratio from the value  $A/B = 1$. Qualitavely the same picture seems to be taken place by  the quantum treatment of   Dyson effect, although the given task is not considered at present. So, we can insist, that there is the biective mapping between the ordinate axes in both the Figures compared.
On the other hand, the appearance of the dependence itself of the  A/B ratio on the HF-modulation amplitude  by the presence of the relaxation $\tau$-process with very short relaxation times above evaluated, and which according to the aforegoing proof determine the shape of the resonance line in CF registration regime, that is, the $A/B$ ratio  is  possible the only in the case when the velocity of the microwave EM field propagation itself is strongly decresed. It relly takes place, since the fluxons being formed from the  magnetic component of the microwave EM-field are propagating with the velocity $u$, being much smaller [almost in  two order of the value \cite{Goldobin}] in comparison with the  classical velocity, which is near to the light velocity $c$ in vacuum. At the same time the product $\omega_m H_m$ determines the passage rate through the resonance, that is, the penetration depth of fluxons for the time corresonding to passage time through the resonance. It is understandable that the  penetration depth, which corresponds to the abscissa axis in Figure 3 in \cite{Goldobin}, will be decreasing with $H_m$ increasing, that is, we obtain the clear explanation for the opposite directions of the abscissa axes in  the compared Figures. It is shown in \cite{Goldobin}, that with increasing soliton velocity the 
length of the wake grows quickly. The mechanism  of the  soliton velocity increase by the increase of the strength of HF-modulation  field seems to be evident, if to take into account the results of the work \cite{Kim}, where it has been shown that a small magnetic field alternating at audio frequencies causes an amplitude
modulation of the electric field associated with the flux motion oscillating at microwave frequencies. the inhomogeneous damped Mathieu equation was derived for the numerical description of the given effect, however the way of the analitical 
 solution  of damped Mathieu equation has been found for the first time the only in \cite{ME}.

Therefore, the first difference between the dependencies, represented in Figure 34 and in Figure 3 in \cite{Goldobin} consists in that, that in Figure 3 in \cite{Goldobin} the  dependencies are represented by four  fixed soliton velocities, while  the dependence, represented in Figure 34 in the given work indicates on the increase of the soliton velocity with increasing of the amplitude of 100 kHz high frequency modulation. 

Thus, the result, illustrated by Figure 34, indicates on the Cherenkov radiation phenomenon, and consequently the steps in the range [$0.62 H_{1m}$ - $H_{1m}$], Figures 31 to 33  are Cherenkov steps.
Summing all the results, we can insist, that the superconductive state in the anthracite samples studied is really produced  at room temperature in ESR conditions in the accordance with the theory, presented in \cite{QAF}

Practical significance of the results  presented in \cite{QAF} is the following [we will cite the corresponding fragment]. 

	"By the study of the properties of the solids or quantum liquids [for the subsequent using in engeneering] situated in the cavity [for instance, by radio or optical spectroscopy methods] we have to take into account that two discrete $\vec{k}$ spaces will be formed with different discretness extent, determined by interatomic spacing in solids  on the one hand and  by cavity modes on the second hand. At that, they can be incommensurable. In given case, the cavity FA-field phonons can lead by the interaction with electronic subsystem of solids practically the only to exchange by the energy and the impulse between electrons [that is, to the formation of Cooper pairs]. 
Really, the processes of electron scattering with the generation or the absorption of FA-field phonons are impossible, since the probability of the processes of electron scattering with the generation or the absorption of phonons, for instance, for nondegenerated semiconductors is determined by  the known expression
\begin{equation}
\label{eq29}
W^{\pm\vec{q}_j}_{\vec{k}_l\rightarrow\vec{k}_m} = \frac{2\pi}{\hbar}{\mid M^{\pm\vec{q}_j}_{\vec{k}_l\rightarrow\vec{k}_m}\mid}^2 \delta(\mathcal{E}_{\vec{k}_l}-\mathcal{E}_{\vec{k}_m} \mp \hbar\omega_{\vec{q}_j})[N_{\vec{q}_j} + \frac{1}{2} \pm \frac{1}{2}],
\end{equation}
upper signs in which correspond to the phonon emission, lower signs  correspond to the phonon absorption.
Here $\vec{k}_l$, $\vec{k}_m$ are the  quasiwave vectors of the lattice, $l,m = 1, 2, ..., N'$, determined by discrete structure of $\vec{k}$ lattice space with the discretness
\begin{equation}
\label{eq30}
 k_n = \frac{2\pi n}{N'}, n = 1, 2, ..., N', 
\end{equation}
 $N'$ is the number of the atoms in the lattice, $\mathcal{E}_{\vec{k}_l}$ and $\mathcal{E}_{\vec{k}_m}$ are the values of the energy of the electron in the states with  quasiwave vectors $\vec{k}_l$ and $\vec{k}_m$   correspondingly, $\vec{q}_j$ is quasiwave vector of absorbed or emitted phonon of the lattice with the value $q_j$, satisfying to the  relation (\ref{eq30}), by the replacement $k\rightarrow q$, $\omega_{\vec{q}_j}$ is circular phonon frequency, $N_{\vec{q}_j}$ is a number of phonons with  quasiwave vector $\vec{q}_j$,
$M^{\pm\vec{q}_j}_{\vec{k}_l\rightarrow\vec{k}_m}$ is matrix element of the transition $\vec{k}_l\rightarrow\vec{k}_m$, for which the conservation law of the quasiimpulse, satysfying to the relation
\begin{equation}
\label{eq31}
\begin{split}
\hbar\vec{k_l}- \hbar\vec{k_m} \mp \hbar\vec{q}_j = \vec{b}
\end{split}
\end{equation}
takes place. Here $\vec{b}$ is an arbitrary vector of a reciprocal lattice. Let us remember, that the thansitions with $\vec{b} = 0$ are normal transitions, if $\vec{b} \neq 0$, the thansitions are accompanying with Peierls transfer processes. It can be concluded, that FA-field phonons will be not satisfy in the general case to the expression (\ref{eq29}) and consequently they will not contribute to scattering processes. It follows from that that they will have the quasiimpulse distribution obeing to the   relation (eq12a in \cite{QAF})
in  which $\vec{k}$-space discretness is determined by the cavity length in the microwave field propagation direction. It is understandable that in the case of incommensurate cavity and lattice $\vec{k}$ space discretnesses all the generated hypersound phonons will have the energy and quasiimpulse characteristics, which will be not satisfying to the relation (\ref{eq29}). Let us remark, that the mechanism of  the creation of the system of coherent resonance hypersound phonons is described in the case  of the strong interaction of electrons with cavity FA-field phonons and cavity photons (the term "strong" means that practically all conducting electrons in a sample are involved in the given process directly or undirectly and they will not participate in the scattering processes with equilibrium own lattice phonons, see further)
in the work \cite{QFTDST}.

Simultaneously, the processes of electron scattering with the emission or the absorption of own lattice equilibrium phonons will be suppressed in the case considered. Really, a photon flux  incoming in the cavity in the microwave region is very intensive. So, the photon flux, corresponding to the microwave power in 100 mW, that is to the standard power of microwave generators in ESR-spectrometers without an attenuation, is equal to $\approx 10^{22}$ photons/s  of the frequency equaled to 10 GHz. It leads by a strong electron-photon interaction, realised by a spin subsystem of a sample electron system, to a very great number of cavity FA-field hypersound phonons.
At the same time, it is well known, that the  interaction which produces the energy difference
between the normal and superconducting phases in theory of superconductivity
 arises from the virtual exchange of phonons and
the screened Coulomb repulsion between electrons.
Other interactions are thought to be essentially
the same in both states, their effects thus cancelling in
the energy difference.
The Hamiltonian for the the electron-phonon interaction,
which comes from virtual exchange of phonons between
the electrons is the following \cite{BCS}
\begin{equation}
\label{eq32}
\begin{split}
&\mathcal{H}_{e-ph} = \\
&\hbar\sum_{\vec{k}\vec{k'}\sigma\sigma{'} \vec{q}}\frac{\omega_{\vec{q}}|M_{\vec{q}}|^2\hat{c}^{+}(\vec{k'}-\vec{q},\sigma{'})\hat{c}(\vec{k'},\sigma{'})\hat{c}^{+}(\vec{k}+\vec{q},\sigma)\hat{c}(\vec{k},\sigma)}{(\varepsilon_{\vec{k}} - \varepsilon_{\vec{k}+\vec{q}})^2 - (\hbar\omega_{\vec{q}})^2 }
\end{split}
\end{equation}
where $\varepsilon_{\vec{k}}$ is the Bloch energy measured relative to the
Fermi energy, $M_{\vec{q}}$ is matrix element for phonon-electron
interaction, $\hat{c}^{+}(\vec{k},\sigma)$, $\hat{c}(\vec{k},\sigma)$ are fermion  creation and
annihilation operators, based on the renormalized
Bloch states specified by quasiwave vector $\vec{k}$ and spin projection $\sigma$,
which satisfy the usual Fermi commutation relations. 
The most significant transitions are those for which  $|\varepsilon_{\vec{k}} - \varepsilon_{\vec{k}+\vec{q}}| \ll \hbar\omega_{\vec{q}}$. It is evident, that by the generation of the coherent system of resonance hypersound phonons the given condition will be easily fulfilled. The probality of the  given process is proportional to $\sum_{\vec{q}}$ in the expression (\ref{eq32}), that is, to the number of FA-field hypersound phonons wich can be produced in a very large amount. It gives clear proof for the participation of the conducting electron system in the materials with the strong electron-phonon and electron-photon interactions at ESR conditions the only in the virtual exchange processes, excluding the scattering processes.

 At the same time it is well known, that just the electron scattering  processes are the main processes determining the electrical resistance. Thus, it is the direct way for the formation of the superconducting states  in ESR conditions or another magnetic resonance conditions by room temperature and even by more high temperatures in conductors or semiconductors in which  the strong interaction of sample electrons with cavity FA-field phonons  is realized.
The given conclusion has the  experimental confirmation.   Peculiarities of spin-wave resonance observed in carbon nanotubes, produced by high-energy
ion beam modification of diamond single crystals in $\langle{100}\rangle$ direction \cite{Dmitri} 
allowed to insist on the formation in given nanotubes of
s+-superconductivity at room temperature, coexisting with
uncompensated antiferromagnetic ordering just in ESR conditions. The role of hypersound resonance phonons produced in ESR-spectrometer cavity, that is the phonons which can be referred to cavity FA-field quanta, in the formation of the superconducting state in a microwave frequency range is proposed in the cited work to be crucial".

Therefore, we wish to  emphasize once again, that  the observation of ESR analogues  of the ZFSs and  the Cherenkov radiation phenomenon, displaying both by the presence of  the steps in the range [$0.62 H_{1m}$ - $H_{1m}$] in Figures 31 to 33, and wakes in Figure 34 allows to insist, that the superconductivity state in the anthracite samples studied is really produced  at room temperature in ESR conditions in the accordance with the theory, presented in \cite{QAF}.

The additional argument in the favour of the given conclusion is the observation of the memory effects, above decribed.

\section{Conclusions} 

The predictions of new quantum phenomenon by QFT-model for dynamics of spectroscopic transitions were confirmed by
experimental radiospectroscopy studies of anthracites of
medium-scale metamorphism. The coherent system of
the resonance phonons, that is, the phonons with the energy, equaled to resonance photon energy, is really formed
by ESR, that leads to the appearance along with Rabi oscillations, determined by spin-photon coupling with the
frequency $\Omega^{RF}$,  of quantum Rabi oscillations with the frequency $\Omega_{Ph}$, determined by spin-phonon coupling. Its value was
found on the first  damping stage to be equal 920.6 kHz in the samples studiedbeing to be independent on the microwave power level in the range 20 - 6 dB [0 dB corresponds to 100  mW].
The second quantum Rabi oscillation process was  observed by the stationary state of the subsystem - \{microwave EM-field plus 
spin subsystem\} and it was separated, which was achieved by setting up the
high frequency modulation amplitude of the static magnetic
field to be equal zero. 

 By the  subsequent increase of  the microwave power the stepwise transition to the new quantum phenomenon takes place,  just to the  phenomenon of nonlinear quantum  Rabi oscillations, characterised by splitting of the oscillation  group of lines  into two subgroups with doubling of the  total  lines' number and approximately the twofold decrease of linewidth of an  individual oscillation line. It is became to be equal the only  to $0.004 \pm 0.001$ G, that is, oscillation lines in the nonlinear process of the resonance phonon generation are the most narrow lines, being  to be registered by  the stationary ESR spectroscopy at all.  

At nonzeroth 100 kHz modulation amplitude of the static
magnetic field along with the absorption process of EM-field
energy the strongly coherent emission process was observed, at that the absorption process was observed by the
sweep of the static magnetic field from lower to higher values
($\frac{dH_0}{dt} > 0$)) and  the emission process was observed by $\frac{dH_0}{dt} < 0$),
that is, by the sweep of the static magnetic field from higher to
lower values in a regime with an automatic frequency control. The emission process is the realisation of  the quantum acoustic spin resonance.

We accentuate once again,  that all quantum field effects aforegoing are observed in the radiospecroscopy for the first time. Especially interesting,
that quantum acoustic Rabi oscilllations were registered
on  the conventional ESR spectrometer for   stationary signal
studies. 

The creation of a longlived coherent phonon subsystem was achieved in AF registration regime by means
of rather great deviation of resonance frequency, equaled
to $\approx{1}$ MHz, which is appeared when the absorption  is rather strong, it is the necessary condition for the formation of a sample cavity in a measuring cavity. The sufficient condition for the formation of a sample cavity is that, that  the all selective (resonance) photons, incoming in the sample,  are remained in the sample space.
    
It has been found, that the lifetime of the coherent state of the
collective subsystem of resonance phonons in anthracites
of medium-scale metamorphism is very long and, even by the
room temperature, it is evaluated in more than 4.6 minutes.
The value of the spin-phonon coupling constant was found
in anthracite samples studied to be equal $\approx{4}\times{10^3} s^{-1}$.

The evalution of the characteristic times for the formation
of the resonance phonon system gives the value $\lesssim{10^{-10}}$   s.
It seems to be more long (from order up to two orders
of the values) in comparison with $\tau$-process in t-PA \cite{Weinberger}, \cite{Nechtschein} and in carbon NTs \cite{Ertchak}, \cite{Ertchak_Stelmakh}.

Radiospectroscopic analogue of Stokes shift, which is well
known in optical luminescence studies, has been identified in the radiospectroscopy for the first time.

The model of instantons of two new kinds is  argued.
The instantons of the first kind absorb radiospectroscopy
range photons and relax mainly by means of the generation of coherent hypersound phonons. The given prevailing coherent relaxation process determines rather narrow ESR-linewidth (equal to 0.18 G) in structurally disordered anthracite samples studied, being  to be the  direct consequence of  the coherence of  the main relaxation mechanism. The instantons of the second kind absorb hypersound phonons and
relax mainly by the  generation of coherent microwave field  photons. The instantons of both the kinds can free, that is,
without an additional activation energy, "propagate" along
$t$-axis in Minkowski space like to SSH-solitons along $R_3$ Euclidian space $z$-axis. The very
short value of spin-lattice $\tau$-relaxation $\lesssim{10^{-10}}$  s, which
determines the time of the formation of the  resonance phonon
subsystem, is strong experimental support in the favour
of the theoretical model discussed.

The proof that the superconductivity state in the anthracite samples studied is  produced  at the  room temperature in ESR conditions in the accordance with the theory, presented in \cite{QAF} has  experimentally been  obtained.

The phenomenon of the formation of the coherent system of the resonance phonons can find the number of
practical applications, in particular it can be used by the 
elaboration of logic quantum systems including quantum
computers and quantum communication systems.

\section{Acknowledgment}
      
The authors are  grateful to F.Borovik for the help in the work.


\begin{thebibliography}{}
\bibitem{PSPA} Dmitri Yerchuck, Yauhen Yerchak, Alla Dovlatova, Vyacheslav Stelmakh and Felix Borovik, The Photon Concept and the Physics of Quantum Absorption Process,  British Journal of Applied Science and Technology,\textbf{6}, N 5 (2015) 425-462, DOI: 10.9734/BJAST/2015/12244
\bibitem {Dovlatova_Yearchuck} Dovlatova A, Yearchuck D, QED model of resonance phenomena in quasionedimensional multichain
qubit systems, Chem.Phys.Lett., \textbf{511} (2011) 151-155, DOI:10.1016/j.cplett.2011.05.039
\bibitem{Slepyan_Yerchak} Slepyan G.Ya, Yerchak Y.D, Hoffmann A,  Bass F.G, Strong electron-photon coupling in a one-dimensional quantum dot chain: Rabi waves and
Rabi wave packets, Phys.Rev.B, \textbf{81} (2010) 085115 (1-18), DOI:10.1103/PhysRevB.81.085115
\bibitem{Yerchuck_D_Dovlatova_A} Yerchuck D, Dovlatova A, Quantum Optics Effects in Quasi-One-Dimensional and
 Two-Dimensional Carbon Materials, J.Phys.Chem.C, \textbf{116}, N 1 (2012) 63-80, DOI: 10.1021/jp205549b 
\bibitem{Gromov} Moritz K\"{a}lin, Igor Gromov, and Arthur Schweiger, The continuous wave electron paramagnetic resonance experiment revisited, Journal of Magnetic Resonance \textbf{160} (2003) 166–182
\bibitem{Scully} Scully M O, Zubairy M S, Quantum Optics, Cambridge University Press, 1997, 650 pp
\bibitem{Dovlatova_Yerchuck} Alla Dovlatova and Dmitri Yerchuck, Concept of Fully Dually Symmetric Electrodynamics, Journal of Physics: Conference Series, \textbf{343}, 012133, DOI:10.1088/1742-6596/343/1/012133
\bibitem{DAA} Dmitri Yerchuck, Alla Dovlatova, Andrey Alexandrov, Boson Model of Quantized EM-Field and Nature of Photons,  2nd International Conference on Mathematical Modeling in Physical Sciences 2013, 
Journal of Physics: Conference Series, \textbf{490} (2014) 012102-8 DOI:10.1088/1742-6596/490/1/012102
\bibitem{ABE} Dmitri Yerchuck, Yauhen Yerchak, Alla Dovlatova, Steve Robert, Felix Borovik, Quantum Field Theory of Aharonov-Bohm Effect, in press
\bibitem{Slater} Slater J C, Nature, \textbf{113} (1924) 307-308
\bibitem {SSH} Su W.P., Schrieffer J.R, and Heeger A.J, Solitons in Polyacetylene, Phys.Rev.Lett., \textbf{42} (1979) 1698-1701 
\bibitem {SSH_PRB} Su W.P., Schrieffer J.R, and Heeger A.J, Soliton Excitations in Polyacetylene,  Phys.Rev.B, \textbf{22}, (1980) 2099-2111
\bibitem{Heeger_1988} Heeger A.J, Kivelson S, Schrieffer J.R, Su W.-P., Solitons in conducting polymers, Rev.Mod.Phys., \textbf{60} (1988)  781-850
\bibitem{Berman} Berman P R, ed., Cavity Quantum Electrodynamics
Academic, New York, 1994
\bibitem{Nielsen} Nielsen M A and I. L. Chuang I. L., Quantum Computation
and Quantum Information, Cambridge University Press,
Cambridge, 2001
\bibitem{DiVincenzo} Loss D and DiVincenzo D P, Quantum computation with quantum dots, Phys.Rev.A, \textbf{57}   (1998) 120-126 
\bibitem{DiVincenzoD} DiVincenzo D P, The physical implementation of quantum computation, Fortschr.Phys., \textbf{48}  (2000) 771-783 
\bibitem{Tyryshkin} Tyryshkin A M, Morton J J L, Benjamin S C, Ardavan A et al, Journal of
Physics: Condens.Matter, \textbf{18} (2006) 783 
\bibitem{Petta} Petta J R et al, Coherent manipulation of coupled electron spins in
semiconductor quantum dots, Science, \textbf{309} (2005) 2180-2184 
\bibitem{Nowack} Nowack K C et al, Single-shot correlations and two-qubit gate of solid-state
spins, Science, \textbf{333}  (2011) 1269-1272 
\bibitem{Shulman} Shulman M D et al, Demonstration of entanglement of electrostatically
coupled singlet-triplet qubits, Science, \textbf{336} (2012) 202-205 
\bibitem{Koppens} Koppens F H L, Nowack K C, and Vandersypen L M K,
Phys.Rev.Lett., \textbf{100} (2008) 236802 
\bibitem{BluhmH} Bluhm H, Foletti S, Mahalu D, Umansky V and Yacoby A, Enhancing the coherence of a spin qubit by operating it as a feedback loop that controls its
nuclear spin bath, Phys.Rev.Lett., \textbf{105} (2010) 216803 
\bibitem{Maune} Maune B M et al, Coherent singlet-triplet oscillations in a silicon-based
double quantum dot, Nature, \textbf{481} (2012) 344-347 
\bibitem{Bluhm} Bluhm H et al, Dephasing time of GaAs electron-spin qubits coupled to a nuclear bath exceeding 200 $\mu$s, Nature Phys., \textbf{7} (2011) 109-113 
 \bibitem{Gaebel} Gaebel T, Domhan M, Popa I et al, Nat.Phys. \textbf{2} (2006) 408
\bibitem{Bar-Gill} Bar-Gill N, Pham L M, Jarmola A, Budker D,  Walsworth R L, Solid-state 
electronic spin coherence time approaching one second, Nature Commun., \textbf{4}(2013) 1743 (
\bibitem{TyryshkinA} Tyryshkin A M et al, Electron spin coherence exceeding seconds in high-purity 
silicon, Nature Mater., \textbf{11} (2012) 143–147 
\bibitem{Dzurak} M. Veldhorst, J. C. C. Hwang, C. H. Yang, A. W. Leenstra, B. de Ronde, J. P. 
Dehollain, J. T. Muhonen, F. E. Hudson, K. M. Itoh, A. Morello, A. S. Dzurak, An addressable quantum dot qubit with fault-tolerant control-fidelity, Nature Nanotechnology (2014) doi:10.1038/nnano.2014.216
\bibitem{Morello} Juha T. Muhonen, Juan P. Dehollain, Arne Laucht, Fay E. Hudson, Rachpon Kalra, 
Takeharu Sekiguchi, Kohei M. Itoh, David N. Jamieson, Jeffrey C. McCallum, 
Andrew S. Dzurak, Andrea Morello, Storing quantum information for 30 seconds in 
a nanoelectronic device, Nature Nanotechnology, 2014, DOI: 10.1038/nnano.2014.211 
\bibitem{Witzel} Witzel, W. M., Carroll, M. S., Morello, A., Cywinski, L.  Das Sarma, S, 
Electron spin decoherence in isotope-enriched silicon, Phys.Rev.Lett., 105 (2010)
187602 
\bibitem{Steger} Steger, M. et al, Quantum information storage for over 180 s using donor spins 
in a 28Si ‘semiconductor vacuum’, Science, 336 (2012) 1280–1283  
\bibitem{Saeedi} Saeedi, K. et al, Room-temperature quantum bit storage exceeding 39 minutes 
using ionized donors in silicon-28, Science, 342 (2013) 830–833 
\bibitem{Alvarez} Alvarez, G. A,  Suter, D, Measuring the spectrum of colored noise by dynamical 
decoupling, Phys.Rev.Lett., 107, 230501 (2011) 
\bibitem{QFTDST} Alla Dovlatova and Dmitri Yerchuck,  ISRN Optics, 2012, Article ID 390749, 10 pages, DOI:10.5402/2012/390749
\bibitem{Erchak} Erchak D.P, Efimov V.G, Zaitsev A M, Stelmakh V.F,Penina N.M, Varichenko V.S, Tolstykh V P, Nucl.Instr.Meth in Phys.Res.B, \textbf{69} (1992) 443-451 
\bibitem{Dyson} Dyson F D, Phys.Rev.,\textbf{98} (1955) 349-359
\bibitem{Jaynes_Cummings} Jaynes E T and Cummings F W, Proc.IEEE \textbf{51}, (1963) 89 
\bibitem{Tavis} Tavis M,  Cummings F W, Phys.Rev., \textbf{170}(2) (1968) 387
\bibitem{Ertchak_J_Physics_Condensed_Matter} Ertchak D.P, Kudryavtsev Yu.P, Guseva M.B, Alexandrov A.F, Evsyukov S.E, Babaev V G, Krechko L.M, Koksharov Yu.A, Tikhonov A.N, Blumenfeld L.A, v.Bardeleben H J,  J.Physics: Condensed Matter, \textbf{11}, N3 (1999) 855-870 
\bibitem{Ertchak} Ertchak D.P, Efimov V.G, Stelmakh V.F, J.Appl.Spectroscopy, \textbf{64}, N 4 (1997) 433-460
\bibitem{Ertchak_Stelmakh} Ertchak D.P, Efimov V.G, Stelmakh V.F,  Martinovich V.A, Alexandrov A.F, Guseva M.B, Penina N.M, Karpovich I.A, Varichenko V.S, Zaitsev A M, Fahrner W.R,
Fink D, Phys.Stat.Sol.b, \textbf{203} (1997) 529-547
\bibitem{Yearchuck_Yerchak_Dovlatova} D.Yearchuck, Y.Yerchak, A.Dovlatova,  Optics Communications, \textbf{283} (2010) 3448-3458
\bibitem{Yearchuck_Doklady} Yearchuck D, Yerchak Y, Red'kov V, Doklady NANB, \textbf{51}, N 5 (2007) 57-64
\bibitem{Yearchuck_PL} Yearchuck D, Yerchak Y, Alexandrov A, Phys.Lett.A,  \textbf{373}, N 4 (2009) 489-495
\bibitem{Cummings} Cummings F W,  Phys.Rev.,\textbf{140} (1965) A1051
\bibitem{Eberly} Eberly J H, Narozhny N B, Sanchez-Mondragon J J, Phys.Rev.Lett., \textbf{44} N 20 (1980) 
1323-1326
\bibitem{Erchak_Zaitsev_Stel'makh} Erchak D P, Zaitsev A M, Stel'makh V F, Tkachev V D, Phys.Tekhn.Polupr.,  \textbf{14}, N 1 (1980) 139 -143, Sov.Phys.Semicond., USA, \textbf{14}, N 1 (1980) 79 - 82
\bibitem{Yerchak_Stelmakh} Fedoruk G G, Rutkovskii I.Z, Yerchak D.P, Stelmakh V.F,  Zh.Experiment.Teor.Fiz., \textbf{80}, N 5 (1981)  2004-2009, Fedoruk G G, Rutkovskii I.Z, Erchak D.P, Stel'makh
V.F, Sov.J.Exp.Theor.Physics, USA, \textbf{53}, N 5 (1981)
1042-1044
\bibitem{Weger} Weger M, The Bell System Technical Journal, July 1960, 1014-1112
\bibitem{Radjaraman} Radjaraman R, Solitons and Instantons in Quantum Field Theory, M., 1985
\bibitem{Abragam} Abragam A.I, The Principles of Nuclear Magnetism,
Oxford Univ.Press, London, 1961, Nuclear Magnetism,
Iz.In.Lit., M., 1963, 551 pp
\bibitem{Weinberger} Weinberger B R, Ehrenfreund E, Pron A, Heeger, A.J,
MacDiarmid A G, J.Chem.Phys., \textbf{72} (1980) 4749-4755
\bibitem{Nechtschein} Nechtschein M, Devreux F, Greene R I, Clarke T C,
Street, G B Phys.Rev.Lett., \textbf{44} (1980) 356-359
\bibitem{Ph.Enc.} Physical Encyclopedia,(Editor in-Chief Prokhorov A.M) v.5, p.339, M., 1998, 688 pp
\bibitem{Born_M} Born M,  Z.f.Phys. \textbf{38} (1926) 803-827; Born M, Das Adiabatenprinzip in der Quantenmechanik, \textbf{40} (1926) 167; Born M,  Experiment and Theory in Physics, Cambridge University Press, 1943, p.23
\bibitem{PQMD} Dmitri Yerchuck,  Alla  Dovlatova, Felix Borovik, Yauhen Yerchak, Vyacheslav Stelmakh, To Principles of Quantum Mechanics Development,  International Journal of
Physics, \textbf{2} N 5 (2014) 129-145, DOI: 10.12691/ijp-2-5-2
\bibitem{Shaltiel} D.Shaltiel, H.A.Krug von Nidda, B.Ya.Shapiro, A.Loidl, T. Tamegai, T.Kurz, B.Bogoslavsky, B.Rosenstein
and I.Shapiro, Investigating Superconductivity with Electron Paramagnetic Resonance (EPR)
Spectrometer, pp 63-82, In: Superconductors - Properties, Technology, and Applications, Dr.Yury Grigorashvili (Ed.), InTech, 2012, 436 pp
\bibitem{D_Shaltiel} D.Shaltiel, H-A.Krug von Nidda, B.Rosenstein, B.Ya.Shapiro, M.Golosovsky,
I.Shapiro, A.Loidl, B.Bogoslavsky, T. Fujii, T. Watanabe, and T. Tamegai, Field cooling memory effect in Bi2212 and Bi2223 single crystals,
arXiv:0907.5484v1 [cond-mat.supr-con] 31 Jul 2009
\bibitem{Struik} L. C. E. Struik, Physical Aging in Amorphous Polymers and Other Materials, Elsevier, Amsterdam, (1978); A. Boutet de Monvel, A.Bovier, Spin Glasses: Statics and Dynamics, (Eds.),
Summer School, Series: Progress in Probability, 61, Paris (2007);
H. Nishimori, Statistical Physics of Spin Glasses and Information Processing: An Introduction
(International Series of Monographs on Physics, N 111), Schpringer (2009);
 L. Lundgren, P. Svedlindh, P., Nordblad, O. Beckman, Phys.Rev.Lett., \textbf{51} 1983) 911; 
M. Lederman, R. Orbach, J.M. Hammann, M. Ocio,  E. Vincent, Phys.Rev.B, \textbf{44} (1991) 7403–7412;
G. F.Rodriguez, G. G. Kenning, and R. Orbach, Phys.Rev.Lett.,\textbf{91} (2003) 037203; 
J.P.Bouchaud, J.Phys.(Paris) I 2, 1705 (1992);
 R. G.,Palmer, D. L., Stein, E. Abrahams, P.W. Anderson, Phys.Rev.Lett., \textbf{53} (1984) 958;
 \bibitem{Repain} V. Repain, M. Bauer, J.P. Janet, J. Ferre, A. Moujin, C. Chapert, and H Bernas, Europhys.Lett. \textbf{68} (2004) 460 
\bibitem{Vaknin} A. Vaknin and Z. Ovadyahu, Phys.Rev.Lett.\textbf{84} (2000) 3402; V. Orlyanchik and Z. Ovadyahu,
Phys.Rev.Lett., \textbf{92} (2004) 066801 
\bibitem{Du} X. Du, G.H. Li, E.Y. Andrei, M. Greenblatt and P. Shuk, Nature Physics, \textbf{3} (2007) 111;   
Z. L.Xiao and E. Y. Andrei, and M. J. Higgins, Phys.Rev.Lett., \textbf{83} (1999) 1664;
\bibitem{Henderson} W. Henderson, E. Y. Andrei, M. J. Higgins, and S. Bhattacharya, Phys.Rev.Lett., \textbf{77} (1998) 2077; W. Henderson, E. Y. Andrei, and H. J. Higgins, Phys.Rev.Lett., \textbf{81} (1998) 2352;
N.R. Dilley, J. Herrmann, S. H. Han, and M. B. Maple, Phys.Rev.B, \textbf{56} (1997) 2379; 
G.Ravikumar, V. C. Sahni, P. K. Mishra, T. V. Chandrasekhar Rao, S. S. Banerjee, A. K.Grover, S. Ramakrishnan, and S. Bhattacharya, M. J. Higgins, E. Yamamoto, Y. Haga, M.Hedo, Y. Inada, and Y. Onuki, Phys. Rev. B \textbf{57} (1998) R11069; 
M.N. Kunchur, S.J. Poonand M.A. Subramanian, Phys. Rev. B, \textbf{41} (1990) 4089 
\bibitem{Valenzuela} S.O. Valenzuela and V. Bekeris, Phys.Rev.Lett., \textbf{84} (2000) 4200 
\bibitem{Blatter} G. Blatter, M. V. Feigel’man,V.B. Geshkenbein, A.I. Larkin,  V.M. Vinokur, Rev. Mod.Phys., \textbf{66} (1994) 1125-1388 
\bibitem{Barone} A. Barone and G. Paterno, Pinning and Applications of the Josephson Effect, John Wiley and Sons, New York, 1981, p.279
\bibitem{Ustinov} A.V. Ustinov, Solitons in Josephson junctions, Physica D, \textbf{123} (1998) 315-329
\bibitem{Parmentier} R.D. Parmentier, in: K. Lonngren, A.C. Scott (Eds.), Solitons
in Action, Academic Press, New York, 1978, p.173
\bibitem{McLaughlin} D.W. McLaughlin, A.C. Scott, Phys.Rev.A, \textbf{18} (1978) 1652
[\bibitem{Pedersen}] N.F. Pedersen, in: S.E. Trullinger, V.E. Zakharov, V.L.
Pokrovsky (Eds.), Solitons, Elsevier, Amsterdam, 1986,
p. 469.
\bibitem{ParmentierR} R.D. Parmentier, in: H, Weinstock, R.W. Ralston (Eds.),
The New Superconducting Electronics, Kluwer, Dordrecht,
1993, p. 221.
\bibitem{Fulton} T.A. Fulton, R.C. Dynes, Sol.St.Commun., \textbf{12} (1973) 57
\bibitem{Kivshar} Yu.S. Kivshar, B.A. Malomed, Phys.Rev.B, \textbf{37} (1988) 9325
\bibitem{Mints} R.G. Mints, I.B. Snapiro, Phys.Rev.B, \textbf{52}  (1995) 9691
\bibitem{Kurin} V.V. Kurin, A.V. Yulin, Phys.Rev.B\textbf{55}  (1997) 11659
\bibitem{Goldobin} E. Goldobin, A. Wallraff, N. Thyssen, A.V. Ustinov, Cherenkov Radiation in Coupled Long Josephson Junctions, Phys.Rev.B, \textbf{57} (1998) 130-133
\bibitem{Kim} Y.W. Kim, A.M. de Graaf, J.T. Chen, E.J. Friedman, and S.H. Kim, Phase Reversal and Modulated Flux Motion in Superconducting Thin Films,
Phys.Rev.B, \textbf{6}, N 3   (1972) 887-893
\bibitem{ME} Dmitri Yerchuck, Alla Dovlatova, Yauhen Yerchak, Felix Borovik
Analytical Solution of Homogeneous Damped Mathieu Equation,
International Journal of Physics, \textbf{2}, N 5 (2014)  165-169
DOI:10.12691/ijp-2-5-6
\bibitem{QAF} Dmitri Yerchuck, Felix Borovik, Alla Dovlatova
and Andrey Alexandrov, Concept of Free Acoustic Field,  British Journal of Applied Science and Technology
\textbf{4} (15) (2014) 2169-2180
\bibitem{Dmitri} Dmitri Yerchuck, Yauhen Yerchak, Vyacheslav Stelmakh, Alla Dovlatova, Andrey Alexandrov, J.Supercond.Nov.Magn.,(2013) DOI 10.1007/s10948-013-2300-7
\bibitem{BCS} Bardeen J, Cooper L N, Schrieffer J.R, Phys.Rev., \textbf{108}, N 5 (1957)
1175–1204 
\end{thebibliography}
\end{document}